\documentclass[11pt]{article}
\usepackage[utf8]{inputenc}
\usepackage{amsmath,amssymb,amsthm}
\usepackage{graphicx}
\usepackage{booktabs}
\usepackage{longtable}
\usepackage{natbib}
\usepackage{xurl}
\usepackage{hyperref} 
\renewcommand{\harvardurl}[1]{\textbf{URL:} \url{#1}}
\hypersetup{colorlinks=true, linkcolor=blue, citecolor=blue, urlcolor=blue}
\usepackage{geometry}
\usepackage{setspace}
\usepackage{caption}
\usepackage{subcaption}
\usepackage{enumitem}
\usepackage{bbm}

\title{\Large{Fairness Drift: \\Assessing Algorithmic Fairness in a Moving Market}}
\author{Meng Liu\thanks{Olin Business School, Washington University in St. Louis. Email: mengl@wustl.edu}}

\date{November 8, 2026}

\begin{document}

\maketitle
\thispagestyle{empty}

\begin{abstract}
\noindent
Flexible models and richer data can reduce systematic valuation errors and improve fairness. Yet algorithmic values stay fixed between updates while the market moves, so a new algorithm gains most against an intact distortion and least where uneven appreciation has eroded or just reversed it. We study property assessment, which sets tax bills and has long over-assessed cheaper homes. Cook County replaced township regressions and manual adjustments with a gradient-boosted model and richer data on its triennial cycle: Chicago in 2021, northern suburbs in 2022, southern in 2023. We estimate difference-in-differences in how the assessment-to-price ratio varies with sale price and neighborhood minority share. In Chicago, the reform halved the relative over-assessment of cheaper homes and, among equally priced homes, weakened the link to neighborhood racial composition. In the suburbs the price gradient was unchanged: the North’s legacy values were already accurate, and the South’s cheap neighborhoods had appreciated faster under frozen values, correcting part of the distortion, then stalled in the reassessment year, leaving the new values too high where prices had gained and accuracy unimproved. Fairness thus drifts with the market between updates, and single-deployment evaluations may credit or blame the algorithm for what the market did.

\end{abstract}


\newpage
\pagenumbering{arabic}
%======================================================================
% NEW (2026-09-17). Revised introduction for the new framing:
%   algorithm x reassessment cycle x market.
% Replaces Section 1 of the May 2026 submission.
%
% Citation keys: all keys match references.bib (updated 2026-09-19).
% Section labels used (rename to yours): sec:literature, sec:setting, sec:data,
%   sec:chicago, sec:robustness_durability, sec:mechanism, sec:market, sec:conclusion.
% Kept from the old introduction: the dollar illustration (with its footnote) and the
%   footnote on New York City and Maricopa County.
% Numbers marked % VERIFY should be checked against the current main tables.
%
% Working title discussed earlier:
%   Algorithmic Fairness in a Changing Market: Evidence from Cook County's
%   Property Assessment Reform
%======================================================================

\section{Introduction}
\label{sec:introduction}

Organizations increasingly use algorithms in setting prices, values, and scores: what a home is worth, how credible a borrower is, what a host should charge. A central promise of this shift is fairness, where a model that predicts more accurately should make fewer of the systematic errors that fall on the same group of individuals time after time. Whether that promise is kept is an open question, and the evidence is divided: some studies find that algorithms reduce disparities when they replace noisier human rules \citep{kleinberg2018human, HoffmanKahnLi2018, zhang2021frontiers, BartlettEtAl2022}; others find that they reproduce or create them \citep{lambrecht2019algorithmic, ObermeyerEtAl2019, fuster2022predictably}.

Most of this evidence answers a yes-or-no question: was the algorithm fairer than the process it replaced? That framing misses an important feature of many real-world algorithms: the algorithmic outputs (e.g., prices or scores) are set on one date and left in force until the next update, while the market they describe keeps moving. Property assessments, essentially a price that determines property tax bills, are revised on fixed yearly or multi-year cycles; insurance premiums and loan terms are set by a model at issuance and held fixed for the policy or loan term; and the credit-score models that U.S. mortgage lenders were required to use dated from the early 2000s and remained in force until their replacements were approved in 2022.\footnote{See FHFA, ``FHFA Announces Validation of FICO 10T and VantageScore 4.0 for Use by Fannie Mae and Freddie Mac,'' October 24, 2022, \url{https://www.fhfa.gov/news/news-release/fhfa-announces-validation-of-fico-10t-and-vantagescore-4.0-for-use-by-fannie-mae-and-freddie-mac}, last accessed on 10/06/2026.} Between updates the algorithmic values stand still, but the market does not. When some segments of the market move faster than others, the gap between assigned values and market values opens or closes on its own, and the next update is trained on the movement that occurred. Algorithmic fairness can therefore drift\footnote{Fairness drift, a term \citet{davis2025fairness} introduce for disparities that emerge after a model is deployed, is a phenomenon anticipated by work on how deployed models fail under shifting data \citep{finlayson2021clinician, singh2021fairness, schrouff2022diagnosing, ghosh2022faircanary}.} after its deployment, in either direction, with no change in the algorithm itself. Similarly, an algorithmic reform can arrive to find its fairness work already done by the moving market, or find that the segment trends it learned from have just turned, so that the values it sets are stale on the day they take effect. We show that \textit{fairness drift} can decide whether an algorithm appears to deliver the fairness it promises.

We study the question in a setting where the price is a tax. The property tax is the largest source of tax revenue for local governments in the United States, and each owner's bill is proportional to the value an assessor assigns to the home. The tax is fair when homes of equal market value receive equal assessed values. In practice, lower-priced homes are assessed at a higher fraction of their market value than higher-priced homes, so their owners pay higher effective tax rates \citep{berry2021reassessing, amornsiripanitch2020residential}, and Black and Hispanic homeowners face higher assessments relative to market value than white homeowners in the same jurisdiction \citep{avenancio2022assessment}. The poor pay more \citep{Caplovitz1963}, and here the higher price is not the result of where people shop, but rather produced by a valuation system.

Cook County, Illinois, is an unusually informative place to study whether a better valuation system fixes this. Until 2018 the county valued homes with township-level linear regressions and extensive, undocumented manual adjustments, and its assessments were shown to be strongly regressive \citep{GrottoEtAl2017, berry2021reassessing}. A new assessor took office in 2018 and introduced a machine-learning valuation system built on gradient-boosted trees, richer property and geographic data, and a documented, publicly released pipeline. Because the county reassesses one third of its territory (a \emph{triad}) each year, the reformed algorithm reached the Chicago triad in 2021, the North triad in 2022, and the South triad in 2023. We therefore observe the same system, run by the same office, deployed three times, in housing markets that differed in how distorted their assessments were and in how their prices were moving. 

%We also observe the two reassessments that the new administration carried out in 2019 and 2020 under its earlier township-level workflow, before the reformed algorithm, which let us separate the algorithm from the administration that introduced it. Our data run through 2025, so we also observe the second round under the reformed algorithm, Chicago in 2024 and the North in 2025.

%The three deployments did not produce the same result. In Chicago 2021, the reform removed about half of the long-standing over-assessment of lower-priced homes, and the gain persisted and survived appeals. In the North and South suburbs, the same system had no detectable effect on price regressivity. This paper explains why. Two conditions outside the algorithm turn out to matter. The first is what the algorithm met when it arrived: the North's assessments were already far more accurate, so there was less to correct. The second is what the housing market did between updates. Assessed values are frozen for three years, and in the South suburbs inexpensive neighborhoods appreciated so much faster than expensive ones during 2018--2022 that their assessment ratios fell before the reform arrived; the same boom shaped the data the reformed algorithm learned from in 2023, and when it ended, the reformed algorithm predicted no better than the reassessment it replaced. The same boom left Chicago's distortion intact and erased part of the South's; the same algorithm then met a full distortion in one place and a partly corrected one in the other.

Our empirical design is built around the object that matters for fairness. The question is not whether the reform changed the average ratio of assessed value to sale price, but whether it changed how that ratio varies with the price of the home and with the racial composition of the neighborhood. We estimate an event-study difference-in-differences (DiD) model in slopes. The outcome is the log ratio of assessed value to sale price, and the objects of interest are the \emph{price regressivity}, the relationship between that ratio and the log sale price, and the \emph{minority slope}, the relationship between that ratio and the neighborhood minority share among homes of the same price.\footnote{Our preferred measure of racial disparity is neighborhood minority composition rather than owner race. The reason is substantive: in a historically segregated housing market, algorithmic valuation is most likely to propagate inequities through place-based patterns across neighborhoods, not through direct treatment of individual owners by race.} A negative price regressivity means that inexpensive homes are over-assessed relative to expensive ones. The specification includes township-by-quarter fixed effects, which absorb all local shifts in the level of assessments, together with township-specific and quarter-specific baselines for both slopes, which allow the price and minority slopes to differ permanently across townships and to move over time in ways common to all townships. The treatment effect is therefore a DiD in slopes: the change in the within-township price slope (or minority slope) in treated triad after the algorithmic reform, relative to the change over the same quarters in the triads that had not yet been reassessed under the reformed algorithm.

We report three sets of findings.

\textbf{First, the algorithmic reform sharply reduced inequality in Chicago, and the reduction lasted.} Before the reform, in 2020, Chicago's price slope was about $-0.43$. The 2021 algorithmic reassessment raised it by 0.216 on average over the following two years ($p = 0.03$ in the joint price-and-race specification and $0.08$ in the price-only specification under our preferred triad-level wild bootstrap; $p<0.01$ with township-clustered standard errors), which removes about half of the pre-reform regressivity. The change appears immediately and does not fade. As an illustration of the effect size, after the algorithmic reform, a \$200,000 home would be assessed about 19.4 percent (or \$39,000) lower relative to a home one log point more expensive (about \$543,000), which at Chicago's 2022 effective tax rate translates into a reduction of the tax bill by about \$660 a year.\footnote{Here we use Chicago's estimated 2022 residential effective property tax rate of 1.69\%. See \url{https://www.civicfed.org/Effective_Property_Tax_2013_2022}, last accessed on 03/25/2026. The figure is an illustration, not an estimate of anyone's bill. Our estimate identifies only the relative change between the two homes, so the illustration attributes all of it to the lower-priced home, holding the higher-priced home's assessment, the tax rate, and all exemptions fixed; if part of the compression came instead from raising the assessments of expensive homes, the saving to the cheaper home would be smaller and the burden would shift upward rather than fall.} The within-price minority slope, negative before the reform, also moved toward zero, although that estimate is less precise under the preferred inference. The estimates are nearly identical when we use post-appeal values, so the appeal process does not undo the fairness gain. The effect is robust to various alternative explanations --- it is not produced by the new administration's other changes, or by the pandemic housing market, or alternative minority measures.

\textbf{Second, the Chicago fairness gain came from better prediction, which removed two forms of pooling in the legacy system.} First, the legacy model estimated township-specific linear regressions and applied the common coefficients to all homes in a township, thereby pulling predictions toward the middle and over-assessing inexpensive homes. Second, the legacy system captured location through coarse neighborhood averages, which means that equally priced homes may exhibit different assessment ratios depending on the respective neighborhood averages they are pooled with, thereby giving rise to the minority slope in the case of a segregated housing market. The reform's more flexible machine learning model and finer geography data together relax both forms of pooling: in Chicago the mean absolute error of assessed values fell from 40.0 to 28.0 percent; fine geography accounts for a much larger share of the variation in assessed values after the reform; and the reductions in error are largest at the ends of the price and minority-share distributions, where pooling toward averages had erred most.

The North show the other side of the same mechanism: its assessments were already far more accurate (a mean absolute error of 17.1 percent) and only half as regressive; the 2022 reassessment lowered the error only to 15.9 percent, and the price slope did not move meaningfully.

\textbf{Third, the same algorithm yielded no fairness gain in the South in 2023, and the moving market explains most of why: uneven appreciation corrected part of the distortion before the reform arrived, and then reversed, turning against the values the algorithm had learned.} In 2017 the South triad was as regressive as Chicago (a price slope of $-0.46$), yet its 2023 reassessment had no detectable effect on the price slope ($-0.01$, standard error 0.03). The market explains most of why. Between 2017 and 2022 the least expensive fifth of South neighborhoods nearly doubled in price while the most expensive fifth rose by about a quarter. With assessed values frozen between reassessments, this lowered the assessment ratios of inexpensive homes: the slope moved from $-0.46$ to $-0.38$ while the 2017 values were still in force, before any reassessment, and the 2020 reassessment under the earlier workflow added only 0.02 more. By the time the reformed algorithm arrived in 2023, part of what it could have corrected had already been corrected. The boom then did a second thing. It stalled in 2023 in the neighborhoods that had been rising fastest. The reformed algorithm values each neighborhood at roughly the level its prices have reached, so its 2023 values stood above what homes in those neighborhoods then sold for: the values are too high exactly where prior appreciation had been fastest. In Chicago, where fast-appreciating neighborhoods kept rising, the same algorithm left them under-assessed instead. The direction of the model's error was set by what the market did next, not by the model. The 2023 reassessment also added no accuracy: the 2020 reassessment under the earlier workflow had already brought the South's mean absolute error to 26.7 percent, close to the 28.0 that Chicago's reformed algorithm reached, and the 2023 values left it there.

This paper's main contribution is to research on whether algorithms reduce or reproduce inequality \citep{kleinberg2018human, lambrecht2019algorithmic, zhang2021frontiers, raymond2023market}. We provide causal evidence that replacing a simple-model-based, discretionary valuation process with a machine-learning system reduced both price-based and race-based inequality in the assignment of a tax burden. In addition, because we observe the same system deployed three times, we can speak to why it did so once and not twice. This turns two findings that the literature has so far reported separately into \textit{one account}: algorithms are fair-making relative to the process they displace \citep{zhang2021frontiers, HoffmanKahnLi2018}, and the market an algorithm operates in can generate or erase disparities without any change in the algorithm \citep{lambrecht2019algorithmic}. 

More specifically, our findings extend the organizing concept, fairness drift, in three ways. First, we identify a source. In a valuation system, fairness drift may simply follow from the fact that values are held fixed while the segments they describe appreciate at different rates, and its sign can therefore be predicted from relative price movements alone. Second, we show that drift and reform interact through the baseline. A favorable drift, as in the South suburbs, lowers the disparity that the next reassessment will be measured against, so that part of what the reform might have accomplished has already happened by the time it is deployed. Third, we show that drift and reform interact through the training sample as well. The price movements that generate the drift are the same movements the reformed algorithm learns from, and the values it sets are right only if those movements persist: where they reversed, as in the South, the fresh values over-assessed the neighborhoods that had been rising fastest, and where they continued, as in Chicago, the same algorithm under-assessed them. The second and third extensions together mean that the measured effect of one deployment reflects the state of the market as much as the quality of the algorithm, and a study that observes only one deployment cannot separate the two. We do not claim that these conditions fully determine an algorithm’s effect on fairness. In this setting we measure the first channel, and the tilt the second channel leaves in the values, directly at the level of neighborhoods; why the 2023 reassessment added no accuracy is supported by reconstructed models rather than identified. Together the two channels account for how a large effect in one place became none in another.

The mechanisms travel to any setting that shares three features of ours: a model assigns values or prices to heterogeneous segments; the values it sets are evaluated against, and trained on data shaped by, the values and prices in force when it arrives; and its outputs are held fixed between refreshes while the segments move at different speeds. Automated home valuation by lenders and online platforms, insurance and credit pricing, platform pricing tools, and the customer-value and propensity scores that firms re-estimate on a schedule all have this structure. Our findings suggest that the fairness effect of adopting a model will depend on the inequality in force when the model arrives, that between refreshes the direction of any drift in who is over- and under-valued will be set by the moving market rather than by the model, and that a model which prices segments at the level they have reached will err in opposite directions depending on whether their trends persist.

These findings also bear on how algorithms are kept in check. The audit regimes now written into law index their timing to the calendar or to changes in the system: New York City requires an annual bias audit of hiring algorithms\footnote{New York City Local Law 144 of 2021, in force from January 1, 2023 and enforced from July 5, 2023; an automated employment decision tool may be used only if an independent bias audit was conducted within the prior year. See \url{https://regulations.ai/regulations/RAI-US-NY-NYCL1XX-2021}, last accessed on 10/12/2026.}; Colorado's 2024 act requires an impact assessment at deployment, annually, and within 90 days of a substantial modification (its 2026 replacement drops the assessment altogether)\footnote{Colorado Senate Bill 24-205, C.R.S. 6-1-1703, as enacted in 2024 (\url{https://content.leg.colorado.gov/sites/default/files/2024a_205_signed.pdf}). It was repealed and replaced by Senate Bill 26-189, signed May 14, 2026 and effective January 1, 2027, which removes the impact-assessment and risk-management duties and centers on notice, explanation of adverse outcomes, and human review (\url{https://leg.colorado.gov/bills/sb26-189}). Both last accessed on 10/21/2026.}; and the European Union demands a new conformity assessment only when the system itself is substantially modified, alongside a provider's post-market monitoring plan that collects performance data over the system's lifetime without specifying what should trigger a review\footnote{Regulation (EU) 2024/1689 (AI Act), Article 43(4) on new conformity assessments after substantial modification and Article 72 on post-market monitoring. See \url{https://ai-act-service-desk.ec.europa.eu/en/ai-act/article-43} and \url{https://ai-act-service-desk.ec.europa.eu/en/ai-act/article-72}, last accessed on 10/21/2026.}. Court-ordered remedies range from one-time validation of a new score\footnote{\emph{Louis v. SafeRent Solutions}, No. 1:22-cv-10800 (D. Mass.), settlement approved November 20, 2024; any new screening score must be validated by a third party before use. See \url{https://clearinghouse.net/case/45888/}, last accessed on 10/12/2026.} to compliance reports every four months\footnote{\emph{United States v. Meta Platforms}, No. 1:22-cv-05187 (S.D.N.Y.), settlement of June 2022; Meta reports to an independent reviewer every four months against numeric compliance metrics, with court oversight through June 2026. See \url{https://www.justice.gov/usao-sdny/pr/united-states-attorney-implements-groundbreaking-settlement-meta-platforms-inc-formerly}, last accessed on 10/12/2026.}. None of these indexes audit timing to movement in the market the algorithm prices; the closest, the European Union's post-market monitoring, asks for continuous data collection but leaves the trigger for a review unspecified. Our evidence shows what that omission costs: in the South suburbs the price slope moved by nearly a fifth of its level inside one cycle in which no assessed value changed, and an audit of the same algorithm would have found a large fairness gain in Chicago in 2021 and none in the suburbs in the two following years. Overall, our findings suggest that fairness audits should run on the market’s clock, not the calendar’s: once low- and high-priced segments appreciate at different rates, fairness has moved, whether or not the algorithm has.

%Although the setting is a tax, the problem is one marketing has studied in other forms: a valuation is a price set by a model, the model is refreshed on a schedule, and the question is who pays more between refreshes. Customer-value and propensity scores, dynamic pricing tools, and automated home valuations share this structure, and the two-pooling mechanism and the two roles of the market apply to each.\footnote{Cook County's experience is not unique: other large assessment jurisdictions are already piloting or deploying related technologies. In 2025, New York City's Department of Finance awarded a six-month pilot to explore AI-based valuation of residential condominium properties (see \url{https://a856-cityrecord.nyc.gov/RequestDetail/20250304005}, last accessed on 03/25/2026). Maricopa County, Arizona has already rolled out---and continues to refine---Assessment Analyst GAMA, a cloud-based mass appraisal system that supports more than 1.8 million property assessments annually (see \url{https://resources.esri.ca/customer-stories/maricopa-county-goes-live-with-assessment-analyst-gama}, last accessed on 03/25/2026).}

\section{Literature Review}
\label{sec:literature}

This paper connects three bodies of work: research on the distributional effects of algorithms, including cases where the market around an algorithm, rather than the algorithm, produces an apparent bias; research on algorithms deployed in markets that keep moving, where fairness must be maintained after deployment rather than established once; and research on segmentation, targeting, and who pays more, in consumer markets and under the property tax.

\textbf{1. Algorithms' distributional effects and (apparent) algorithmic bias.}
A growing literature shows that algorithms can improve both the quality and the equity of decisions when they replace noisier or more distorted human rules \citep{kleinberg2018human, HoffmanKahnLi2018, BartlettEtAl2022, ascarza2022eliminating, raymond2023market, rambachan2024identifying, fu2025unequal, li2026hiring}. \citet{BartlettEtAl2022} find that algorithmic lenders discriminate substantially less than face-to-face lenders, and \citet{howell2024lender} find that automating loan processing raised lending to Black-owned firms, most where lenders' animus had been strongest. \citet{fu2021crowds} show that a machine-learning lender outperforms crowd investors and is less biased against disadvantaged borrowers, though not unbiased. \citet{zhang2021frontiers} find that Airbnb's pricing algorithm cut the revenue gap between white and Black adopting hosts, because demand for Black hosts' listings responded more to its lower prices, while the gap among all hosts widened because Black hosts adopted less. 

%\citet{gui2025llm} find that the adoption of large language models shrinks certain gaps in online consumer activism but not others.

A parallel literature shows that algorithms can also create or perpetuate disparities when they learn from distorted data, optimize a narrow objective, or operate in unequal environments, producing an apparent algorithmic bias, even when protected characteristics are never used \citep{lambrecht2019algorithmic, ObermeyerEtAl2019, KleinbergEtAl2018Fairness, levy2021algorithms, fuster2022predictably, zhang2024measuring, arnold2025building, bohren2025systemic, elzayn2025measuring}. \citet{lambrecht2019algorithmic} show that a gender-neutral job advertisement reached fewer women because advertising to women costs more; \citet{lambrecht2024apparent} show that search ads appeared more often after searches for Black names because the algorithm learns more slowly where data are sparser, a disparity that fades as data accumulate. \citet{cook2021gender} show that a gender-blind dispatch and pricing algorithm still produced a gender gap in rideshare earnings, entirely through where, when, and how fast drivers chose to drive. \citet{blattner2021costly} trace disparities in credit scores to noisier data for minority and low-income borrowers rather than to the model. \citet{ascarza2022eliminating} show that dropping protected attributes from a targeting model does not remove the disparity, because proxies carry it; \citet{chen2026algorithmic} show the same for pricing algorithms, where banning protected characteristics can leave the disparity in place while a parity requirement need not; and \citet{ahmadi2025protected} measure how much of a personalized score the protected attributes and their proxies explain.

We contribute to and extend this growing literature in two ways. First, these studies share a common structure: one deployment and one baseline. We observe one algorithm deployed against three baselines in three years. In Chicago it met the distorted legacy values that this literature expects an algorithm to correct; in the North it met values with little left to correct; and in the South it met values that the market had already partly corrected and then turned against. Importantly, the comparison of an algorithm with the process it is designed to replace has a third party, the market, and we find that the market movement between deployments decides what the comparison will find. Second, existing studies are mostly cross-sectional: the market around the algorithm shapes the disparity at a point in time. We put the same insight on a time axis. In our setting the market changes the disparity while the algorithm's output is literally frozen, and it also sets the direction of the error the next model makes, because the trend the model learned from its training sales either continues or reverses after the valuation date. We show that a deployed algorithm's signed error by neighborhood is predictable from what the market did between the end of its training data and the sale, in opposite directions in two regions of the same county.

\textbf{2. Algorithmic fairness as a maintenance problem in a moving market.}
Our paper also relates to recent studies of algorithms whose environment does not stand still. Housing is the leading example. Algorithmic home buyers expanded quickly during the 2020--2021 boom and then withdrew, in Zillow's case after large losses when prices stopped rising as its model expected; research on these firms emphasizes how hard it is to value heterogeneous homes quickly and how algorithmic valuations interact with market conditions \citep{BuchakEtAl2020, fu2022does, raymond2023market}. \citet{troncoso2026asymmetric} show that this visible failure of a valuation algorithm to anticipate a turn in prices changed how sellers used its estimates, and \citet{fu2025unequal} show that the same estimates are least accurate, and most valuable, in low-income neighborhoods. In marketing analytics, \citet{simester2020targeting} show in two field experiments that the advantage of flexible machine-learning targeting over simple methods disappears when the deployment period differs from the training period. \citet{brown2023competition} treat the frequency with which a pricing algorithm updates as a strategic choice with welfare consequences.

The moving market, while mattering greatly for efficiency and profitability of algorithms, also carries important implications for algorithmic fairness. Recent studies, mostly in computer science and medicine, show that fairness established at deployment need not persist, so that keeping an algorithm fair is a maintenance problem on top of a design problem. Static fairness criteria can harm the group they protect once outcomes feed back into the data \citep{liu2018delayed, damour2020fairness}, predictions change the distribution on which they are later evaluated \citep{perdomo2020performative}, and fairness guarantees that hold in the training population fail to transfer when that population shifts \citep{schrouff2022diagnosing}. \citet{davis2025fairness} follow clinical models over eleven years and name the phenomenon \emph{fairness drift}: disparities across groups that emerge after deployment through differential shifts in the data, and that population-level model updates can either restore or widen.

To our knowledge, ours is the first field measurement of fairness drift against realized market prices as the ground truth, where the yardstick is observed continuously, the drift's sign is predictable from relative price movements alone, and the market that produces the drift also corrects part of the disparity that the next deployment is measured against.

\textbf{3. Segmentation, targeting, and who pays more.}
Residential property assessment is a segmentation problem in a highly heterogeneous housing market. Marketing and economics have long treated segmentation and targeting as central to improving allocation decisions, and a broad literature shows that richer customer data and more flexible algorithms can improve targeting, ad delivery, and personalized pricing by identifying finer-grained heterogeneity across customers and submarkets \citep{rossi1996value, DubeMisra2023, smith2023optimal, narang2025privacy, wernerfelt2025estimating}. But that same literature also makes clear that segmentation is not automatically benign because outcomes depend on how segments are defined, what signals are used, and what objective the algorithm is asked to optimize \citep{dellavigna2019uniform, lambrecht2019algorithmic, iyer2024competitive, chen2026algorithmic}, which may widen a disparity that is already well documented: low-income households and some demographic groups often pay more for the same goods \citep{Talukdar2008, OrhunPalazzolo2019, jaravel2019unequal}.

Consumers can punish a price they find unfair by switching sellers \citep{dai2025online}; the property tax allows no such exit, and its one recourse, an appeal, is used least by lower-income and Hispanic owners \citep{nathan2025taxes}. Its fairness therefore rests on the valuation that sets it, and a large literature in public and urban economics shows that valuations violate the principle that homes of equal value bear equal tax. Assessments are regressive in price \citep{berry2021reassessing, amornsiripanitch2020residential, McMillenSingh2020}, part of the measured regressivity is mechanical when sale prices are noisy \citep{mcmillen2022measures}, and Black and Hispanic homeowners carry higher assessments relative to market value within the same jurisdiction, at every stage from valuation to appeal \citep{avenancio2022assessment, kahrl2024black, ihlanfeldt2021explaining, ihlanfeldt2023beyond, holz2024racial}. Two studies tie this inequity to the market's movement: Detroit's assessments turned sharply regressive as values fell faster than assessments \citep{hodge2017assessment}, and Philadelphia's move to frequent reassessment improved equity, unevenly across neighborhoods \citep{hou2025assessment}.

Most recently, \citet{smith2026tradeoffs} and \citet{candogan2023achieving} ask which assessment model is fairer on a fixed set of sales, the latter for the Cook County Assessor's own 2022 model, and \citet{berry2025evaluation} audits the outcomes of the Cook County reforms without a counterfactual. Our contribution is neither to document assessment inequality nor to propose a fairer model. It is to estimate causally whether a concrete change in the valuation system reduced inequality, and to show that the gain a more accurate model delivers depends on what the market did to the baseline before deployment and to the learned trend after it.

%%%%%%%%%%%%%%%%%%%%%%%%%%%%%%%%%%%%%%%%
\section{Institutional Setting, Data, and Model-free Evidence}
\label{sec:setting_data}
%%%%%%%%%%%%%%%%%%%%%%%%%%%%%%%%%%%%%%%%

%%%%%%%%%%%%%%%%%%%%%%%%%%%%%%%%%%%%%%%%
\subsection{Institutional Setting}
\label{sec:setting}
\subsubsection{The Legacy Assessment System (through 2018)}
\label{sec:old_system}

Prior to the algorithmic transition studied in this paper, Cook County employed a traditional assessment work flow: statistical models estimated on property sales were combined with staff review and a formal appeals process to produce assessed values. The core valuation step used linear hedonic regression models that relate recent transaction prices to observed housing characteristics and location attributes (e.g., size, age, and geography), a standard approach in both real estate economics and assessment practice \citep{rosen1974hedonic, wang2019mass, international2011standard}. The estimated coefficients—representing average marginal values such as price per square foot or value per bedroom—were then applied to unsold properties to generate
assessed values. 

However, regression models estimated on broad strata (e.g., large geographic groupings or broad property classes) can smooth over neighborhood-specific heterogeneity and help create and/or amplify systematic valuation errors: the model effectively used coefficients that reflected township-wide averages, which were disproportionately influenced by higher-valued properties. When these uniform coefficients were applied to lower-valued properties in economically distressed neighborhoods, the result was systematic overassessment—a key driver of the regressivity documented by the Chicago Tribune investigation \citep{GrottoEtAl2017}.

Another distinguishing feature of the legacy system was its heavy reliance on undocumented manual adjustments. After the initial model-based valuations were computed, assessors performed manual case-by-case adjustments to individual property assessments, without properly maintaining statistics on the frequency or magnitude of these adjustments, and the office declined Freedom of Information Act requests seeking documentation of the hand-check methodology \citep{GrottoEtAl2017}.

%\footnote{\citet{ChicagoTribuneFOIA2018} reports that roughly 3--5\% of properties were flagged as outliers and that analysts made changes to approximately 20--25\% of parcel-level valuation reports during the review stage.} 

Taken together, the legacy system provides an empirically important baseline: a standard regression-based valuation model that uses limited set of housing characteristics and spatial features, complemented by staff review and appeals, with distributional performance and transparency largely assessed through external studies and investigations.

%%%%%%%%%%%%%%%%%%%%%%%%%%%%%%%%%%%%%%%%
\subsubsection{The New Algorithmic Assessment System (2021+)}
\label{sec:newsystem}
\label{sec:new_algorithm}
%%%%%%%%%%%%%%%%%%%%%%%%%%%%%%%%%%%%%%%%

Beginning in 2019, the Cook County Assessor’s Office (CCAO) rebuilt its residential valuation workflow around an open-source automated valuation model (AVM) and a version-controlled production pipeline. The transition was gradual. For the 2019 (North) and 2020 (South) reassessments, the office's documentation records a township-level workflow in the predecessor codebase, with ``linear regression or GBM model per township'' \citep{CCAOModelResAVM}; these reassessments were conducted by the new administration but not with the reformed algorithm. By the 2021 reassessment cycle, first-pass values for single- and multi-family residential parcels were generated using a gradient-boosted decision tree ensemble (LightGBM).

\textbf{Model and data.} The core methodological change is the use of gradient-boosted decision trees \citep{friedman2001greedy,ke2017lightgbm}, which capture nonlinearities and interactions among characteristics and location without pre-specified functional forms, combined with a much richer feature set: standard parcel and building characteristics plus neighborhood and proximity measures such as Census/ACS covariates, walkability, flood risk, airport noise, and distance to transit.\footnote{See the ``Features Used'' and ``Data Sources'' sections in \citet{CCAOModelResAVM}.}

\textbf{Pipeline, post-modeling, and transparency.} The AVM runs as an end-to-end pipeline from data ingestion to training, evaluation, and reporting, with a limited set of documented post-modeling adjustments and a hand-review step before values are finalized \citep{CCAOModelResAVM,IAAO2025RatioStudies}.\footnote{The documented adjustments cover multi-card handling, prorated parcels, complex-level averaging for certain housing types, and rounding; see the ``Post-Modeling'' section of \citet{CCAOModelResAVM}.} The 2021 model also applied two post-modeling adjustments that the 2022--2025 models did not; we refer to them together as the 2021 post-adjustment step. The first divided the parcels of each township, property class, and number of units into quintiles of predicted value and, in each quintile with at least 15 sales in the previous three years covering at least 9 percent of its parcels, multiplied every parcel's prediction by one plus the median percentage gap between sale price and prediction among the sold parcels, capped at 40 percent; the second set the prediction of any parcel sold in the previous three years whose ratio of prediction to sale price fell below 0.7 or above 2 to that bound.\footnote{The grouping follows the valuation script on the 2021 branch of the office's repository and the \texttt{postval\_model} function of its R package, which group by township code, property class, and number of apartments; the README describes the same step as operating within ``neighborhood, modeling group, and estimated value quantile.''} The office describes these adjustments as designed to ``prevent regressivity in assessment'' \citep{CCAOModelResAVM}. Because the first shifts whole quintiles of predicted value within a township, it acts directly on the within-township price gradient that this paper estimates, and it is part of the 2021 package whose effect we measure. We cannot separate its contribution with the production system's stage-level predictions, which are not public. Section~\ref{sec:robustness_durability} instead uses Chicago's 2024 reassessment, which ran the reformed algorithm without the post-adjustment step, to ask whether it is necessary for the Chicago result. The office has published its modeling code and data since 2019 \citep{CCAO2019Release,GovTech2019OpenCode,CCAOOpenDataStory} and has added public tools that explain individual valuations \citep{CCAO2025Tool}.

\textbf{Implications for identification.} For our empirical design, the relevant ``treatment'' is not only a change in statistical learner, but a bundled modernization of (i) model form (LightGBM), (ii) feature engineering and data/pipeline infrastructure, (iii) documented post-modeling rules, including the post-adjustment step applied only in 2021, and review practices, and (iv) expanded transparency and reporting. Because these components move together within reassessment cycles, our estimates should be interpreted as the combined effect of this package rather than, for example, the isolated effect of switching from linear regression to LightGBM. Consistent with this interpretation, an external evaluation of Cook County’s residential assessments during 2019--2024 reports substantial reductions in regressivity relative to earlier years while noting remaining areas for improvement \citep{Berry2025KaegiEvaluation}.

\subsubsection{Other Important Institutions}

\textbf{Cook County's Triennial Reassessment Cycle} Cook County employs a triennial reassessment cycle, dividing the county into three geographic \textit{triads}: (1) Chicago (8 townships), (2) North suburbs (13 townships), and (3) South \& West suburbs (17 townships; hereafter ``South''). Each triad undergoes full reassessment every three years on a staggered schedule—Chicago in years divisible by 3 (e.g., 2015, 2018, 2021), North suburbs one year later (2016, 2019, 2022), and South/West suburbs two years later (2017, 2020, 2023). In non-reassessment years, assessed values carry forward from the most recent reassessment year within each property's three-year cycle.

\noindent \textbf{The COVID-19 adjustment (Tax Year 2020)}
In tax year 2020, the Cook County Assessor’s Office (CCAO) applied a one-time ``COVID-19 Adjustment'' in response to the anticipated economic and market effects of the COVID-19 pandemic \citep{CCAO2020CovidSouth,CCAO2021CovidNorthCity}. For Class 2 residential properties (that our sample fits into), the CCAO used estimated COVID-related changes in unemployment at the Census-tract level---constructed by combining industry-specific unemployment shocks with each tract’s industry employment mix---and mapped this analysis into neighborhood- and property-type-specific percentage reductions in estimated market value (and therefore assessed value) \citep{CCAO2020CovidSouth}. In the South triad (the 2020 triennial reassessment triad), the resulting reductions for single-family homes and condominiums had a median of 10.3\% and ranged from 8.0\% to 12.2\% \citep{CCAO2020CovidSouth}. For Chicago and the North suburbs, the CCAO reports reductions for single-family homes and condominiums ranging from 7.5\% to 12.3\%, with an average of 9.9\% \citep{CCAO2021CovidNorthCity}.\footnote{The COVID adjustment was computed and applied to tax-year 2020 assessed values. Because Cook County assessments are set on a triennial cycle, the resulting 2020 certified values can serve as the baseline carried forward in subsequent non-reassessment years (absent property-specific updates). This means that: (1) For the South triad that was scheduled for full triennial reassessment in 2020, the COVID adjustment was also reflected in the 2021 and 2022 assessment. (2) For the North triad that was reassessed in 2019, the COVID adjustment would change the 2020 and 2021 assessment values from the 2019 values, until it got reassessed again in 2022. (3) For the Chicago triad, the COVID adjustment was only applicable for the year 2020 because it got reassessed in 2021 with the reformed algorithm. }

%%%%%%%%%%%%%%%%%%%%%%%%%%%%%%%%%%%%%%%%
\subsection{Data and Model-free Evidence}
\label{sec:data}
%%%%%%%%%%%%%%%%%%%%%%%%%%%%%%%%%%%%%%%%
\subsubsection{Data and Sample}

Our analysis combines six data sources, all from the CCAO or the Census Bureau: (1) annual assessed values for all residential parcels, at three stages (mailed, certified after assessor-stage appeals, and Board of Review);\footnote{See \url{https://datacatalog.cookcountyil.gov/Property-Taxation/Assessor-Assessed-Values/uzyt-m557/about_data}, last accessed on 02/27/2026.} (2) sales transactions with prices, dates, and buyer and seller names;\footnote{See \url{https://datacatalog.cookcountyil.gov/Property-Taxation/Assessor-Parcel-Sales/wvhk-k5uv/about_data}, last accessed on 02/27/2026.} (3) year-specific property characteristics;\footnote{See \url{https://datacatalog.cookcountyil.gov/Property-Taxation/Assessor-Single-and-Multi-Family-Improvement-Chara/x54s-btds/about_data}, last accessed on 03/08/2026.} (4) appeals records with mailed and certified values, outcomes, and professional representation;\footnote{See \url{https://datacatalog.cookcountyil.gov/Property-Taxation/Assessor-Appeals/y282-6ig3/about_data}, last accessed on 02/27/2026.} (5) neighborhood-level COVID adjustments;\footnote{See \url{https://public.tableau.com/app/profile/ccao/viz/COVID19Dashboard_15906776254570/COVID19AdjustmentMap}, last accessed on 03/03/2026.} and (6) racial composition (\% Black and \% Hispanic) for each census block, block group, and tract, linked to parcels through the CCAO crosswalk.\footnote{See \url{https://datacatalog.cookcountyil.gov/Property-Taxation/Assessor-Parcel-Universe/nj4t-kc8j/about_data}, last accessed on 02/27/2026.} Our primary outcome uses mailed values, the direct output of the assessment model before appeals.

For the main analysis, we focus on properties that were ever sold in the sample period because our analyses require the sale price as the ground truth to compare the assessed value to.  The sample consists of arm's-length sales of single-family homes with valid assessed values in the year of sale, priced at least \$10,000 and below the 99.5th percentile, at the sale level (property by year-quarter); Appendix~\ref{app:sample} gives the construction steps. The observed sale price is specific to the property and quarter, while the assessed value is specific to the property and three-year reassessment cycle.

Our primary outcome variable is the log assessment ratio: $y_{it} = \ln\left(\frac{\text{Mailed Assessed Value}_{it} \times 10}{\text{Sale Price}_{it}}\right)$, where $i$ indexes properties and $t$ indexes time (year-quarter). The factor of 10 converts Cook County's 10\% assessment level to a full market value basis, making $y_{it}$ interpretable as the log ratio of estimated market value to the actual market value. This measure captures proportional assessment error: $y_{it} = 0$ indicates perfect accuracy, $y_{it} > 0$ indicates overassessment, and $y_{it} < 0$ indicates underassessment. The log ratio symmetrizes assessment errors and follows assessor practice and prior research \citep[e.g.,][]{avenancio2022assessment}. To prevent extreme values from affecting our results, we remove values below the 0.5th percentile or above the 99.5th percentile.

The main analysis sample (for the Chicago 2021 rollout study) contains 176,567 sales from 2018--2022\footnote{We choose 2018-2022 for the study of the Chicago 2021 effect because (1) 2018 is the start of the 3-year reassessment cycle of 2018-2020 for the Chicago triad, giving us one full cycle before the focal algorithm adoption in 2021; (2) We stop at 2022 because the South triad started to adopt the algorithm in 2023; (3) We further drop 2022 for North triad because it adopted the algorithm in 2022.} in 38 townships and 845 neighborhoods (Appendix Table~\ref{tab:summary_stats}). The mean log assessment ratio is $-0.151$ and the mean absolute assessment error is 28.8 percent. Chicago has the highest neighborhood minority share (59.5 percent) and the highest absolute error (33.5 percent); the North suburbs have the highest prices, the lowest minority share (18.4 percent), and the lowest error (20.4 percent); the South suburbs have the lowest prices and the most observations.

\subsubsection{Model-Free Evidence on Price and Race Regressivity}
\label{sec:modelfree}

\begin{figure}[!t]
\centering

% column headers
\begin{minipage}{0.49\textwidth}\centering \small\textbf{By Sale Price Quintile}\end{minipage}\hfill
\begin{minipage}{0.49\textwidth}\centering \small\textbf{By \% Minority Quintile}\end{minipage}

\vspace{0.4em}

\begin{subfigure}{0.49\textwidth}
    \centering
    \includegraphics[width=\linewidth]{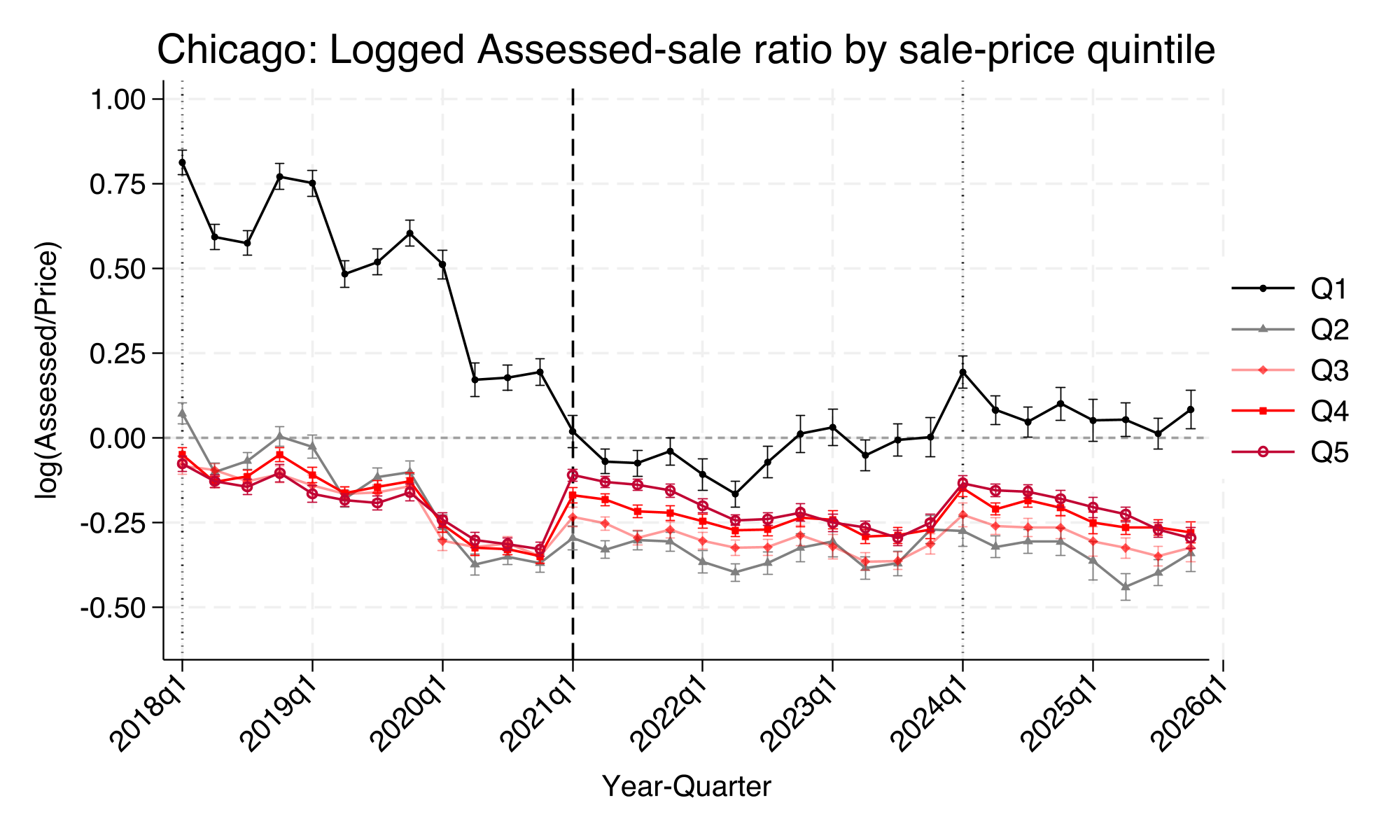}
    \vspace{-1.6em}
    \caption{Chicago Triad: price}
    \label{fig:modelfree_price_chicago}
\end{subfigure}\hfill
\begin{subfigure}{0.49\textwidth}
    \centering
    \includegraphics[width=\linewidth]{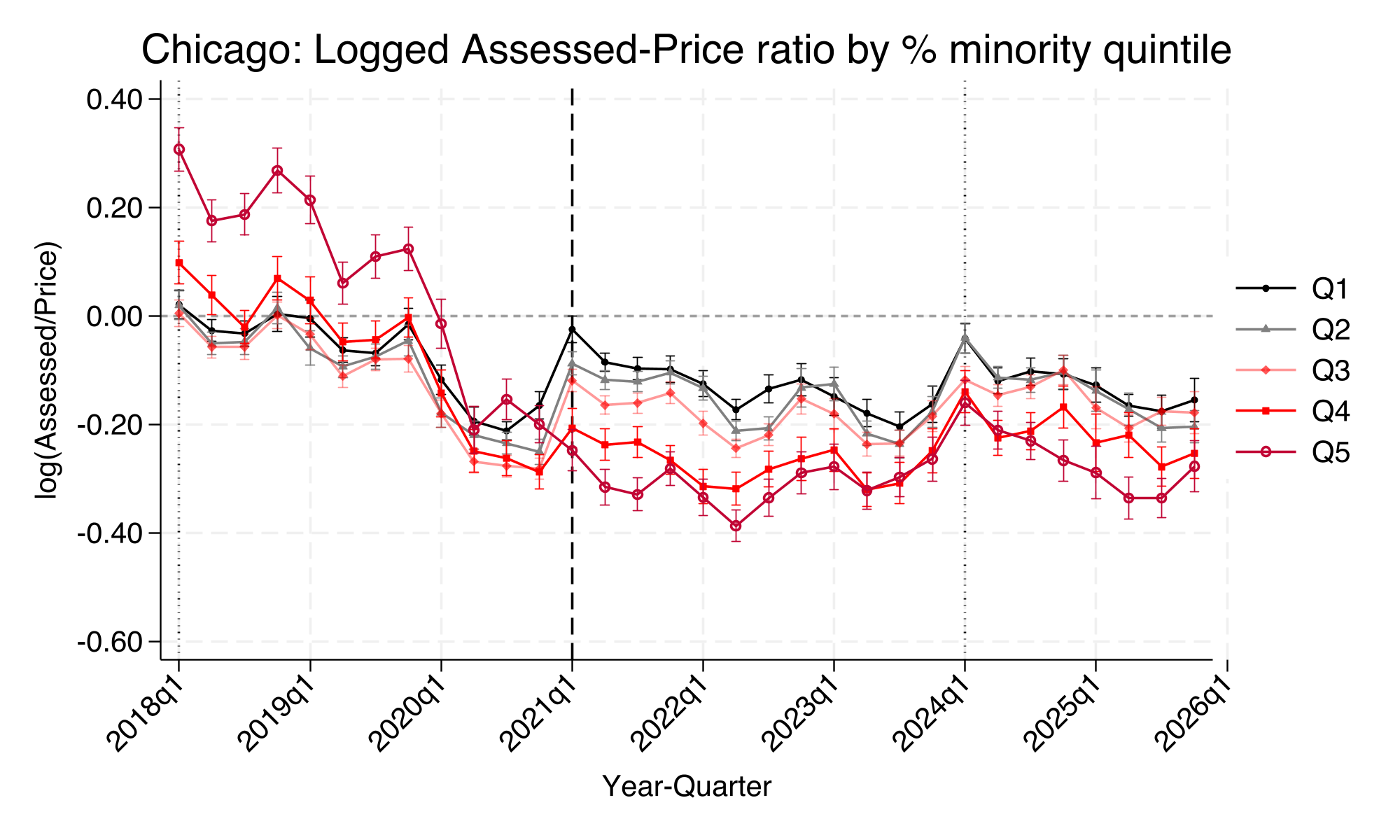}
    \vspace{-1.6em}
    \caption{Chicago Triad: race}
    \label{fig:modelfree_race_chicago}
\end{subfigure}

\vspace{0.6em}

\begin{subfigure}{0.49\textwidth}
    \centering
    \includegraphics[width=\linewidth]{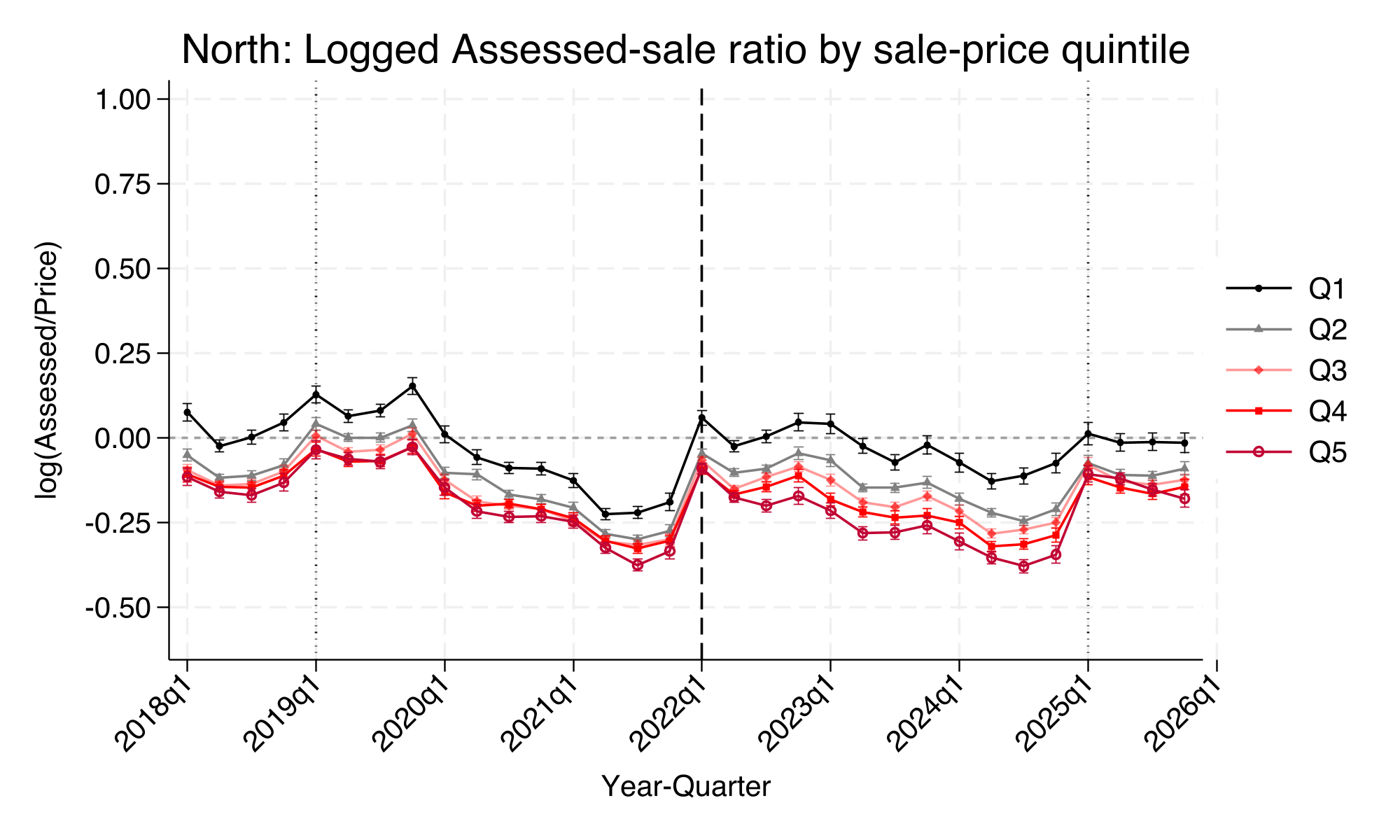}
    \vspace{-1.6em}
    \caption{North Triad: price}
    \label{fig:modelfree_price_north}
\end{subfigure}\hfill
\begin{subfigure}{0.49\textwidth}
    \centering
    \includegraphics[width=\linewidth]{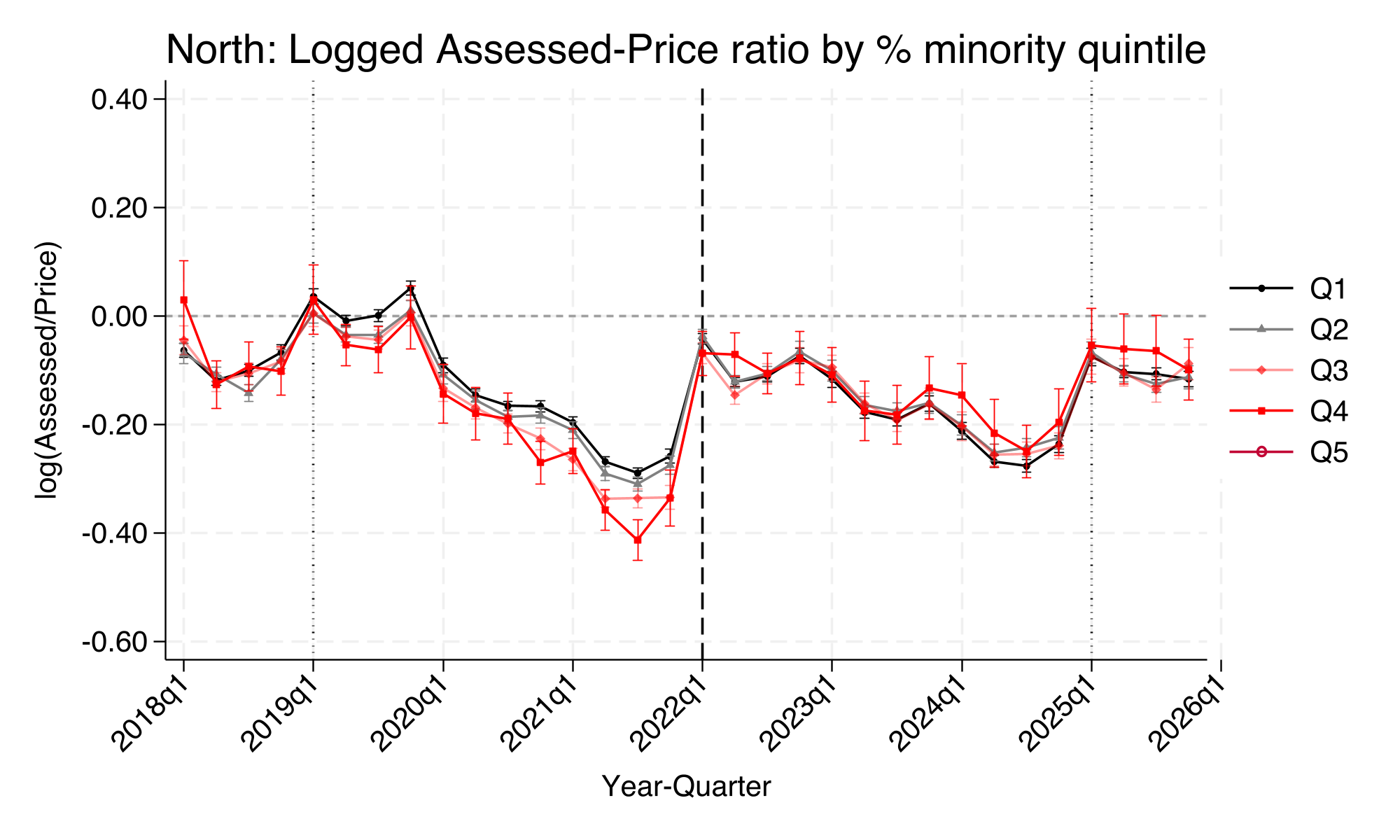}
    \vspace{-1.6em}
    \caption{North Triad: race}
    \label{fig:modelfree_race_north}
\end{subfigure}

\vspace{0.6em}

\begin{subfigure}{0.49\textwidth}
    \centering
    \includegraphics[width=\linewidth]{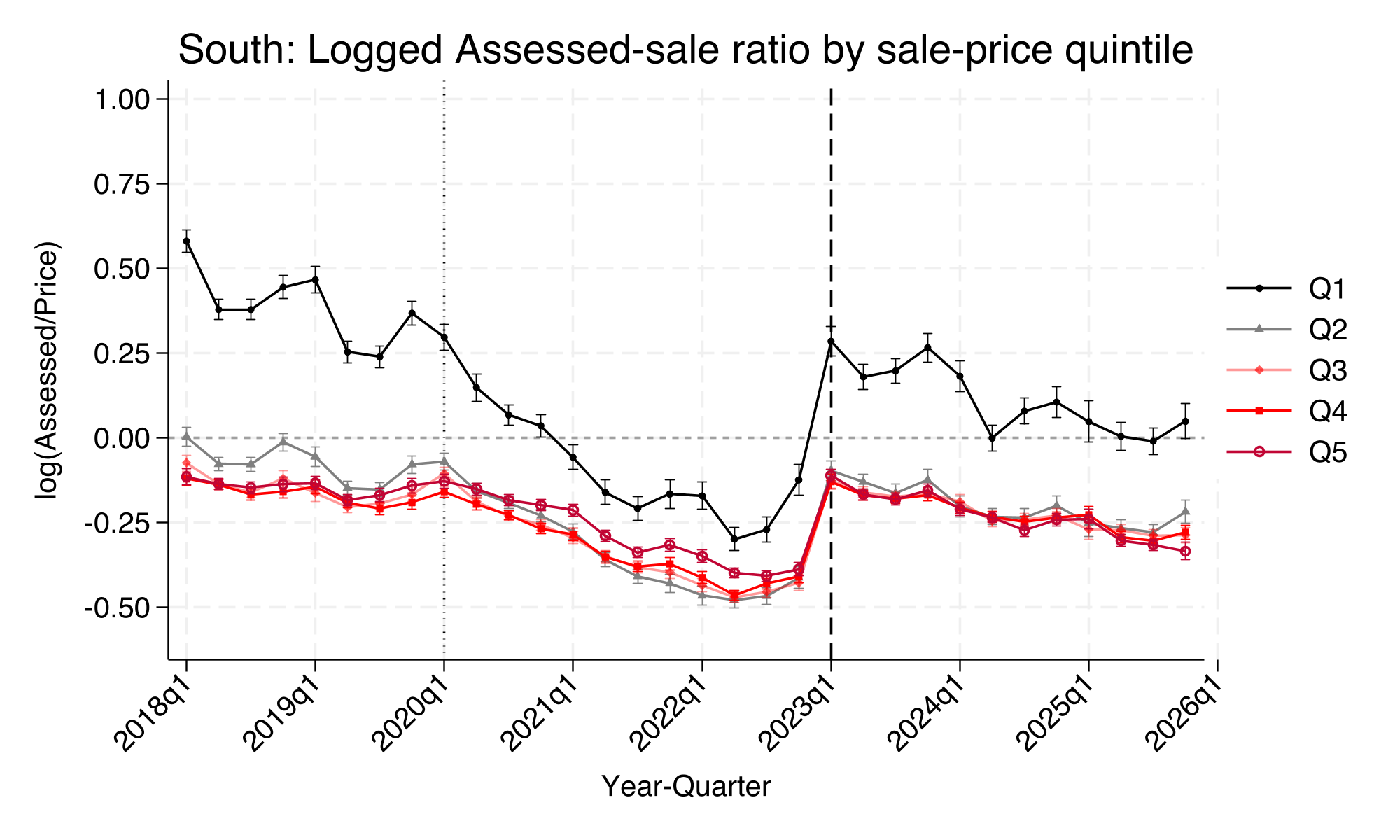}
    \vspace{-1.6em}
    \caption{South Triad: price}
    \label{fig:modelfree_price_south}
\end{subfigure}\hfill
\begin{subfigure}{0.49\textwidth}
    \centering
    \includegraphics[width=\linewidth]{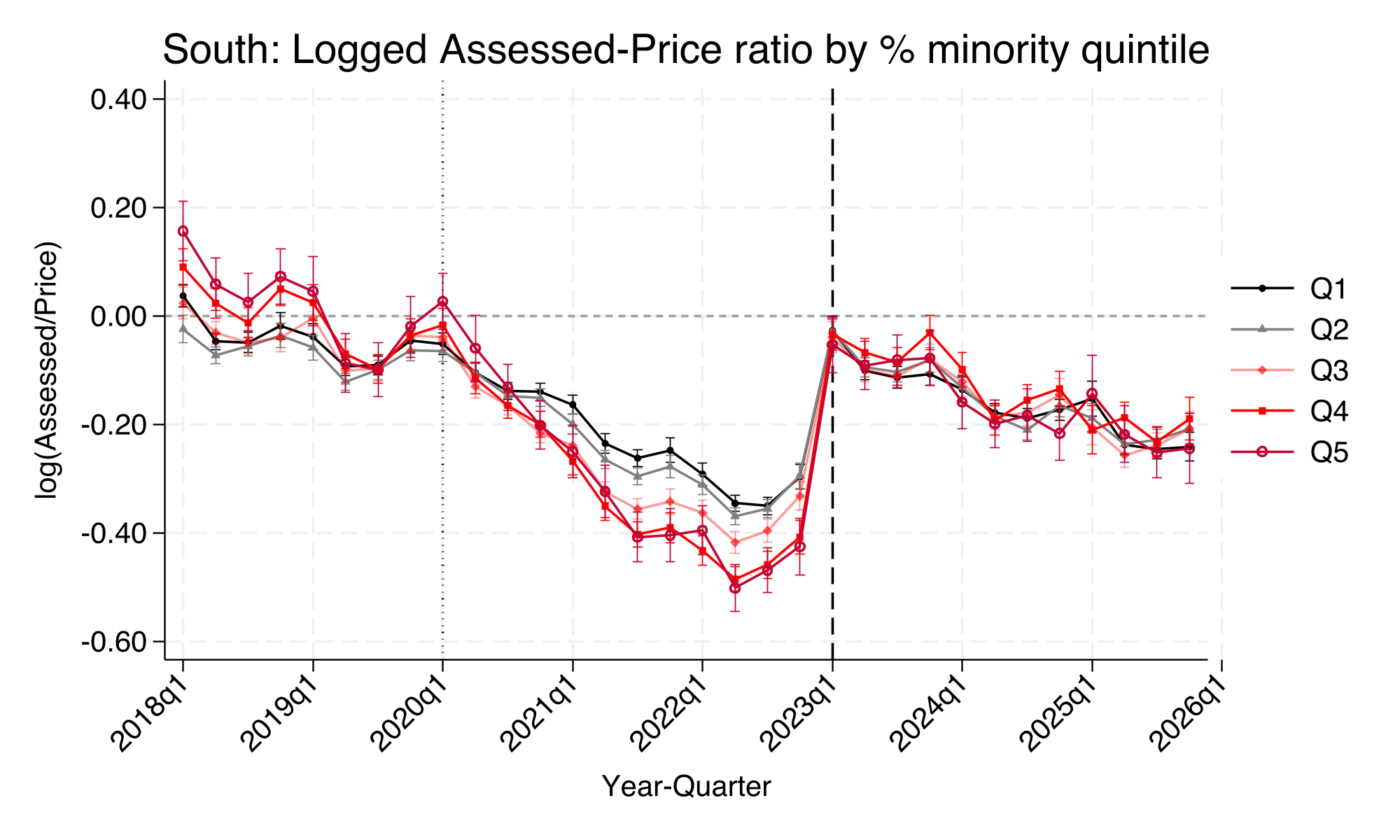}
    \vspace{-1.6em}
    \caption{South Triad: race}
    \label{fig:modelfree_race_south}
\end{subfigure}

\vspace{0.5em}
\caption{Model-Free Evidence: Assessment Regressivity by Sale Price and \% Minority Quintile}
\label{fig:modelfree}

\vspace{0.3em}
\begin{minipage}{\textwidth}
    \scriptsize \textit{Notes:} Each panel plots the mean of log(assessed/sale price) by quintile for arm's-length sales in a triad and year-quarter. Rows are ordered by the timing of each triad's algorithmic reassessment (Chicago 2021, North 2022, South 2023). Left column: sale price quintiles (Q1 = lowest 20\%, Q5 = highest 20\%), computed within each triad-quarter. Right column: quintiles of census block group \% minority (Q1 = lowest, Q5 = highest). Vertical dotted lines indicate reassessment years for each triad. The black dashed line (e.g., 2021 for Chicago) marks algorithmic reform adoption. Price quintiles are re-formed within each triad-quarter, so the set of homes in a price quintile changes across quarters. Minority quintiles are fixed: each CBG is assigned once to a quintile of the countywide CBG minority-share distribution, and the series track the homes sold in those CBGs each quarter. In both columns the series describe the homes sold in a quarter, not a fixed set of homes.
\end{minipage}
\end{figure}

Before turning to our DiD design, Figure~\ref{fig:modelfree} provides model-free evidence on how assessment ratios evolved across the Cook County reassessment cycles and around the 2021 algorithmic reform. Each point is the mean of $\log(\text{Assessed}/\text{Sale Price})$ for a group of arm's-length sales in a given triad and year-quarter. In the left column, sale-price quintiles are defined within triad-by-quarter cells, so that properties sold across years within the same triad are comparable. In the right column, we categorize Census Block Groups (CBGs) into quintiles of the countywide distribution of CBG minority share; a CBG's quintile is fixed over time.

\paragraph{Price-based patterns.}
Figure~\ref{fig:modelfree} (a), (c) and (e) show clear descriptive evidence of price regressivity in assessment ratios: within a triad-quarter, lower-priced sales generally have higher log assessed-to-sale-price ratios than higher-priced sales. In Chicago, this gradient is steep in 2018--2020, then compresses sharply around 2021 and remains materially smaller thereafter. Although the post-2021 pattern is not perfectly flat, the gap between the bottom and top of the price distribution is much narrower than in the earlier period, consistent with a substantial reduction in price-based overassessment.

The suburban triads exhibit the same basic price gradient, but their major breaks seem to more align with their own reassessment timing than Chicago's 2021 reform. In the South triad, the spread narrows after a pronounced shift through 2020--2022 and remains smaller after the 2023 reassessment. In the North triad, the most visible changes are level shifts at the 2022 and 2025 reassessments, which move all quintiles together; the spread across quintiles changes little. This timing pattern is consistent with the interpretation that the sharp post-2021 compression in Chicago reflects Chicago-specific institutional change rather than a common countywide break in 2021.\footnote{Vertical lines mark reassessment-cycle boundaries visible in the 2018--2025 window; the thick dashed line marks the first cycle under the algorithmic reform for that triad: Chicago in 2021, North in 2022, and South in 2023.}

The marked movements around 2020 are visible across multiple quintiles and all triads, which likely reflect the broad countywide COVID adjustments. As discussed in Section \ref{sec:regressivity}, we address these level shifts using fixed effects at appropriate levels. 

\paragraph{Race-based patterns.}
Figure~\ref{fig:modelfree} (b), (d), and (f) show a parallel descriptive pattern by neighborhood racial composition. In Chicago, higher-minority neighborhoods have substantially higher assessment ratios in 2018--2019, but this gradient compresses sharply by 2020, likely reflecting the COVID adjustments that favored minority neighborhoods that are considered more impacted by COVID. The pattern reverses from 2021 onward: the highest-minority quintile generally lies below the lowest-minority quintiles, implying that the post-reform pattern is \emph{race-progressive} in the descriptive sense that higher-minority neighborhoods face lower assessment ratios on average. This unconditional ordering also reflects the price composition of minority neighborhoods, which are concentrated at the lower end of the price distribution; the DiD estimates below condition on price and therefore measure a different, narrower object.

The South triad shows a similar, though more gradual, decline in racial dispersion, with much less separation across minority-share quintiles after the 2023 reassessment. By contrast, the North triad displays smaller and noisier racial gradients throughout, with less stable separation across quintiles. Taken together, the descriptive evidence aligns with the main DiD results shown in later sections: the largest and most temporally distinctive compression in both price and race gradients occurs in Chicago around the 2021 reform, while the suburban triads mainly shift at their own reassessment dates.\footnote{The weaker racial gradients in the North triad are also consistent with more limited variation in neighborhood minority share and greater quarter-to-quarter noise in the plotted series.}

%%%%%%%%%%%%%%%%%%%%%%%%%%%%%%%%%%%%%%%%
\section{Effect of the Algorithmic Reform: Chicago 2021}
\label{sec:regressivity}
%%%%%%%%%%%%%%%%%%%%%%%%%%%%%%%%%%%%%%%%

%%%%%%%%%%%%%%%%%%%%%%%%%%%%%%%%%%%%%%%%
\subsection{Difference-in-Differences Design: Estimating Changes in Slopes}
\label{sec:did_design}
%%%%%%%%%%%%%%%%%%%%%%%%%%%%%%%%%%%%%%%%

The model-free figures highlight two empirical challenges for identifying the effects of Cook County’s assessment reforms. First, assessment ratios exhibit substantial \emph{level} movements at triad reassessment boundaries and during the COVID period, and these level shifts are not synchronized across space (Figure~\ref{fig:modelfree}). Second, the patterns of interest are not primarily changes in average assessment ratios, but changes in \emph{gradients}: how the log assessment ratio varies with sale price (price regressivity) and with neighborhood racial composition (race regressivity). These considerations motivate a DiD design that isolates changes in slopes while absorbing local level shocks.

\textbf{Outcome and gradients.}
For each arm’s-length sale $i$ in township $j$ and quarter $t$, we study $y_{it} \equiv \log\!\left(\frac{A_{it}}{P_{it}}\right)$, where $A_{it}$ is the assessment (i.e., mailed assessed value $\times$ 10) and $P_{it}$ is the transaction price. We define $p_{it}\equiv \log(P_{it})$ and let $m_{i}$ denote the minority share (the Black plus Hispanic share of the population, expressed as a fraction between 0 and 1) of the property’s CBG, so that a slope coefficient on $m_i$ is the change in the log assessment ratio associated with moving from an all-white to an all-minority neighborhood.\footnote{Our baseline uses CBG-level minority share, but the main results are robust to coarser and finer geographies (Census tract and Census block, respectively).} A negative relationship between $y_{it}$ and $p_{it}$ indicates price regressivity (lower-priced sales having higher assessment ratios), while a positive relationship between $y_{it}$ and $m_{it}$ indicates higher assessment ratios in higher-minority neighborhoods.

\textbf{Identification.}
Let $\text{Treat}_j$ indicate the treated geography (i.e., property sales in Chicago townships) and let $\{\mathbbm{1}[\tau=t-k]\}_{k}$ denote event-time indicators relative to the reform quarter (normalized so that $k=0$ is the first post-reform quarter and $k=-1$ is the omitted reference period). The central estimating equation models \emph{changes} in the price and race gradients as:
\begin{align}
y_{it} \;=\;& \alpha_{jt} 
+ \big(\beta_j + \beta_t\big)\,p_{it}
+ \big(\phi_j + \phi_t\big)\,m_{i} \nonumber\\
&+ \sum_{k\neq -1} \delta_k\;\text{Treat}_j \cdot \mathbbm{1}[\tau=t-k]\cdot p_{it}
+ \sum_{k\neq -1} \rho_k\;\text{Treat}_j \cdot \mathbbm{1}[\tau=t-k]\cdot m_{i}
+ \varepsilon_{it}.
\label{eq:did_slopes}
\end{align}

We include township-by-quarter fixed effects $\alpha_{jt}$ to absorb all local level shocks to assessment ratios (including reassessment-cycle jumps and COVID-era level shifts), so that any slope estimates should be thought of as the regressivity across homes within the township. This choice of within-township regressivity, aligning with other studies (e.g., \citealp{avenancio2022assessment}), is reasonable because homes within the same township face the same public amenities and many local tax regulations are made at the township level. We also allow the \emph{baseline} price and race gradients to vary across townships ($\beta_j,\phi_j$) and across calendar quarters ($\beta_t,\phi_t$), so identification comes from deviations in slope dynamics for treated townships relative to controls, not from assuming common gradients across space or time. 

The key identifying assumption is a \emph{parallel trends} condition for slopes: absent the reform, the evolution of the price gradient and race gradient in treated townships would have followed the same path as in control townships, after accounting for flexible township- and quarter-specific baseline slopes. The event-study coefficients $\{\delta_k\}$ and $\{\rho_k\}$ trace the dynamic effect of the reform on the price slope and the race slope, respectively.\footnote{In the tables we also report the pre-reform (reference-period) slopes estimated in 2020 as summary rows to aid interpretation, since $\delta_k$ and $\rho_k$ are changes relative to the omitted quarter.} We assess the parallel trends assumption using event-study leads (pre-period $\delta_k$ and $\rho_k$). Because the comparison triads' values are frozen over the window, the estimate is the change the reassessment made to the treated triad's slope, net of any change in the comparison triads' slopes over the same quarters. Movements in slopes before the window, which set the regressivity each reassessment met, enter the estimate only through that baseline; Section~\ref{sec:market} returns to this. 

\textbf{Why we use neighborhood minority instead of owner-level race.} Our preferred race measure is the minority share at the CBG level rather than owner-level race, because the main object of interest is neighborhood-level disparate impact in a historically segregated housing market, not individual-level disparate treatment. Different from \cite{avenancio2022assessment}, this neighborhood-level minority is the more relevant margin in our setting: the reformed assessment process is largely automated and leaves limited scope for case-by-case human intervention, but automated valuations can still reproduce spatial inequities embedded in segregated housing markets.

\textbf{Why we estimate both price-only and joint specifications.} We begin with the price-only specification because price regressivity is the paper's primary outcome. At the same time, a key observation from the model-free evidence is that price and neighborhood racial composition are tightly intertwined in Cook County's segregated housing market. A joint specification that includes both price and minority-share interactions is therefore useful for a different purpose: it asks whether the flattening of the price gradient survives once neighborhood racial composition is separately partialled out, and it also recovers the within-price minority gradient. We do not emphasize a race-only specification as a main estimand because, in this setting, it would mechanically combine neighborhood racial composition with differences in property values.

\textbf{Inference with triad-level treatment assignment.}
Inference is nonstandard in our setting because treatment varies at the triad level and the estimation sample contains only three triads. Standard cluster-robust asymptotics are therefore poorly justified for the treatment variation of interest. Our preferred inference accordingly uses a \emph{triad-level wild cluster bootstrap} with Webb weights and 9,999 replications, reported in Columns (1) and (2) of Table~\ref{tab:regressivity_effect}. This procedure is designed for few-cluster settings and is better aligned with the effective level of treatment assignment. Because the bootstrap is nonetheless based on only three treated-assignment clusters, we interpret it as a small-cluster robustness procedure rather than a literal large-sample approximation. For transparency, Columns (3) and (4) also report township-clustered standard errors as a secondary benchmark. Reassuringly, the two approaches produce the same dynamic patterns; the substantive conclusions do not depend on the choice of inference, although statistical precision is mechanically greater under township clustering. We therefore treat the wild-bootstrap columns as primary and the township-clustered columns as supplementary.

As an additional design-based check, Appendix~\ref{app:rob_ri} reports exact randomization inference that reassigns treatment across the three triads; with three admissible assignments its p-values are necessarily coarse (0.333), but the Chicago estimate is the largest under every assignment.

\begin{table}[!htbp]
\footnotesize
\centering
\caption{Main DiD: Price Regressivity and Within-Price Neighborhood Minority Gradient}
\label{tab:regressivity_effect}
\begin{tabular}{lcccc}
\toprule
 & \multicolumn{2}{c}{Preferred inference:} & \multicolumn{2}{c}{Secondary inference:} \\
  & \multicolumn{2}{c}{triad-level wild bootstrap} & \multicolumn{2}{c}{township-clustered SE} \\

  & \multicolumn{2}{c}{\emph{(p values in parentheses)}} & \multicolumn{2}{c}{\emph{(s.e. in parentheses)}} \\
\cmidrule(lr){2-3} \cmidrule(lr){4-5}
 & (1) & (2) & (3) & (4) \\
 & Price & Price + Race & Price & Price + Race \\
\midrule
Treat $\times$ Pre4 $\times$ ln(price) &    -0.034 (0.134) &    -0.022 (0.580) &    -0.034 (0.021) &    -0.022 (0.030) \\
Treat $\times$ Pre3 $\times$ ln(price) &     0.023 (0.344) &     0.018 (0.374) &     0.023 (0.021) &     0.018 (0.031) \\
Treat $\times$ Pre2 $\times$ ln(price) &     0.018 (0.148) &     0.013 (0.146) &     0.018 (0.014) &     0.013 (0.017) \\
Treat $\times$ Post0 $\times$ ln(price) &     0.154 (0.113) &     0.187 (0.122) &     0.154*** (0.040) &     0.187*** (0.039) \\
Treat $\times$ Post1 $\times$ ln(price) &     0.182** (0.029) &     0.200** (0.036) &     0.182*** (0.026) &     0.200*** (0.028) \\
Treat $\times$ Post2 $\times$ ln(price) &     0.200 (0.122) &     0.223* (0.100) &     0.200*** (0.027) &     0.223*** (0.033) \\
Treat $\times$ Post3 $\times$ ln(price) &     0.234 (0.137) &     0.272 (0.160) &     0.234*** (0.035) &     0.272*** (0.038) \\
Treat $\times$ Post4 $\times$ ln(price) &     0.181 (0.129) &     0.216 (0.159) &     0.181*** (0.047) &     0.216*** (0.050) \\
Treat $\times$ Post5 $\times$ ln(price) &     0.187 (0.142) &     0.218 (0.128) &     0.187*** (0.032) &     0.218*** (0.035) \\
Treat $\times$ Post6 $\times$ ln(price) &     0.160 (0.147) &     0.185 (0.138) &     0.160*** (0.044) &     0.185*** (0.047) \\
Treat $\times$ Post7 $\times$ ln(price) &     0.184 (0.151) &     0.226 (0.145) &     0.184*** (0.050) &     0.226*** (0.055) \\
Treat $\times$ Pre4 $\times$ minority &  &    -0.035 (0.570) &  &    -0.035 (0.068) \\
Treat $\times$ Pre3 $\times$ minority &  &    -0.034 (0.126) &  &    -0.034 (0.042) \\
Treat $\times$ Pre2 $\times$ minority &  &     0.001 (0.696) &  &     0.001 (0.032) \\
Treat $\times$ Post0 $\times$ minority &  &     0.098 (0.125) &  &     0.098*** (0.031) \\
Treat $\times$ Post1 $\times$ minority &  &     0.110 (0.140) &  &     0.110** (0.044) \\
Treat $\times$ Post2 $\times$ minority &  &     0.137 (0.140) &  &     0.137** (0.056) \\
Treat $\times$ Post3 $\times$ minority &  &     0.151 (0.145) &  &     0.151*** (0.042) \\
Treat $\times$ Post4 $\times$ minority &  &     0.150 (0.397) &  &     0.150** (0.064) \\
Treat $\times$ Post5 $\times$ minority &  &     0.125 (0.419) &  &     0.125*** (0.036) \\
Treat $\times$ Post6 $\times$ minority &  &     0.094 (0.506) &  &     0.094** (0.042) \\
Treat $\times$ Post7 $\times$ minority &  &     0.188 (0.395) &  &     0.188*** (0.050) \\
\midrule
2020 price slope &    -0.362* (0.081) &    -0.430* (0.092) &    -0.362*** (0.042) &    -0.430*** (0.048) \\
2020 race slope &  &    -0.415 (0.117) &  &    -0.415*** (0.067) \\
\midrule
Pooled post effect on ln(price) slope &     0.186* (0.083) &     0.216** (0.029) &     0.186*** (0.033) &     0.216*** (0.037) \\
Pooled post effect on minority slope &  &     0.130 (0.122) &  &     0.130*** (0.038) \\
Joint post p-value: ln(price) slope &  0.084 &  0.064 &  0.000 &  0.000 \\
Joint post p-value: minority slope &  &  0.130 &  &  0.001 \\
\midrule
Observations &      176,567 &      176,567 &      176,567 &      176,567 \\
R-squared &     0.466 &     0.525 &     0.466 &     0.525 \\
Town-by-quarter FE & YES & YES & YES & YES \\
Town-specific slopes & YES & YES & YES & YES \\
Quarter-specific slopes & YES & YES & YES & YES \\
\bottomrule
\end{tabular}

\begin{minipage}{0.98\textwidth}
\footnotesize \textit{Notes:} This table presents the main DiD (Equation \ref{eq:did_slopes}) estimates examining how the 2021 Cook County reform affected price regressivity and the within-price neighborhood minority gradient in assessments. Across specifications, the outcome variable is ln(Assessed/Sale Price), where Assessed is the initial assessed value. The reference period is $\tau=-1$ (2020 Q4). Columns (1) and (2) are the preferred inference because treatment varies at the triad level; they use wild cluster bootstrap method with Webb weights and 9999 replications, and the numbers in parentheses are bootstrap p-values. The estimation sample contains 3 triads, so these p-values should be interpreted as small-cluster robustness checks rather than as asymptotic approximations. Columns (3) and (4) report township-clustered standard errors in parentheses for comparability with the baseline specification and show that point estimates are similar even though precision differs. For all specifications, we do not report here the dynamic treatment effects for the periods before -4 to save space but they can be seen in event study plots (see Figure \ref{fig:event_studies_combined}). All specifications include township-by-quarter fixed effects, town-specific slopes, and quarter-specific slopes. *** p$<$0.01, ** p$<$0.05, * p$<$0.1.
\end{minipage}
\end{table}

\subsection{Main results: effects of the algorithmic reform on price and race gradients}
\label{sec:main_results}

\begin{figure}[t!]
\centering
\caption{Dynamic Treatment Effects: Price Gradient and Neighborhood Minority Gradient Jointly Estimated}
\label{fig:event_studies_combined}

\begin{subfigure}[t]{0.49\textwidth}
    \centering
    \includegraphics[width=\linewidth]{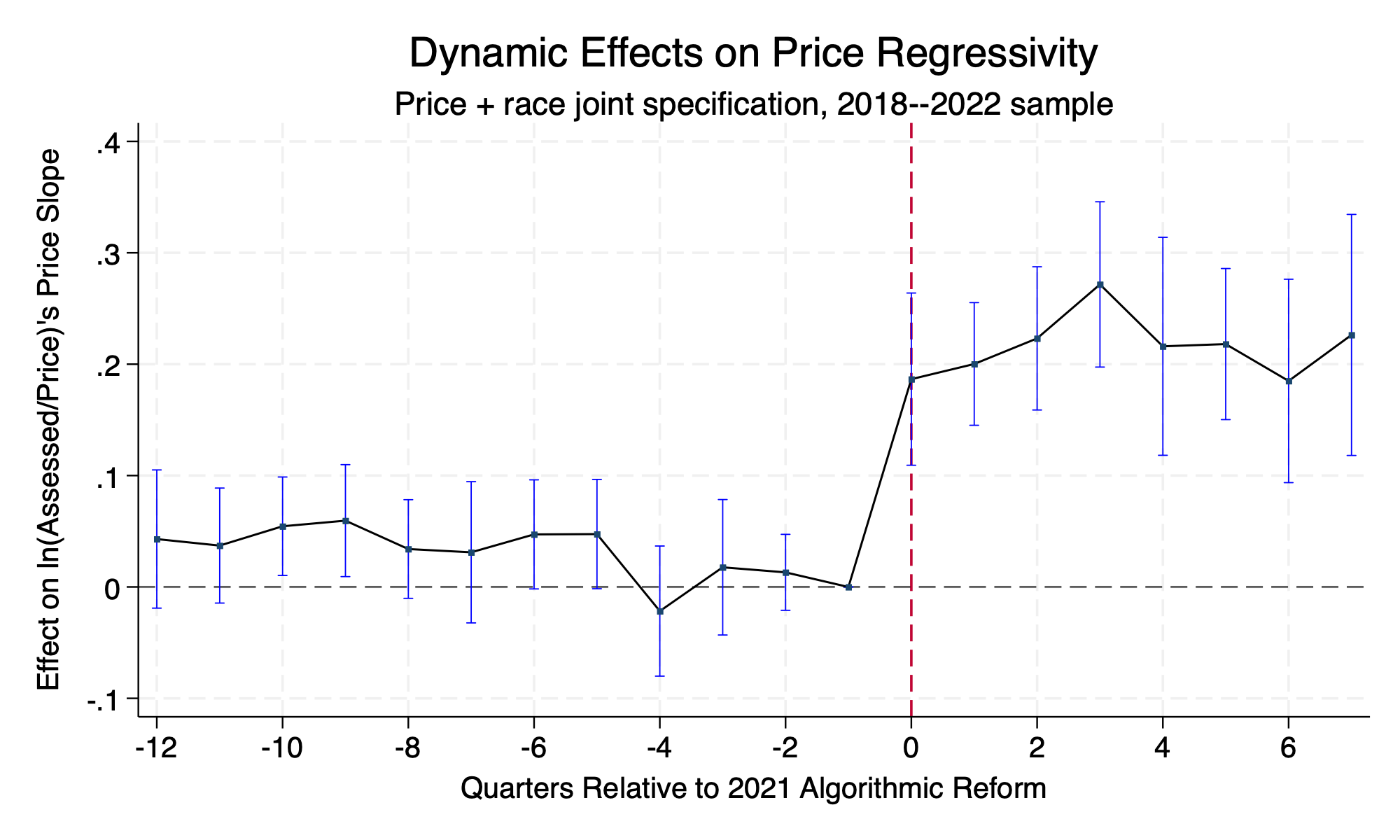}
    \caption{Price-Based Regressivity}
    \label{fig:price_regressivity_event_study_race_controlledfor}
\end{subfigure}\hfill
\begin{subfigure}[t]{0.49\textwidth}
    \centering
    \includegraphics[width=\linewidth]{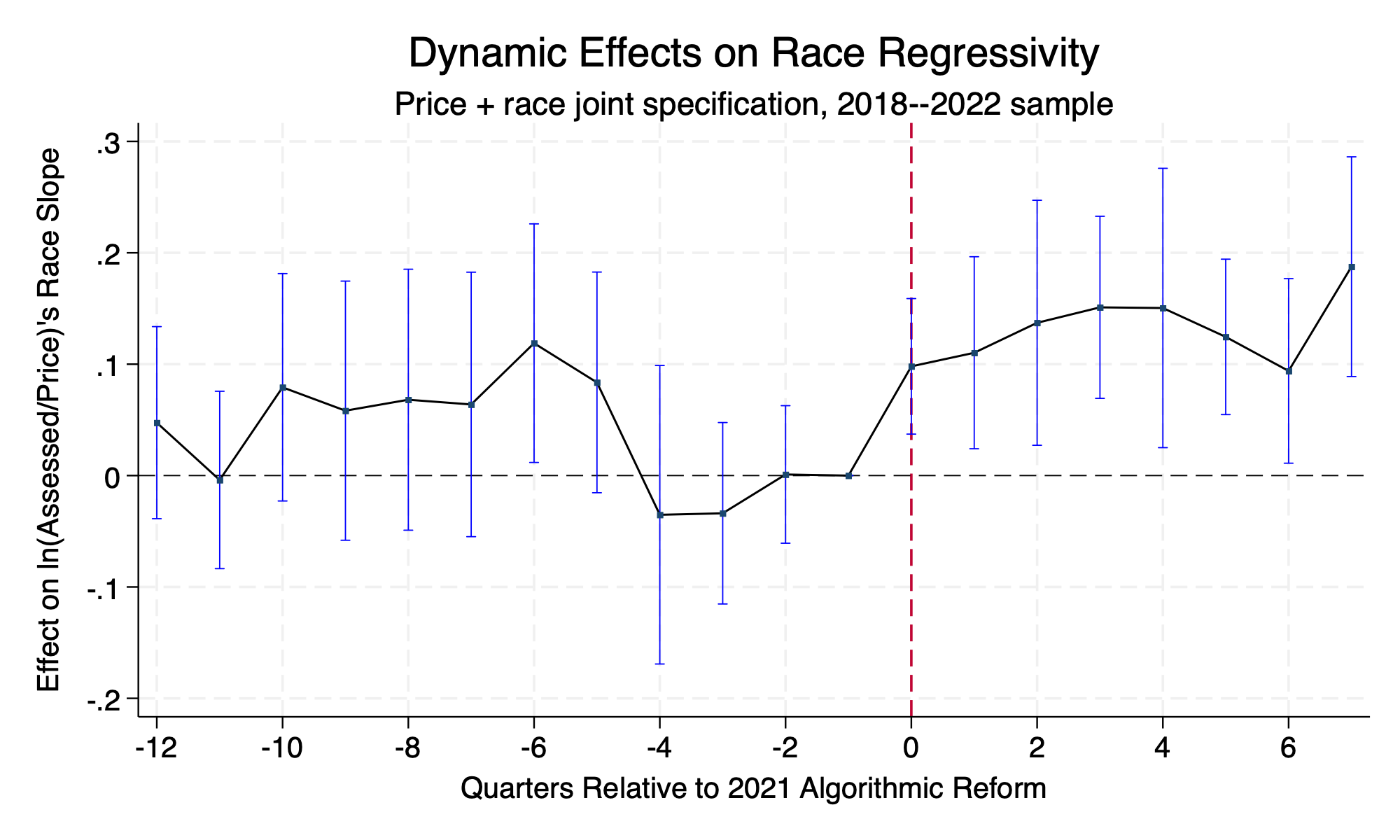}
    \caption{Within-Price Neighborhood Minority Gradient}
    \label{fig:race_regressivity_event_study_price_controlledfor}
\end{subfigure}

\vspace{0.5em}
\begin{minipage}{\textwidth}
    \footnotesize \textit{Notes:} Each point represents the estimated coefficient $\delta_k$ (panel a) or $\rho_k$ from equation \eqref{eq:did_slopes}, from the joint specification in Table \ref{tab:regressivity_effect} Column (4): panel (a) plots the treatment effect on price-based regressivity and panel (b) the treatment effect on the within-price neighborhood minority gradient. The 95\% confidence intervals are constructed using the standard errors reported in Table \ref{tab:regressivity_effect} Column (4). Event time $\tau = 0$ corresponds to Chicago's Q1 2021 reassessment.
\end{minipage}
\end{figure}

Table~\ref{tab:regressivity_effect} reports our main DiD estimates. Columns (1) and (3) estimate the price-only specification, which is our most direct reduced-form measure of within-township price regressivity. Columns (2) and (4) estimate the joint specification that includes both price and race interactions; these columns ask whether the same price-gradient result survives once neighborhood racial composition is separately controlled for, and they also recover the within-price minority gradient. Columns (1) and (2) report our preferred triad-level wild-bootstrap inference, while Columns (3) and (4) report township-clustered standard errors for comparison. For all specifications, we omit the event-study coefficients before Pre4 to save space; they are shown in Figure~\ref{fig:event_studies_combined}. Note that the numbers in parentheses are wild-cluster-bootstrap p-values in Columns (1)--(2) and township-clustered standard errors in Columns (3)--(4).

\textbf{Price regressivity.} The main result is a sharp and persistent reduction in price regressivity after Chicago's 2021 reform. In the price-only specification, the pre-treatment event-study coefficients on the price slope are small and statistically indistinguishable from zero. After treatment, the coefficients turn positive immediately and remain positive throughout the post period. The same pattern appears in the joint price-plus-race specification. Relative to the 2020 baseline price slope of $-0.362$ in Column (1), the first post-reform estimate of $+0.154$ implies an immediate reduction of roughly 43\% in the magnitude of price regressivity. This effect represents a meaningful economic significance.  As an illustration, take a \$200{,}000 Chicago home and a \$543{,}000 home one log point higher in sale price. The pooled post-reform estimate of 0.216 in the joint specification means that the log assessment ratio of the lower-priced home falls by 0.216 relative to the higher-priced home, which is a reduction of $1-e^{-0.216}\approx 19.4\%$ in its assessed market value, or about \$39{,}000. Using Chicago's estimated 2022 residential effective property tax rate of 1.69\%, that reduction corresponds to about \$660 per year in tax liability for the lower-valued home. Two assumptions sit behind the number. The slope estimate pins down only how the two homes' assessments moved relative to each other, so the calculation assigns the whole relative change to the cheaper home and leaves the dearer home's assessment where it was; and it holds the tax rate and exemptions at their 2022 values. If the compression came partly from assessing expensive homes higher, the cheaper home's saving would be smaller and part of the adjustment would show up as a higher bill for the expensive one. 

In the joint specification, the corresponding comparison is relative to a 2020 baseline of $-0.430$ in Column (2), and the first post-reform estimate of $+0.187$ implies a similarly large compression. Thus, the main conclusion does not depend on whether neighborhood racial composition is separately partialled out.

The pooled post estimates reinforce this conclusion. In the preferred wild-bootstrap columns, the average post-reform effect on the price slope is 0.186 in the price-only specification (bootstrap $p = 0.083$) and 0.216 in the joint specification ($p = 0.029$). A separate joint test that all eight quarter-specific post-period coefficients are zero rejects at $p = 0.084$ and $0.064$, respectively. Under township-clustered inference, the same pooled effects are estimated even more precisely. Taken together, these results indicate economically large and persistent declines in price-based regressivity after algorithm adoption.

\textbf{Within-price minority slope.} The race results are also substantively important, although they must be interpreted carefully because our estimand differs from the one emphasized in much of the prior literature. In the joint specification, the pre-treatment race-slope coefficients are close to zero, suggesting no obvious differential pre-trend in within-price neighborhood disparities. Post-treatment coefficients are uniformly positive, ranging from 0.098 at Post0 to 0.188 at Post7. Since the 2020 baseline within-price minority slope in Column (2) is negative ($-0.415$), these positive post-treatment coefficients imply that the reform made that conditional gradient less negative.

This result should not be read as the system becoming pro-white. Our object is the \emph{within-price} race slope, and its pre-reform value is negative: conditional on sale price, higher-minority neighborhoods had \emph{lower} assessment ratios, the signature of a system that pooled homes with a (lower) neighborhood average. The reform moves that gradient toward zero. Section~\ref{sec:mechanism} develops this point.

At the same time, the race results are not as statistically decisive as the price results. The pooled post effect on the minority slope is 0.130 in the preferred specification, but the joint post-period p-value under wild-bootstrap inference is 0.13. By contrast, the township-clustered columns report substantially smaller standard errors and reject strongly for both the pooled post effect and the joint post test. Two further facts temper the race result. Positive minority-slope effects of 0.05 to 0.12 also appear at the South 2020, North 2022, and South 2023 reassessments (Table~\ref{tab:five_events}), so unlike the price result it is not specific to Chicago 2021. We therefore view the evidence on the within-price minority gradient as meaningful and directionally consistent with the mechanism, while treating the reduction in price regressivity as the paper's clearest and strongest result.

%Two features of Table~\ref{tab:regressivity_effect} are worth emphasizing. First, the inferential choice affects precision more than point estimates: across columns, the estimated dynamic effects are extremely stable. Second, the price-only and joint specifications tell a consistent substantive story. Overall, the estimates imply that the reform sharply reduced price-based assessment inequity, with additional but more modest evidence of reduced within-price neighborhood disparities.

\textbf{Event-study evidence.}
Figures~\ref{fig:price_regressivity_event_study_race_controlledfor} and \ref{fig:race_regressivity_event_study_price_controlledfor} report event-study estimates from the joint price-plus-race specification with township-clustered standard errors (Column (4)), with $\tau=-1$ (2020 Q4) normalized to zero. In both panels, the pre-treatment coefficients are generally close to zero and show no clear systematic trend, although a few individual pre-period estimates are statistically significant. Because the plots use township-clustered standard errors, which are mechanically more precise than our preferred few-cluster inference, we interpret them as visual diagnostics rather than as the primary basis for inference. At $\tau=0$, both series shift upward and remain positive throughout the post period. For the price slope, this implies an immediate and persistent reduction in price regressivity after the 2021 reform. For the within-price minority slope, the same dynamic pattern is directionally consistent with smaller neighborhood racial disparities conditional on property value.

%The price estimates are more precise, but the qualitative dynamics are similar across outcomes: little evidence of anticipation, a discrete break at treatment, and persistent post-reform effects. The event studies therefore strengthen the DiD interpretation of Table~\ref{tab:regressivity_effect} by showing that the estimated improvements in assessment equity are not simply the continuation of pre-existing trends, but instead coincide closely with the timing of the reform.

\subsection{Alternative Explanations, Robustness, and Appeals}
\label{sec:robustness_durability}

This subsection asks two questions about the Chicago 2021 estimates. First, could something other than the algorithmic reform have produced them? Second, do the equity gains survive the appeals process that converts mailed values into the values that result from assessor-stage appeals?\footnote{Cook County has two appeal stages. Appeals to the Assessor's Office yield the assessor's \emph{certified} value; owners may then appeal to the Board of Review, whose value is the basis for the tax bill. Our appeals analysis uses certified values and therefore captures the first stage; Board of Review outcomes are not analyzed here.} We summarize the evidence here and report the full analysis in Appendix~\ref{app:robustness} and Appendix~\ref{app:appeals}.

\subsubsection{Alternative Explanations and Robustness}
\label{sec:robustness_checks}

Table~\ref{tab:robustness_summary} summarizes four checks on the Chicago 2021 estimate by listing the concern each check addresses, what we do, what we find, and where the full analysis appears. Appendix~\ref{app:robustness} presents the checks with their design and estimates in more detail.

\begin{table}[t!]
\centering
\footnotesize
\caption{Alternative Explanations and Robustness Checks for the Chicago 2021 Effect}
\label{tab:robustness_summary}
\renewcommand{\arraystretch}{1.15}
\begin{tabular}{p{0.12\textwidth} p{0.23\textwidth} p{0.24\textwidth} p{0.24\textwidth} p{0.09\textwidth}}
\toprule
Check & Concern & What we do & What we find & Where \\
\midrule
1. A new administration
& The assessor who took office in December 2018 changed staff, data, and review along with the algorithm. The Chicago effect could be caused by the new administration, rather than the reformed algorithm.
& Run the same event study around the two reassessments the new administration conducted before the reformed algorithm (North 2019, South 2020), each against triads not reassessed in the window.
& North 2019 moves the slope by 0.00 and South 2020 by 0.03 to 0.05, against 0.22 for Chicago 2021 on the same window. A reassessment, or this administration without its algorithm, produces at most a quarter of the Chicago effect.
& App.~\ref{app:rob_admin} \\
\addlinespace
2. The 2021 post-model adjustment
& The 2021 reform includes a post-model adjustment that shifted whole quintiles of predicted value within each township toward a median sales ratio of one, a step that flattens the within-township slope by design; the 2022--2025 models dropped it.
& Compare Chicago's 2024 reassessment that is run without the step with 2021 and 2018: fresh-value slopes and accuracy, the jump on the same sales, and the event study around 2024.
& 2024 reproduces the accuracy gain (28.7 vs.\ 28.0 percent; 40.0 in 2018) and seven tenths of the flattening ($-0.29$ vs.\ $-0.21$; $-0.45$ in 2018). The post-model adjustment is not a necessary component to compress the price gradient.
& App.~\ref{app:rob_step}; Table~\ref{tab:five_events}; Table~\ref{tab:slope_decomposition}\\
\addlinespace
3.COVID-era housing market
& If lower-priced Chicago homes appreciated faster after 2020, the ratio's denominator would flatten the slope with no change in assessments.
& Check the timing of the effect; add neighborhood-by-quarter fixed effects and neighborhood-specific slopes; let baselines and effects vary with tract-level gentrification.
& The effect appears as a discrete break in the reassessment quarter, not a gradual trend; it is 0.16 within neighborhood-quarters and 0.22 allowing for gentrification.
& App.~\ref{app:rob_covid}; Tables~\ref{tab:regressivity_effect_robustness_nbhd}, \ref{tab:regressivity_effect_robustness_gentrification} \\
\addlinespace
4.Alternative minority measures
& The race result could depend on measuring minority share at the block-group level.
& Replace it with census-block minority share and with owner race inferred from the seller's surname.
& Price effect 0.22 and 0.17; race effect 0.13 and 0.09. Same sign and similar size; less precise with the owner-level proxy.
& App.~\ref{app:rob_minority}; Tables~\ref{tab:regressivity_effect_robustness_block_minority}, \ref{tab:regressivity_effect_robustness_seller_minority} \\
\bottomrule
\end{tabular}

\vspace{0.3em}
\begin{minipage}{\textwidth}
\footnotesize \textit{Notes:} Price effects are average post-period effects on the ln(price) slope from Equation~\eqref{eq:did_slopes} or the stated variant of it; race effects are the corresponding effects on the within-price minority slope. The Chicago 2021 benchmark is 0.216 in Table~\ref{tab:regressivity_effect}, or 0.220 on the four-quarter windows matched to the other events (Table~\ref{tab:early_reassess_did}).
\end{minipage}
\end{table}

%%%%%%%%%%%%%%%%%%%%%%%%%%%%%%%%%%%%%%%%
\subsubsection{Do Appeals Undo the Algorithm's Equity Improvements?}
\label{sec:appeals}
%%%%%%%%%%%%%%%%%%%%%%%%%%%%%%%%%%%%%%%%
Our main outcome is the mailed value, the assessor's initial output. Owners can appeal to the Assessor's Office, which issues a certified value after those appeals; owners may then appeal to the Board of Review, whose value is the basis for the tax bill. Appeals could reverse the reform's gains if they mainly benefit owners of expensive homes. The appeals records show that this concern is reasonable. Owners of more expensive homes, as measured by higher assessed values, are more likely to make an appeal and get a reduced assessment.  Among those who appeal, owners of more expensive homes are more likely to have a higher success rate and a larger reduction. Homes in higher-minority neighborhoods are less likely to appeal and less likely to have professional representation, although those who do appeal obtain somewhat larger reductions (Appendix Table~\ref{tab:appeals_regs}).

Even so, appeals leave the reform's effect intact at both stages. When we re-estimate Equation~\eqref{eq:did_slopes} with the certified value, the average post-reform effect on the price slope is 0.219, compared with 0.216 for mailed values; with the Board of Review value, the value on which tax bills are based, it is 0.216. The effect on the minority gradient is 0.137 and 0.139, compared with 0.130. The pre-period coefficients remain close to zero and the post-period path is the same quarter by quarter (Appendix Tables~\ref{tab:regressivity_effect_by_stage}). Appeals are socioeconomically patterned, but the equity gain reaches the tax bill.

%%%%%%%%%%%%%%%%%%%%%%%%%%%%%%%%%%%%%%%%
\subsection{Mechanism: From Coarse Pooling to Fine Geography}
\label{sec:mechanism}
%%%%%%%%%%%%%%%%%%%%%%%%%%%%%%%%%%%%%%%%

The 2021 reform flattened the price gradient in assessment ratios and moved the within-price minority gradient toward zero. Our explanation is that the legacy system relied on \emph{coarse pooling}, rigid mappings applied across heterogeneous properties, neighborhoods, and price segments, and that the reformed algorithm's finer geography and more flexible interactions separate submarkets the legacy system had lumped together. This section states the two forms of pooling and tests their two implications: after the reform, assessed values should depend less on coarse neighborhood averages and more on fine geography, and the largest accuracy gains should appear where coarse pooling was most costly.

\subsubsection{Intuition: a two-pooling framework for price and race regressivity}
\label{subsec:twopooling_intuition}

Any mass-appraisal system learns a mapping from property characteristics and location to value. When the mapping is too coarse, unlike properties share too much of the same prediction rule, which generates systematic error in a segmented market even when the model is estimated separately by township. Two forms of pooling matter.

\textbf{Cross-price pooling.} A linear model with common coefficients for all homes in a township averages across submarkets. If the relationship between characteristics and value differs across low-, middle-, and high-value segments, a single mapping pulls predictions toward the middle, imputing too much value to the bottom of the market and too little to the top. In assessment-ratio terms, this is price regressivity.

\textbf{Within-price pooling.} Two homes of similar price can differ in micro-location, school zone, flood or noise exposure, and style. If the model captures location mainly through a coarse neighborhood intercept, similarly priced homes are assessed differently depending on the neighborhood average they are pooled with. In a segregated market, homes in minority neighborhoods pool with a lower neighborhood average than similarly priced homes in white neighborhoods, which produces the negative within-price race slope.

The two channels interact: neighborhood intercepts can partly offset price regressivity while worsening within-price distortions, because homes are still valued through coarse local averages. A reform that replaces those averages with finer geography can therefore reduce price regressivity and move the within-price minority gradient toward zero at the same time. Two predictions follow. After the reform, fine geography should explain more of the residual variation in assessed values and assessment ratios than coarse neighborhood averages do, and the largest error reductions should occur in the parts of the market furthest from the pooled middle, where coarse rules fit worst.

\subsubsection{Direct evidence: from coarse neighborhood averages to fine geography}
\label{subsec:twopooling_evidence}

We test the first prediction with partial-$R^2$ decompositions that compare the incremental explanatory power of coarse neighborhood fixed effects and of fine geography, holding a rich set of housing characteristics fixed.\footnote{For each year, we estimate a ``full'' regression and a ``reduced'' regression on the same sample and compute partial $R^2 \equiv 1-\text{RSS}_{\text{full}}/\text{RSS}_{\text{reduced}}$. A higher partial $R^2$ for a block of variables indicates that block explains more residual variation net of the controls held fixed. The fine-geography block includes school-zone fixed effects (constructed from elementary and secondary school identifiers) and micro-location variables (e.g., flood risk and airport noise exposure). Neighborhood fixed effects capture coarse location averages.} Table~\ref{tab:mechanism_partial_r2} reports them for 2018 and 2021 in Chicago and the two comparison triads, for all assessed parcels with outcome $\log(A)$ (Panel A) and for the sales sample with outcome $\log(A/P)$ and 100 within-year price bins (Panel B), so that Panel B measures variation within narrow price ranges.

\begin{table}[!htbp]\centering\footnotesize
\caption{Mechanism Evidence: Partial $R^2$ Decompositions}
\label{tab:mechanism_partial_r2}
\begin{tabular}{lcccccc}
\toprule
 & \multicolumn{2}{c}{Chicago} & \multicolumn{2}{c}{North} & \multicolumn{2}{c}{South} \\
\cmidrule(lr){2-3}\cmidrule(lr){4-5}\cmidrule(lr){6-7}
 & 2018 & 2021 & 2018 & 2021 & 2018 & 2021 \\
\midrule
\multicolumn{4}{l}{\textbf{Panel A}}\\
\multicolumn{4}{l}{\textbf{All homes, Outcome : $\log(A)$}}\\
Nbhd FE $|$ X + fine geo & 0.1864 & 0.1248 & 0.2907 & 0.2726 & 0.2457 & 0.2650 \\
Fine geo $|$ X + nbhd FE & 0.1792 & 0.2434 & 0.0073 & 0.0116 & 0.0066 & 0.0072 \\
\midrule
\multicolumn{4}{l}{\textbf{Panel B}}\\
\multicolumn{4}{l}{\textbf{Sales only, Outcome : $\log(A/P)$}}\\
Nbhd FE $|$ X + price bins + fine geo & 0.2216 & 0.1297 & 0.3412 & 0.3522 & 0.3481 & 0.3702 \\
Fine geo $|$ X + price bins + nbhd FE & 0.2286 & 0.2761 & 0.0079 & 0.0091 & 0.0269 & 0.0155 \\
Price bins $|$ X + nbhd FE + fine geo & 0.8879 & 0.7248 & 0.7063 & 0.6564 & 0.8673 & 0.8189 \\
\bottomrule
\end{tabular}
\begin{minipage}{\textwidth}
\footnotesize \textit{Notes:} Panel A reports partial $R^2$ decompositions for logged assessed values. Panel B reports partial $R^2$ decompositions for ln(assessed/sale price), allowing flexible nonparametric relationship between assessment ratios and sale price by using the price bins that are 100 within-year quantiles of $\log(P)$. $X$ denotes a rich set of housing characteristics (beds/baths, year built, square footage and lot size, basement/attic/garage indicators, exterior materials, heating/cooling, design plan, property class indicators). ``Fine geo'' includes school-zone fixed effects and micro-location variables (flood exposure and airport-noise measures). 
\end{minipage}
\end{table}

In Chicago the informational basis of valuation changes sharply. In 2018, neighborhood fixed effects and fine geography have comparable explanatory power for assessed values (partial $R^2$ of 0.186 and 0.179); by 2021 the ordering reverses, with fine geography at 0.243 and neighborhood effects at 0.125. For assessment ratios within price bins, fine geography rises from 0.229 to 0.276 while neighborhood effects fall from 0.222 to 0.130, and the price bins themselves explain less (0.888 to 0.725): less of the residual structure is absorbed by coarse price and neighborhood averages and more by granular geography.\footnote{Neighborhood effects do not become irrelevant; their importance relative to fine geography falls.} Neither comparison triad shows this pattern: in the North and South, neighborhood effects remain dominant and fine geography modest in both years. The contrast is what one expects if Chicago's reform changed how assessments were produced rather than continuing a countywide trend. The reformed algorithm is unpooling properties the legacy system had treated too similarly, which is why the price gradient flattened and why the reform has distributional consequences in a segregated market.

\subsubsection{Where the gains appear: heterogeneous error reduction}
\label{subsec:hetero_main}

The second prediction is that accuracy gains concentrate where coarse pooling was most costly: the tails of the price and minority-share distributions, far from the pooled middle. We test it on a census-tract-by-year panel for 2018 and 2021, the years in which fresh values were set, restricted to tracts with at least eight sales in each year (74 percent of tracts, 96 percent of sales).\footnote{Tracts rather than block groups avoid sparsity; other thresholds give similar results.} The outcome is the tract's mean absolute assessment error. A tract-level difference-in-differences regression with tract and year fixed effects compares Chicago tracts with the rest of the county before and after the reform, and interacts the treatment effect with indicators for the top and bottom terciles of tract median price and of tract minority share, so that the middle terciles are the baseline; Appendix~\ref{app:hetero} gives the specification and the full estimates.

Errors fell in the middle-tercile tracts after the reform, and fell by more at the tails: the interactions for high-price, low-price, high-minority, and low-minority tracts are all negative, although the low-price interaction is imprecise. Separate regressions by tercile make the pattern plain. Absolute errors fell by 32 percent in low-price tracts and 39 percent in high-price tracts, against 21 percent in middle-price tracts, and by 40 percent in low-minority tracts and 32 percent in high-minority tracts, against 16 percent in middle-minority tracts (Appendix Table~\ref{tab:error_reduction_where}). A coarse, middle-oriented rule fits middle-tier neighborhoods comparatively well and fails where neighborhoods differ most from that middle, which is where the reformed algorithm gained most.

The North suburbs show the other side of this mechanism. Before its 2022 reassessment the North's assessments were already far more accurate than Chicago's had been (a mean absolute error of 17.1 percent against 40.0) and only half as regressive (a price slope of about $-0.22$ against $-0.43$), and Table~\ref{tab:mechanism_partial_r2} shows no shift from coarse neighborhood averages to fine geography there. The same event-study design run around the North's 2022 reassessment, with the not-yet-treated South as control, returns a price-slope effect of 0.03 to 0.05, not distinguishable from zero, and the reassessment lowered the North's error only to 15.9 percent (Appendix~\ref{app:hetero}, Table~\ref{tab:north_did_2022}; Table~\ref{tab:slope_decomposition}). Where there was little error to remove, there was little regressivity to remove: the fairness gain follows from the accuracy gain rather than being a separate property of the reformed algorithm.

Together, the partial-$R^2$ and error-reduction results describe one mechanism: the reform moved valuation from coarse neighborhood averaging toward finer geography and richer interactions, and the gains were largest where coarse pooling had been most distortionary. This is why the reform produced a large decline in price regressivity and why it also changed the distribution of assessment burdens across neighborhood racial composition.

%%%%%%%%%%%%%%%%%%%%%%%%%%%%%%%%%%%%%%%%%%%%%%%%%%%%%%%%%%%%%%%%%%%%%%%%
% SECTION 5, revised 2026-11-03. Replaces lines 609--805 of main.tex.
% New figure files to upload:
%   figures/stacked_fig_tiers_median_vs_repeat_2015.png        (Figure 3, replaces stacked_fig_tiers_level_growth_2015.png)
%   figures/stacked_fig_tilt_by_prior_growth_wins_5yr.png      (new Figure 5)
%   figures/stacked_fig_tilt_by_prior_growth_wins_4yr.png      (appendix)
%   figures/stacked_fig_tilt_by_prior_growth_wins_6yr.png      (appendix)
%   figures/stacked_fig_tiers_growth_median_vs_repeat_2015.png (appendix)
% Figure 4 (stacked_fig_slope_path_components.png) is unchanged.
% Appendix F (app:drift) is revised separately: appendixF_revised_2026-11-03.tex.
%%%%%%%%%%%%%%%%%%%%%%%%%%%%%%%%%%%%%%%%%%%%%%%%%%%%%%%%%%%%%%%%%%%%%%%%

\section{When the Market Moves First: The South 2023 Reassessment}
\label{sec:market}
\label{sec:south_null}

Section~\ref{sec:regressivity} showed that the reformed algorithm cut price regressivity sharply in Chicago, and Section~\ref{sec:mechanism} traced the gain to the distortions the legacy process had accumulated. This section turns to the South 2023 reassessment, where the same algorithm, run by the same agency on the same three-year cycle, produced no detectable improvement. 

Table~\ref{tab:five_events} collects the seven reassessments on which the paper's design can be run, in the order they occurred. The South 2023 row is the case this section explains: an average post effect on the price slope of $0.007$ (standard error $0.033$). Appendix Table~\ref{tab:south_did_2023} provides robustness of this null effect by varying the comparison group and the window; because Chicago and the North had already been reassessed under the reformed algorithm, they are already-treated comparisons, so the estimate is reported against each separately and against both pooled, with eight and with four post-quarters, on a window (2022q1--2024q4) that the North's 2025 reassessment never enters. The estimate lies between $-0.029$ and $+0.007$ throughout, with every 95 percent confidence interval below $0.08$, about a third of Chicago's $0.216$. Pre-trend tests pass except when the North, whose 2022 reassessment falls inside the pre-window, is the sole comparison. The within-price minority slope, $+0.118$ ($0.064$), is positive as in the other cohorts.

\begin{table}[t!]
\centering
\small
\caption{The Same Design Applied to All Seven Reassessments, 2019--2025}
\label{tab:five_events}
\resizebox{\textwidth}{!}{%
\begin{tabular}{llllccc}
\toprule
Event & Valuation workflow & Comparison triad(s) & Window & Price slope & Minority slope & Pre-trend $p$ \\
\midrule
North 2019   & earlier, township-level & South + Chicago            & 2018q1--19q4 & 0.002 (0.022)      & $-0.006$ (0.047)  & 0.066 \\
South 2020   & earlier, township-level & Chicago                    & 2019q1--20q4 & 0.052** (0.021)    & 0.118* (0.067)    & 0.903 \\
Chicago 2021 & reformed algorithm     & North + South              & 2020q1--21q4 & 0.220*** (0.035)   & 0.138*** (0.041)  & 0.165 \\
North 2022   & reformed algorithm     & South                      & 2021q1--22q4 & 0.042 (0.029)      & 0.084* (0.044)    & 0.489 \\
South 2023   & reformed algorithm     & Chicago + North (treated)  & 2022q1--24q4 & 0.007 (0.033)      & 0.118* (0.064)    & 0.161 \\
Chicago 2024 & reformed algorithm     & North (treated)            & 2023q1--24q4 & $-0.051$ (0.061)   & $-0.135$ (0.104)  & 0.179 \\
North 2025   & reformed algorithm     & South (treated)            & 2024q1--25q4 & 0.105*** (0.028)   & 0.054 (0.035)     & 0.796 \\
\bottomrule
\end{tabular}}

\vspace{0.3em}
\begin{minipage}{\textwidth}
\footnotesize \textit{Notes:} Each row reports the average post-period effect on the ln(price) slope and on the within-price minority slope from the joint specification of Equation~\eqref{eq:did_slopes}, estimated around the indicated event with the reference period one quarter before it. Township-clustered standard errors in parentheses; the pre-trend $p$-value tests that all pre-period price-slope interactions are zero. The rows are not a common design: the first four windows average four post-quarters (so the Chicago row differs slightly from the eight-quarter average of 0.216 in Table~\ref{tab:regressivity_effect}) and use comparison triads not reassessed within the window; the South 2023 window averages eight post-quarters, and its comparison triads had already been reassessed under the reformed algorithm (Chicago in 2021, North in 2022), with Chicago reassessed again in 2024 inside the window. The Chicago 2024 and North 2025 rows average four post-quarters against the one triad not reassessed inside the window, which had itself been reassessed under the reformed algorithm earlier (the North in 2022, the South in 2023); against both other triads the estimates are $-0.066$ (0.057) for Chicago 2024 and $0.112$ (0.028) for North 2025. Full specifications, alternative comparison groups, and triad-level wild-bootstrap inference for the first five rows are in Tables~\ref{tab:early_reassess_did}, \ref{tab:north_did_2022}, and \ref{tab:south_did_2023}. *** p$<$0.01, ** p$<$0.05, * p$<$0.1 based on township-clustered inference.
\end{minipage}
\end{table}

We provide two more robustness checks of the null effect that do not depend on the choice of control. First, on the 2023 sales, the frozen 2022 values give a within-township slope of $-0.39$ and the reassessment's own values $-0.43$ (Appendix Table~\ref{tab:slope_decomposition}): this suggests that on the same homes, the new values were no flatter than the old. Second, a stacked regression on windows of equal length puts the difference between the Chicago 2021 and South 2023 effects at $0.238$ (s.e. $0.033$; $p<0.01$ under township clustering, $0.11$ under the three-cluster bootstrap; Appendix Table~\ref{tab:chi_vs_south}).

The null is not necessarily evidence that the algorithm failed. The reform arrived after the sharpest segment-divergent price movement in our sample, and that movement did two things, which we document in turn. First, while values stood frozen, it removed the between-neighborhood part of the South's regressivity before the reform could: the market acted as a competing corrector, and the reset the algorithm performs found little left to flatten. Second, the neighborhood trends the model had learned from its training data reversed in the valuation year, so the fresh values, set at the level each neighborhood had reached, landed too high in exactly the neighborhoods whose run had just ended. Both are visible in the data, and the staggered rollout lets us watch the same algorithm succeed where neither condition held (as tightly).

\subsection{Segment-divergent appreciation}
\label{sec:fanning}

\begin{figure}[h!]
\centering
\includegraphics[width=0.75\textwidth]{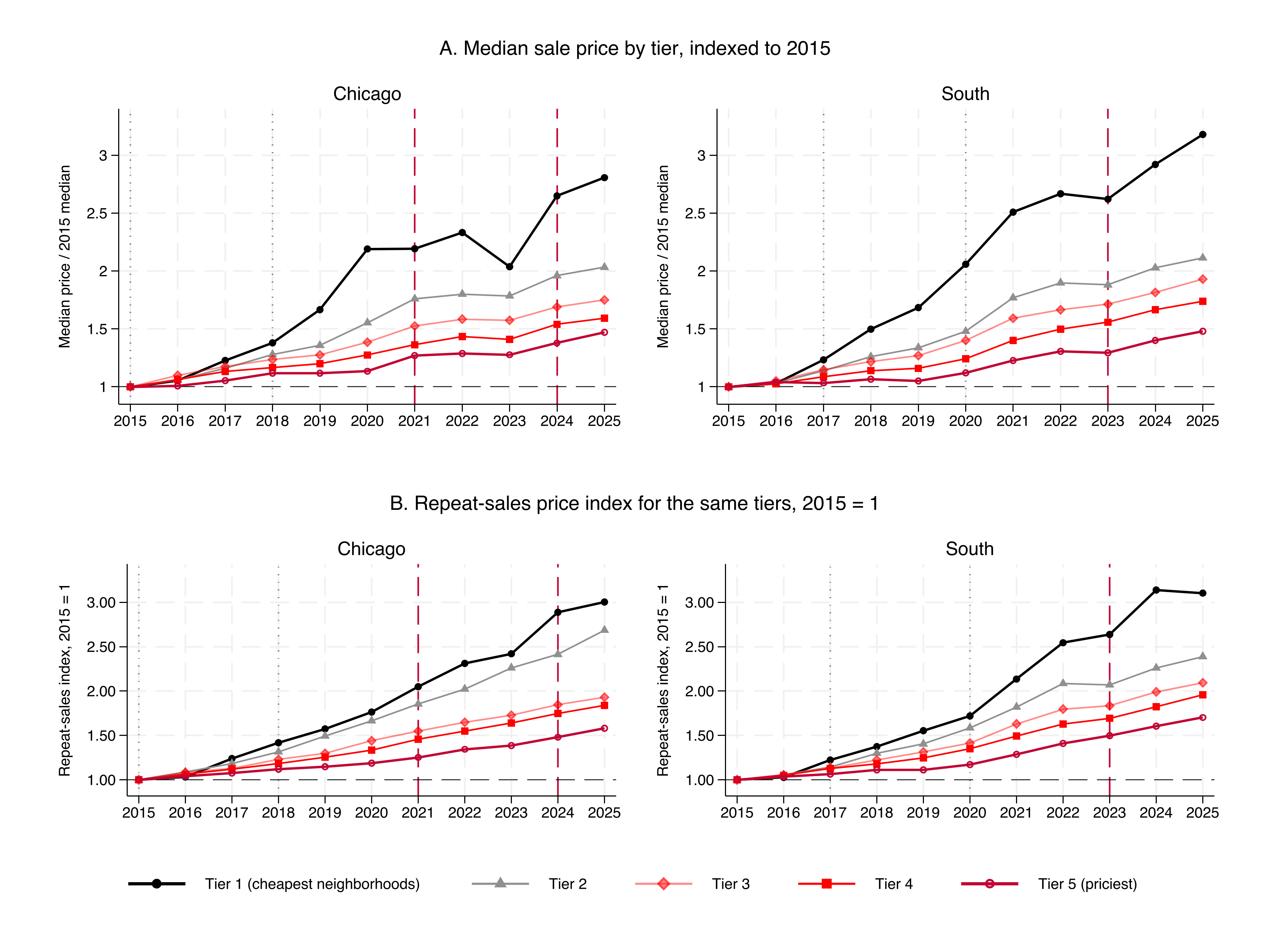}
\caption{Neighborhood price tiers in Chicago and the South, 2015--2025: median prices and same-home prices.}
\label{fig:competing}
\begin{minipage}{0.96\textwidth}
\footnotesize \textit{Notes:} Neighborhoods with at least eight arm's-length sales in every year 2015--2025 (145 in Chicago, 234 in the South), ranked once by their 2015 median sale price into quintiles within triad. Panel A: each tier's annual median sale price divided by its 2015 median. Panel B: a repeat-sales index for the same tiers, from consecutive arm's-length sales of the same parcel with both sales in 2015--2025, at least four quarters apart, and an annualized log price change within $\pm 0.35$ (3,300 to 4,900 pairs per triad-tier); the index is the exponential of the year coefficients from a regression of the log price change on indicators for the resale year minus indicators for the purchase year, 2015 $=1$. Gray dotted vertical lines mark reassessments before the reformed algorithm, red dashed lines reassessments under it.
\end{minipage}
\end{figure}

Market trends differ across segments at the level of neighborhoods. Figure~\ref{fig:competing} ranks the neighborhoods of Chicago and the South into five tiers by their 2015 median sale price and follows each tier's prices, first as the tier's median sale price (Panel A) and then as a repeat-sales index built from consecutive sales of the same homes (Panel B), which removes any change in the mix of homes sold.\footnote{The panel is the neighborhoods with at least eight arm's-length sales in every year from 2015 to 2025 (145 in Chicago, 234 in the South). Panel B uses every pair of consecutive arm's-length sales of the same parcel in those neighborhoods with both sales in 2015--2025, at least four quarters apart, and an annualized log price change within $\pm 0.35$; the last restriction removes rehabilitation flips, which are 40 to 50 percent of pairs in the cheapest tier and would otherwise dominate it. The index is the Bailey--Muth--Nourse regression of the log price change on year dummies, by triad and tier, 2015 $=1$. Appendix~\ref{app:drift} reports the year-over-year growth of both series (Figure~\ref{fig:tier_growth_repeat}).} In both triads the cheapest fifth of neighborhoods pulled away from the rest for years at a time, and the same homes confirm it. In the South, the cheapest tier's repeat-sales index grew by 11 to 16 percent a year from 2017 to 2020 and by 24 and 19 percent in 2021 and 2022, while the priciest tier's never grew by more than 10, so that by 2022 the same cheap homes had reached 2.5 times their 2015 price and the priciest 1.4 times. Then the divergence stopped: in 2023 the cheapest tier grew 4 percent and the priciest 6, the first year since 2016 in which the cheapest tier grew no faster than the priciest. In Chicago the divergence slowed after 2021 but never reversed before 2024: the cheapest tier's index grew 16 percent in 2021 against 6 for the priciest, 13 against 7 in 2022, and 5 against 3 in 2023. The median series in Panel A tell the same story in the South, but in Chicago they are distorted by the mix of homes sold, showing no growth in the cheapest tier in 2021 and a fall in 2023 that the same homes do not show.

Assessed values are frozen between reassessments. A segment-divergent market movement like this therefore changes regressivity on its own; it also affects the prediction accuracy of a model fitted to the years before the reversal happens. The next two subsections take these in turn.

\subsection{Mechanism 1: the market as a competing corrector}
\label{sec:competing}

In this section, we provide evidence that the segment-divergent appreciation lifted the baseline regressivity for the South 2023 reform, in part explaining its null effect. 

\begin{figure}[h!]
\centering
\includegraphics[width=0.8\textwidth]{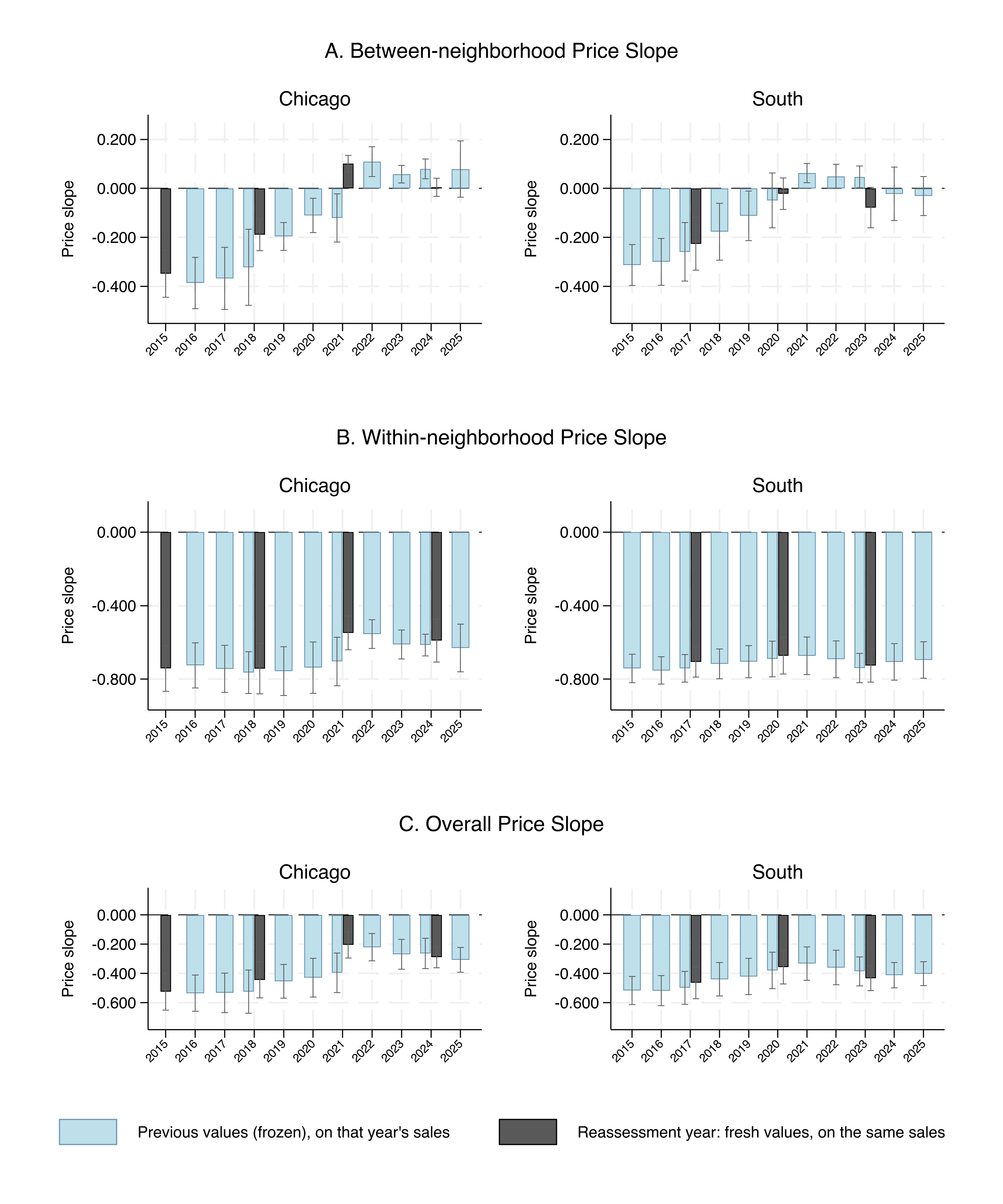}
\caption{The between-neighborhood, within-neighborhood, and overall price slopes year by year, Chicago and the South, 2015--2025.}
\label{fig:slopepath}
\begin{minipage}{0.96\textwidth}
\footnotesize \textit{Notes:} Each bar is the slope of ln(Assessed/Sale Price) on ln(sale price) estimated on the triad's sales in that year with the mailed values in force that year; caps are 95 percent confidence intervals from township-clustered standard errors. In a reassessment year the light bar applies the previous year's values to that year's sales and the dark bar applies the new values to the same sales. Panel A regresses neighborhood-year mean log ratios on neighborhood-year mean log prices with township fixed effects, weighted by sales. Panel B uses neighborhood-by-quarter fixed effects. Panel C uses township-by-quarter fixed effects and is the slope used throughout the paper. Appendix Table~\ref{tab:slope_decomposition} reports the values with standard errors.
\end{minipage}
\end{figure}

Our main regressivity estimate, the within-township slope of ln(Assessed/Sale Price) on ln(sale price), blends two things: how assessment ratios differ across neighborhoods of different price levels, which we call \textit{between-neighborhood} regressivity, and how they differ among homes in the same neighborhood, which we call \textit{within-neighborhood} regressivity.\footnote{We measure the first as the slope of neighborhood-year mean log ratios on neighborhood-year mean log prices, with township fixed effects and weights equal to the number of sales, and the second as the sale-level slope with neighborhood-by-quarter fixed effects. Appendix Table~\ref{tab:slope_decomposition} reports both, with the paper's slope, at every reassessment in all three triads, with standard errors.} When cheap neighborhoods appreciate faster than expensive ones under frozen values, their ratios fall faster, and between-neighborhood regressivity declines with no action by the assessor. Figure~\ref{fig:slopepath} shows the two parts, and then the paper's slope, year by year for Chicago and the South, each estimated on that year's sales with the assessed values in force that year. In a reassessment year there are two bars on the same sales: the previous values (i.e., the assessment values at Year$-3$), then the new ones (i.e., the fresh assessment values at current year). Every change from one light bar to the next is \textit{drift}, the work of prices and the mix of homes sold while assessment values stood still; every change from a light bar to the dark bar beside it is \textit{jump}, or the reassessment itself.

Panel A shows how the market works. With the 2017 values frozen, the South's between-neighborhood slope rose from $-0.23$ to $-0.05$ by 2020; the 2020 reassessment under the earlier workflow added $+0.03$; and with the 2020 values frozen it rose to $+0.05$ by 2022. By the time the reformed algorithm arrived in 2023, the market had removed the South's between-neighborhood regressivity. Chicago's slope drifted the same way, from $-0.19$ in 2018 to $-0.12$ on the eve of its reform, but the drift was smaller and the 2021 reform found a distortion still in place.

What a reassessment then does to this part is a reset, not a continuation of the drift. A reassessment does not adjust the old values; it discards them and predicts every home afresh, so the between-neighborhood regressivity of the new values does not depend on what the old values carried. In all seven reassessments since 2019, under the earlier workflow and the reformed algorithm alike, the fresh values landed within about $0.1$ of zero on this slope, whether the values they replaced stood at $-0.12$ or $+0.08$; only the legacy values of 2018 did not (Appendix Table~\ref{tab:slope_decomposition}). The jump a reassessment produces in this part is therefore close to the negative of the regressivity in force when it arrives. Chicago's 2021 reform found $-0.12$ and lifted the between-neighborhood slope by $0.22$; the South's 2023 reform found $+0.05$ and, landing at $-0.08$, lowered it by $0.13$. These numbers suggest that the market, particularly the segment-divergent appreciation since around 2017, preempted the fairness gain of the reformed algorithm in South 2023 --- Had the South's values still carried their 2017 between-neighborhood regressivity when 2023 arrived, the same reset would have registered as a gain of about $0.2$. 

It is important to note that the starting points (-0.12 for Chicago 2021 and +0.05 for South 2023) do not explain the whole Chicago2021-South2023 contrast, because the landing points differed too, $+0.10$ in Chicago against $-0.08$ in the South; of the $0.35 (=0.22 + 0.13)$ gap between the two jumps, about half comes from where the reforms started and half from where they landed. Section \ref{sec:degradation} shows what the South's landing point was made of. 

Panel B of Figure~\ref{fig:slopepath} plots the within-neighborhood slope, which reflects a natural rate of regressivity that idiosyncratic sale prices produce with no bias in the values \citep{mcmillen2022measures}. Specifically, a sale price is the home’s value plus a component specific to that transaction, such as the buyer’s eagerness, the seller’s urgency, or the timing of the sale, and an assessment can only aim at the value. A home that sold unusually high then looks expensive and shows a low assessment ratio, and one that sold unusually low looks cheap and shows a high ratio, so the ratio falls with the sale price even when every assessment is exactly right on average. Within a neighborhood, where homes are alike and much of the spread in prices is this transaction-specific component, that mechanical slope is steep. This part of the total price slope is harder to tackle for \textit{any} assessors, because it really requires an algorithm that predicts individual homes better. The Chicago 2021 reform seems to have improved on this front to some extent, and the South 2023 seems to have done little. We provided a deeper discussion on this Chicago-South difference in Section \ref{sec:degradation}, which may also have contributed to the South's null effect. 

Panel C shows the slope the paper uses, which combines both the betwee-neighborhood regressivity and the within-neighborhood regressivity.\footnote{The overall slope is not the arithmetic sum of the two components. With common fixed effects it is a variance-weighted combination of a between- and a within-neighborhood slope, and the between-neighborhood measure here uses annual neighborhood means with township fixed effects, so the two panels are complementary measures that together account for the overall pattern, not an exact decomposition.} The paper's main event-study estimate is essentially the jump net of any drift in the comparison triads inside the same time window. One may argue that for the South 2023 reform, its baseline at 2020 is mainly affected by the pre-reform earlier workflow by the new administration. Figure~\ref{fig:slopepath} Panel A and C offer direct evidence againt this argument: the drift in South started since 2017 and raised the between-neighborhood regressivity (Panel A) and the total price slope (Panel C) quickly, and by 2020 they almost reached the same levels as those set by the fresh assessment values in 2020. The drift that mattered came earlier, and it entered not the estimate but the baseline the reform met. 

\subsection{Mechanism 2: an algorithm trained on the segment-divergent appreciation meets a market reversal}
\label{sec:degradation}
A valuation model is estimated on past transactions and applied to a valuation date after its training data end. What it carries forward from those data is, among other things, each neighborhood's price trend, and the question is what happens to that when the trend changes direction at the valuation date. In this section, we document the second mechanism through which the market affects the algorithm's fairness gain. Specifically, for the null effect of South 2023, we hypothesize that the reformed algorithm set each neighborhood's values at the price level it had reached at the end of the segment-divergent trend in its training data, and that the trend then reversed. The direct testable implication is that the neighborhoods that appreciated fastest in the training window are the ones the 2023 values over-assessed the most.

\begin{figure}[!t]
\centering
\includegraphics[width=0.96\textwidth]{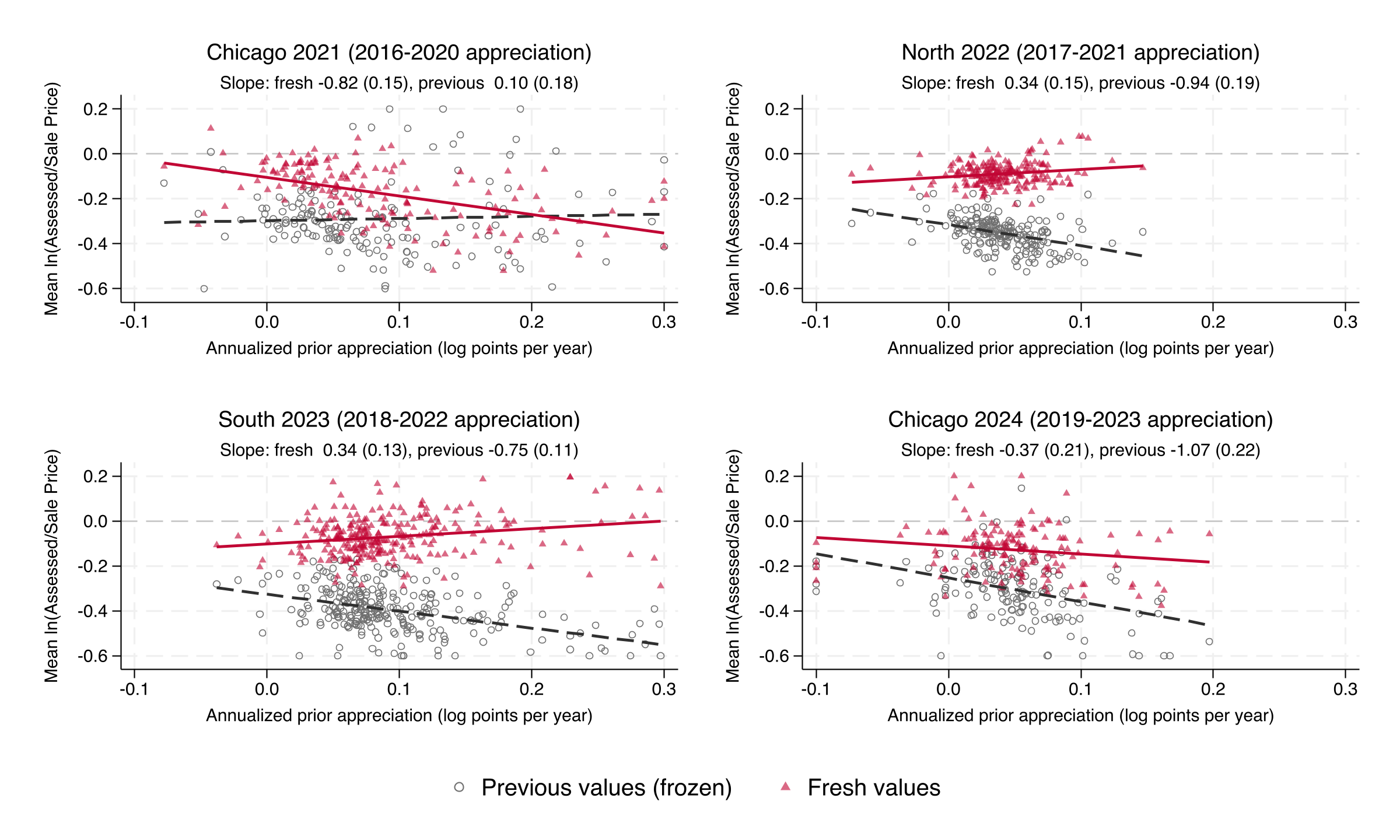}
\caption{Which neighborhoods the fresh values over- or under-assess: assessment ratios against prior appreciation, by neighborhood, at four deployments of the reformed algorithm.}
\label{fig:tilt}
\begin{minipage}{0.96\textwidth}
\footnotesize \textit{Notes:} Each marker is a neighborhood. The vertical axis is the mean of ln(Assessed/Sale Price) over the neighborhood's arm's-length sales in the reassessment year, computed with the values in force the year before (previous values, circles) and with the reassessment's own values (fresh values, triangles), winsorized at the 0.5th and 99.5th percentiles within panel. The horizontal axis is the change in the neighborhood's log median sale price over the five calendar years ending the year before the reassessment (Chicago 2021: 2016 to 2020; North 2022: 2017 to 2021; South 2023: 2018 to 2022; Chicago 2024: 2019 to 2023), divided by four. Neighborhoods enter if they have at least eight sales in the reassessment year, the year before, and the first year of the window (161, 206, 275, and 154 neighborhoods). Lines are unweighted OLS fits; subtitles report their slopes with heteroskedasticity-robust standard errors. Points outside the plotted range are shown at the frame edge; fits use all neighborhoods. Appendix Figures~\ref{fig:tilt_4yr}--\ref{fig:tilt_6yr} report four- and six-year windows for the training data period.
\end{minipage}
\end{figure}

Figure~\ref{fig:tilt} answers it directly, by showing the correlation between the neighborhood-level price appreciation over the 5 years before the respective reassessment and the neighborhood average logged assess-price ratios which measures the signed error. For each of the four deployments of the reformed algorithm (Chicago 2021 and 2024, North 2022, and South 2023), we describe two correlation --- one using the frozen values in force the year before the respective reassessment (circles) and the other using the fresh values (triangles) on the same sales. The slopes of the fitted lines are reported in each panel.

The circles show drift. Wherever a cycle contained real divergence, the previous values under-assess the neighborhoods that appreciated fastest, with slopes of $-0.75$ in South 2023, $-0.94$ in North 2022, and $-1.07$ in Chicago 2024: a neighborhood that appreciated 10 log points a year faster than another over the window carries, by the end of it, an assessment ratio 8 to 11 log points lower. This is the between-neighborhood drift of Figure~\ref{fig:slopepath} seen at the neighborhood level, and the first thing every reassessment does is remove it: the jump from circle to triangle rises one for one with prior appreciation in every panel.

The triangles show where the fresh values land, and here the four deployments split by triad. In Chicago, in 2021 and again in 2024, the fresh values \emph{under}-assess the neighborhoods that had appreciated fastest, with slopes of $-0.82$ and $-0.37$. In the South in 2023 and the North in 2022, the fresh values \emph{over}-assess them, with slopes of $0.34$ in both: a neighborhood that had appreciated 10 log points a year faster over the training window is assessed about 3.4 log points higher relative to what its homes sold for. The same slope means more in the South, where neighborhoods differ twice as much in how fast they had been appreciating: a slope of 0.34 separates the ratios of the fastest- and slowest-appreciating neighborhoods by about 4 log points in the South against 2 in the North. We read this as consistent with our hypothesis that the trend the South 2023 algorithm learned did not persist, leaving its values highest relative to price where that trend had been steepest, which contributed to the absence of a fairness gain. Consistent with this, within every price tier the 2023 values are between one and two percentage points more accurate than the 2020 values, where Chicago's twelve-point gain came from the cheapest tiers; the South's aggregate is unchanged because the cheapest tier's share of sales rose from 21 to 28 percent and its error is twice the average (Appendix Table~\ref{tab:error_by_tier}).

This evidence also aligns with the patterns in Figure~\ref{fig:slopepath} Panel A. The between-neighborhood slope of $-0.08$ that the 2023 values set is consistent with the over-assessment of the fast-appreciating cheap neighborhoods in Figure~\ref{fig:tilt}, and it is the only deployment of the reformed algorithm that landed below zero on that slope. 

Finally, the market-affected training data may also help explain why the within-neighborhood regressivity in Panel B of Figure~\ref{fig:slopepath} improved so little. Recall that the within-neighborhood slope is steeper than the between-neighborhood slope to begin with, in Chicago and the South alike, because idiosyncratic sale prices among the similar homes of one neighborhood produce regressivity even when the assessment is unbiased on average \citep{mcmillen2022measures}. This within-neighborhood part of total price regressivity is the natural outcome of any system that predicts value out of noise, and it is correspondingly hard to reduce. It can be reduced, however: Chicago's 2021 reform raised the within-neighborhood slope from $-0.70$ to $-0.55$ on the same sales. The South's 2023 reform did not. It left the slope at $-0.73$ and the error at $26.7$ percent, essentially where the frozen 2020 values already stood (Appendix Table~\ref{tab:slope_decomposition}). Part of the reason is that there was less accuracy to add: the 2020 reassessment under the earlier workflow had already cut the South's error from $35.7$ to $26.7$ percent, close to the $28.0$ that Chicago's reformed algorithm reached in 2021.

We do not identify why the South's 2023 gain was smaller than Chicago's 2021 gain, but we can offer a plausible account. We reconstructed models of our own on the assessor's archived training data for the South, holding features, training windows, and evaluation sales fixed and varying only the learner: a gradient-boosted tree of the family the reform adopted against a simple hedonic regression. From 2019 through 2022, when the price trend in the training window continued past the valuation date, the flexible model matched or slightly beat the simple one. In 2023, when the trend broke, it fell behind, by $0.7$ percentage points of error on the January 1 valuation the assessor targets ($p<0.01$; Appendix~\ref{app:models}, Table~\ref{tab:ml_vs_hedonic}). This is what one expects if a flexible learner tracks the recent trend more closely and therefore loses more when the trend ends. These are not the production models, and the exercise is a proof of concept rather than a replication of the 2023 reassessment, so we read it as a plausible account of the missing accuracy gain rather than a measurement of it.

\subsection{Both at once}
\label{sec:bothatonce}

The two mechanisms are consequences of one event. The same divergence in neighborhood price paths that, under frozen values, compressed the South's between-neighborhood regressivity before the reform could, also ended at the valuation date, so that the fresh values, set at the level each neighborhood had reached, stood above the prices of the cheap neighborhoods that had been rising fastest. The first removed the gain the reset would have produced; the second turned the reset itself against the neighborhoods the reform was meant to help; and the accuracy gain that in Chicago had converted into equity gains did not arrive. The staggered rollout shows the same algorithm in the opposite situation. Chicago's reform arrived while its neighborhood trends were still running. The values it set were below the prices that fast-appreciating neighborhoods went on to sell at, and since those neighborhoods were the cheap ones, this under-assessment lowered regressivity. The South's reform arrived after the market had corrected much of the legacy distortion and after its trends had turned. The values it set were above the prices that its fast-appreciating cheap neighborhoods went on to sell at, and this over-assessment failed to improve the regressivity.

The implication for evaluating algorithmic interventions is direct. A null effect in a moving environment is not informative about the intervention on its own, because the environment is a rival source of the improvement being measured, and because a model that learns from the past is wrong in one direction when the market keeps going and in the other when it turns. Identifying which of these an estimate reflects requires seeing the same system deployed into markets moving at different speeds and in different phases, which is what the rollout provides.

%%%%%%%%%%%%%%%%%%%%%%%%%%%%%%%%%%%%%%%%%%%%%%%%%%%%%%%%%%%%%%%%%%%%%%%%
% CONCLUDING REMARKS
%%%%%%%%%%%%%%%%%%%%%%%%%%%%%%%%%%%%%%%%%%%%%%%%%%%%%%%%%%%%%%%%%%%%%%%%

\section{Concluding Remarks}
\label{sec:conclusion}
This paper asks what happens to the fairness of an algorithmic valuation when the market keeps moving between the moments at which values are set, in a setting where the answer has direct consequences for what people pay. Cook County replaced a legacy assessment process with one machine-learning valuation system and deployed it in three regions in three consecutive years. In Chicago, the reform removed about half of the long-standing over-assessment of lower-priced homes, made assessments among homes of the same price less dependent on neighborhood racial composition, and did so durably: the effect appeared in the first quarter, did not fade, and survived both stages of appeal. In the North and South suburbs, the same system had no detectable effect on price regressivity. The three deployments differ in what each met. Chicago’s legacy values carried large, systematic errors that a more accurate model removed. The North’s were already far more accurate and only half as regressive, leaving less to correct. The South’s had been as regressive as Chicago’s, but the market reached them first: faster appreciation in inexpensive neighborhoods lowered their assessment ratios while values were frozen, and when that appreciation stalled in the reassessment year, the new values, pegged to the level each neighborhood had reached, stood above the prices of the very neighborhoods that had been catching up; and because the 2020 values were already as accurate as Chicago’s reformed ones, the reassessment added no accuracy to convert into a fairness gain.

These findings bear directly on research on tax regressivity and on the adoption of algorithms by assessors, which is spreading. In 2025, New York City’s Department of Finance awarded a pilot to explore AI-based valuation of residential condominiums,\footnote{See \url{https://a856-cityrecord.nyc.gov/RequestDetail/20250304005}, last accessed on 03/25/2026.} and Maricopa County, Arizona, runs a cloud-based mass appraisal system covering more than 1.8 million parcels a year.\footnote{See \url{https://resources.esri.ca/customer-stories/maricopa-county-goes-live-with-assessment-analyst-gama}, last accessed on 03/25/2026.} Our results suggest that the fairness question in these adoptions is a dynamic one: it depends on the technology chosen and on a market that continually resolves or amplifies unequal burdens across segments. The challenge for assessors is therefore not only which model to adopt, but how to \textit{monitor} fairness drift and \textit{maintain} fairness between reassessments.

More broadly, the study offers two lessons for organizations that set values or prices with models, whether governments assessing property or firms pricing goods, loans, or listings. First, the fairness gain to expect from an algorithm should be judged against the values in force when it arrives, which depend on how the previous values were set and on how far the market has since moved them. Where those values are coarse and inaccurate, a documented, data-driven model can remove a large distortion; where they are already accurate, or the market has already corrected them, the same model will do little, and an evaluation that ignores the baseline will conclude that the algorithm does not work. Second, a changing market is a fairness problem, not only a forecasting problem. When values are updated periodically and segments move at different speeds, the values in force between updates become too high for some segments and too low for others, and the market, not the model, determines which. A model that values each segment at the level it has reached is then wrong in one direction if the segment's trend continues and in the other if it turns, as the same algorithm was in Chicago and in the South. 

The challenge for the regulators writing audit requirements for algorithms is therefore one of timing. Fairness maintenance calls for audits whose frequency rises with the speed of the market, but every audit costs money and attention, and a fixed cycle, the cadence most regimes have chosen, will be too slow in a fast market and more than is needed in a slow one. Our results suggest one way to think about the problem rather than a tested solution. The ratio of assigned to market value by segment is a measurement of the current distribution; in Cook County it can be computed from sales records at any point between reassessments, and used as a trigger it would call a full audit when segments diverge. We have not shown that such a trigger outperforms a calendar audit. What the evidence here establishes is narrower: an audit timed to the calendar or to changes in the system can miss movements that a market-indexed measurement would record.

Several limitations point to future work. Our estimates capture a bundled reform: a new model, new data, and a documented process introduced together. The comparison with the new administration’s earlier reassessments separates the reformed algorithm from the administration, but we cannot assign the Chicago gain to any single component. Chicago's 2024 reassessment, which ran the reformed algorithm without the post-adjustment step that the 2021 model alone applied, reproduces the accuracy gain and most of the flattening (Section~\ref{sec:robustness_durability}), so a flattening of this size does not require the step, but its share of the 2021 gain is not identified and the stage-level predictions that would settle it are not public. The reconstructed models in Section~\ref{sec:market} are our own, not the production model, whose intermediate predictions are not public. And we observe one changing market; the mechanism is general, but its size elsewhere depends on how fast segments diverge and how often values are refreshed.

The paper's key insight is that algorithmic fairness, in a system whose outputs are set periodically, is not a just property the algorithm has at deployment but more of a trajectory it shares with the market: the same relative price movements that erode or improve fairness while values are frozen also decide what the next reform will find left to fix, what it will have learned from, and whether what it learned still holds on the day its values take effect. For evaluation, this implies that a fairness audit of one single deployment is only going to yield a snapshot of a moving target, whose verdict is affected by both the algorithm and when the market was sampled. For governance, it means that the object to be held accountable is not the algorithm but the algorithm together with its refresh policy, because an algorithm cannot be made fair once; it can only be kept fair. In a moving market, the algorithm never acts alone.

%%%%%%%%%%%%%%%%%%%%%%%%%%%%%%%%%%%%%%%%
% Bibliography
%%%%%%%%%%%%%%%%%%%%%%%%%%%%%%%%%%%%%%%%
\begingroup
\setstretch{1.0}
\setlength{\bibsep}{3pt}
\bibliographystyle{apsr}
\bibliography{references}
\endgroup

%***************************************************************
%********************     APPENDIX     ***********************
%***************************************************************

\newpage
\appendix

\begin{center}
\Large \textbf{Online Appendix}
\end{center}

\setcounter{table}{0}
\renewcommand{\thetable}{A\arabic{table}}

\setcounter{figure}{0}
\renewcommand{\thefigure}{A\arabic{figure}}

\section{Sample Construction and Summary Statistics}
\label{app:sample}

We begin with all properties that (1) sold at least once during 2018–2022, (2) are classified as residential single family homes\footnote{We focus on various single family home types as classified by the Assessor's Office: \url{https://prodassets.cookcountyassessoril.gov/s3fs-public/form_documents/Class_codes_definitions_12.16.24.pdf}, last accessed on 02/27/2026. Specifically, we use the following property classes: 202, 203, 204, 205, 206, 207, 208, 209, 210, 234, 278, and 295.}, (3) have valid assessed values in the year of sale\footnote{In the rare event of missing assessed values for the year of sale, we use the assessed value from the previous reassessment cycle before the year of sale, otherwise we drop this sale altogether from the data}, (4) are inferred as arm's-length transactions since prices of non-arm's-length sales are not reflective of the fair market value\footnote{We remove deed types that are plausibly non-arms-length transactions between willing buyers and sellers, such as quit claim deed, sales between related parties, executor deed, etc. We have removed approximately 8\% of the sample that are inferred as non-arms-lengths transactions}, and (5) sold for at least \$10,000\footnote{This is in line with CCAO's practice of removing outliers in sale prices. Sale prices under \$10,000 can be measurement errors and/or non-arms-length transactions, both of which would affect our analyses.}. In addition, we remove outliers in sale prices above the 99.5th percentile.  We construct the base sample at the sale level, where each observation is indexed by the property ID and the year-quarter of the sale. It is important to note that the observed sale price is property-year-quarter specific and the assessed value is property-reassessment-cycle(3 years) specific.

\begin{table}[!htbp]
\centering
\small
\caption{Summary Statistics}
\label{tab:summary_stats}
\begin{tabular}{lcccc|ccc}
\toprule
 & All & All & All  & All & Chicago & North& South\\
 &Mean &SD& Min & Max&Mean & Mean & Mean \\
\midrule
\multicolumn{8}{l}{\textit{Panel A: Property-level variables}} \\
Sale price (\$) & 339,419 & 270,143 & 10,000 & 2,075,000 & 362,813 & 420,040 & 261,242 \\
Mailed assessed value (\$) & 28,914 & 24,135 & 1,500 & 598,385 & 31,251 & 36,564 & 21,394 \\
Certified (assessor-stage) assessed value (\$) & 28,669 & 23,646 & 1,256 & 354,060 & 30,939 & 36,211 & 21,283 \\
Ln(assessment$\times$10/sale price) & -0.151 & 0.364 & -2.439 & 2.114 & -0.128 & -0.137 & -0.180 \\
Absolute assessment error (\%) & 28.8 & 31.5 & 0.0 & 728.4 & 33.5 & 20.4 & 30.6 \\
CBG minority share (\%) & 44.0 & 34.7 & 0.0 & 100.0 & 59.5 & 18.4 & 48.8 \\
\midrule
\multicolumn{8}{l}{\textit{Panel B: Sample coverage}} \\
Observations & \multicolumn{4}{c}{176,567}& 59,541 & 48,841 & 68,185 \\
Townships & \multicolumn{4}{c}{38}& 8 & 13 & 17 \\
Neighborhoods & \multicolumn{4}{c}{845}& 187 & 273 & 385 \\
Census tracts & \multicolumn{4}{c}{1,287}& 753 & 251 & 285 \\
Census block groups &\multicolumn{4}{c}{3,738}& 1,965 & 787 & 988 \\
\bottomrule
\end{tabular}
\begin{minipage}{0.96\textwidth}
%\footnotesize \textit{Notes:} This table summarizes the main analysis sample used in the baseline difference-in-differences estimates. The sample consists of arm's-length residential sales from 2018--2022 in Cook County, excluding North-triad observations in 2022 because that triad begins its own rollout in that year. The mailed, certified, and board assessed values are reported in assessor dollars; multiplying them by 10 converts them to market-value terms. The log mailed assessment ratio is $\ln(10 \times \text{Mailed}/\text{Sale Price})$. Absolute mailed error is $|10 \times \text{Mailed}/\text{Sale Price} - 1| \times 100$, measured in percentage points. Neighborhood minority share and neighborhood Black share are measured at the Census Block Group level.
\end{minipage}
\end{table}

\section{Chicago's Regressivity Effect: Alternative Explanations and Robustness}
\label{app:robustness}

Table~\ref{tab:robustness_summary} in Section~\ref{sec:robustness_checks} summarizes four checks on the Chicago 2021 estimate. This appendix presents them in the same order, giving for each the concern, the design, the estimates, and what they rule out, and closes with a randomization-inference check on the main estimates.

\subsection{A new administration}
\label{app:rob_admin}

The reformed algorithm was introduced by an assessor who took office in December 2018, and the new administration changed more than the valuation model: it changed staff, data practices, and transparency. The Chicago estimate could reflect these changes rather than the algorithm.

The new administration reassessed the North triad in 2019 and the South triad in 2020 under its earlier workflow, township-level linear or gradient-boosted models in the predecessor codebase \citep{CCAOModelResAVM}, before it deployed the reformed algorithm, with its countywide LightGBM model, new data pipeline, and documented post-modeling process, in Chicago in 2021. The office's documentation records ``linear regression or GBM model per township'' for the 2019 and 2020 assessment years and a ``county-wide LightGBM model'' from 2021 \citep{CCAOModelResAVM}, so these two events hold the administration fixed and vary the valuation system; they are reassessments conducted by the new administration without the reformed algorithm, not algorithm-free placebos. For each, we estimate Equation~\eqref{eq:did_slopes} on a two-year window centered on the reassessment. Column (6) of Table~\ref{tab:early_reassess_did} reports the Chicago 2021 estimate on a window of the same length as a benchmark.

\begin{table}[!htbp]\centering
\footnotesize
\caption{The Same Design Applied to the New Administration's Earlier-Workflow Reassessments}\label{tab:early_reassess_did}
\begin{tabular}{lccccc}\toprule
 & (1) & (2) & (3) & (4) & (5) \\
Event & North 2019 & North 2019 & South 2020 & South 2020 & Chicago 2021 \\
Comparison triads & South + Chi. & South + Chi. & Chi. + North & Chi. + North & North + South \\
Sample window & 2018q1--19q4 & 2018q1--19q4 & 2019q1--20q4 & 2019q1--20q4 & 2020q1--21q4 \\
Specification & Price & Price + race & Price & Price + race & Price + race \\
\midrule
Average post effect: ln(price) slope & -0.002 &  0.002 &  0.031 &  0.051** &  0.220*** \\
 & (0.024) & (0.022) & (0.023) & (0.020) & (0.035) \\
\addlinespace
Average post effect: minority slope & & -0.006 & &  0.103 &  0.138*** \\
 & & (0.047) & & (0.065) & (0.041) \\
\addlinespace
Joint pre-trend p: ln(price) slope & 0.452 & 0.066 & 0.742 & 0.923 & 0.165 \\
Observations &    73,397 &    73,397 &    75,431 &    75,431 &    78,449 \\
Town-by-quarter FE& YES & YES & YES & YES & YES \\
Town- and quarter-specific slopes & YES & YES & YES & YES & YES \\
\bottomrule\end{tabular}

\begin{minipage}{0.98\textwidth}\footnotesize \textit{Notes:} Each column estimates Equation \eqref{eq:did_slopes} around the indicated event and reports the average of the four post-quarter Treat $\times$ Post $\times$ slope interactions; the reference period is the quarter before the event. North 2019 and South 2020 are the first reassessments conducted by the administration that took office in December 2018, under its earlier township-level workflow and before the reformed algorithm. In every column the comparison group is the two other triads, as in the main design. Column (5) is a benchmark: the Chicago 2021 algorithmic reassessment estimated on a window of the same length; its inference is reported with the main estimates in Table \ref{tab:regressivity_effect}. The outcome is ln(Assessed/Sale Price) using initial (mailed) values. Township-clustered standard errors in parentheses. *** p$<$0.01, ** p$<$0.05, * p$<$0.1.\end{minipage}
\end{table}

For North 2019, the average post-period effect on the price slope is $-0.002$ (s.e.\ 0.024) and the effect on the minority slope is $-0.006$ (0.047). For South 2020, the effect on the price slope is 0.031 (0.023) in the price-only specification and 0.051 (0.020) in the joint specification, and the effect on the minority slope is 0.103 (0.065). For Chicago 2021, with the reformed algorithm, the effects are 0.220 (0.035) and 0.138 (0.041).

We do not claim that the early reassessments changed nothing. The South estimates are positive, and that triad started with more regressivity than the North. But they are between one tenth and one quarter of the Chicago price effect. The combined evidence here suggests that the new administration by itself does not produces the large, immediate slope change we see in Chicago in 2021.

\subsection{The 2021 post-model adjustment}
\label{app:rob_step}

The second check concerns a part of the 2021 package that was built to do what we measure. After the model produced its predictions, the 2021 pipeline divided each township's parcels into fifths by predicted value and shifted every fifth so that the median sales ratio of its recently sold parcels moved toward one, capped at 40 percent; it also capped the ratios of parcels sold in the prior three years (Section~\ref{sec:newsystem}). The office describes this post-adjustment step as a guard against regressivity, and that is what it is: if a township's cheapest fifth was over-assessed relative to its priciest fifth, the step pushed the two toward the same ratio, which is a flatter within-township price slope. The 2022--2025 models dropped the post-adjustment step. One could reasonably ask whether Chicago's 2021 result was the post-model adjustment rather than the model and the data, with the suburbs showing what the model and data do.

Chicago’s 2024 reassessment offers a direct test, because it used the reformed algorithm without the post-model adjustment. If the 2021 flattening had come from the adjustment, then replacing the 2021 values, which carried it, with 2024 values, which did not, should have steepened Chicago’s price slope again, by roughly the step’s share of the 2021 gain. It did not. On the same 2024 sales, the frozen 2021 values give a slope of $-0.26$ and the new 2024 values $-0.29$, a reassessment jump of $-0.03$ (the Chicago 2024 row of Appendix Table~\ref{tab:slope_decomposition}).

The DiD estimate around the 2024 reassessment, which measures the same change relative to the North, is $-0.05$ with a standard error of 0.06 (Table~\ref{tab:five_events}); it rules out a reversal of the full 2021 gain of 0.22, though not of a part of it. The 2024 values are also as accurate as the 2021 values (a mean absolute error of 28.7 percent against 28.0, down from 40.0 in 2018), and their slope of $-0.29$ keeps about seven tenths of the ground gained between 2018 ($-0.45$) and 2021 ($-0.21$) (Panel C of Appendix Table~\ref{tab:slope_decomposition}). The model without the post-model adjustment  held what the model with it had set.

\subsection{COVID-era housing market}
\label{app:rob_covid}

The third check addresses the concern that our estimates reflect COVID-era housing market dynamics rather than the algorithmic reform. That is, the effects may be mechanically driven by changes in the denominator (i.e. sale price), not the numerator (i.e., assessment). In particular, one might worry that the pandemic increased demand for relatively lower-priced properties in Chicago more than in the comparison triads, raising their sale prices and mechanically reducing measured price regressivity even if assessments did not improve. Under this story, our results would capture differential post-COVID appreciation patterns rather than a change in assessment practice.

The timing of our event-study estimates is inconsistent with this explanation. If COVID-induced demand shifts were gradually increasing sale prices for lower-priced Chicago properties, then we would expect the estimated treatment effect to emerge progressively and grow over subsequent quarters. Instead, we observe a large, discrete break exactly at event time $\tau = 0$, followed by effects that remain relatively stable through $\tau = 1,2,3,4,5,6,$ and $7$. This pattern is much more consistent with an institutional change occurring at the reassessment date than with a smooth trend in relative market appreciation.

We also address this concern with a more demanding specification. Specifically, we conduct a robustness check to the main DiD that allows for neighborhood-by-quarter fixed effects, neighborhood-specific slope baselines, and quarter-specific slope baselines (Table \ref{tab:regressivity_effect_robustness_nbhd}). This specification is useful because it absorbs highly localized time-varying shocks to housing demand and prices. Consequently, the price and race slopes would be interpreted as the within-neighborhood slopes. The price-slope results remain (pooled post effect 0.157), indicating that they are not driven by neighborhood-specific COVID market dynamics. The minority-slope effect falls to 0.057 and is not significant in this specification, which is expected: neighborhood-by-quarter fixed effects absorb most of the variation in block-group minority share, which varies little within a neighborhood.

%%%%%%%%%%%%%%%%%%%%
%% Nbhd FE and slope baselines
%%%%%%%%%%%%%%%%%%%%

\begin{table}[!htbp]
\footnotesize
\centering
\caption{Robustness: Price and Race Slopes --- Neighborhood FE and Neighborhood-specific Slopes}
\label{tab:regressivity_effect_robustness_nbhd}
\begin{tabular}{lcccc}
\toprule
 & \multicolumn{2}{c}{Preferred inference:} & \multicolumn{2}{c}{Secondary inference:} \\
  & \multicolumn{2}{c}{triad-level wild bootstrap} & \multicolumn{2}{c}{neighborhood-clustered SE} \\
    & \multicolumn{2}{c}{\emph{(p values in parentheses)}} & \multicolumn{2}{c}{\emph{(s.e. in parentheses)}} \\
\cmidrule(lr){2-3} \cmidrule(lr){4-5}
 & (1) & (2) & (3) & (4) \\
 & Price & Price + Race & Price & Price + Race \\
\midrule
Treat $\times$ Pre4 $\times$ ln(price) &    -0.044 ( 0.601) &    -0.042 ( 0.605) &    -0.044* (    0.023) &    -0.042* (    0.023) \\
Treat $\times$ Pre3 $\times$ ln(price) &     0.023 ( 0.438) &     0.025 ( 0.430) &     0.023 (    0.020) &     0.025 (    0.020) \\
Treat $\times$ Pre2 $\times$ ln(price) &    -0.005 ( 0.525) &    -0.005 ( 0.376) &    -0.005 (    0.019) &    -0.005 (    0.019) \\
Treat $\times$ Post0 $\times$ ln(price) &     0.127* ( 0.067) &     0.129** ( 0.036) &     0.127*** (    0.028) &     0.129*** (    0.029) \\
Treat $\times$ Post1 $\times$ ln(price) &     0.130 ( 0.100) &     0.129 ( 0.133) &     0.130*** (    0.025) &     0.129*** (    0.025) \\
Treat $\times$ Post2 $\times$ ln(price) &     0.149 ( 0.157) &     0.151 ( 0.164) &     0.149*** (    0.024) &     0.151*** (    0.024) \\
Treat $\times$ Post3 $\times$ ln(price) &     0.224 ( 0.110) &     0.226* ( 0.099) &     0.224*** (    0.025) &     0.226*** (    0.025) \\
Treat $\times$ Post4 $\times$ ln(price) &     0.182 ( 0.151) &     0.185 ( 0.134) &     0.182*** (    0.028) &     0.185*** (    0.028) \\
Treat $\times$ Post5 $\times$ ln(price) &     0.174 ( 0.155) &     0.180 ( 0.118) &     0.174*** (    0.028) &     0.180*** (    0.028) \\
Treat $\times$ Post6 $\times$ ln(price) &     0.100 ( 0.114) &     0.103 ( 0.120) &     0.100*** (    0.030) &     0.103*** (    0.030) \\
Treat $\times$ Post7 $\times$ ln(price) &     0.153 ( 0.114) &     0.153 ( 0.123) &     0.153*** (    0.030) &     0.153*** (    0.030) \\
Treat $\times$ Pre4 $\times$ minority &  &     0.048 ( 0.132) &  &     0.048 (    0.059) \\
Treat $\times$ Pre3 $\times$ minority &  &    -0.022 ( 0.406) &  &    -0.022 (    0.055) \\
Treat $\times$ Pre2 $\times$ minority &  &     0.053 ( 0.165) &  &     0.053 (    0.047) \\
Treat $\times$ Post0 $\times$ minority &  &     0.083 ( 0.390) &  &     0.083 (    0.065) \\
Treat $\times$ Post1 $\times$ minority &  &    -0.010 ( 0.126) &  &    -0.010 (    0.062) \\
Treat $\times$ Post2 $\times$ minority &  &     0.074 ( 0.132) &  &     0.074 (    0.057) \\
Treat $\times$ Post3 $\times$ minority &  &     0.090 ( 0.128) &  &     0.090 (    0.069) \\
Treat $\times$ Post4 $\times$ minority &  &     0.030 ( 0.135) &  &     0.030 (    0.065) \\
Treat $\times$ Post5 $\times$ minority &  &     0.035* ( 0.086) &  &     0.035 (    0.065) \\
Treat $\times$ Post6 $\times$ minority &  &     0.084 ( 0.129) &  &     0.084 (    0.065) \\
Treat $\times$ Post7 $\times$ minority &  &     0.096 ( 0.149) &  &     0.096 (    0.072) \\
\midrule
2020 price slope &    -0.639* ( 0.061) &    -0.640* ( 0.059) &    -0.639*** (    0.017) &    -0.640*** (    0.017) \\
2020 race slope &  &    -0.077* ( 0.094) &  &    -0.077*** (    0.014) \\
\midrule
Pooled post effect on price slope &     0.154* ( 0.077) &     0.157* ( 0.066) &     0.154*** (    0.020) &     0.157*** (    0.021) \\
Pooled post effect on minority slope &  &     0.057 ( 0.144) &  &     0.057 (    0.048) \\
Joint post p-value: price slope &  0.031 &  0.075 &  0.000 &  0.000 \\
Joint post p-value: minority slope &  &  0.421 &  &  0.405 \\
\midrule
Observations &      175,177 &      175,177 &      175,177 &      175,177 \\
R-squared &     0.731 &     0.736 &     0.731 &     0.736 \\
Neighborhood-by-quarter FE & YES & YES & YES & YES \\
Neighborhood-specific slope baselines & YES & YES & YES & YES \\
Quarter-specific slope baselines & YES & YES & YES & YES \\
\bottomrule
\end{tabular}

\begin{minipage}{0.98\textwidth}
\footnotesize \textit{Notes:} This table presents robustness estimates to the main DiD by allowing neighborhood-by-quarter fixed effects, neighborhood-specific slope baselines, and quarter-specific slope baselines. Across specifications, the outcome variable is ln(Assessed/Sale Price), where Assessed is the initial assessed value. The reference period is $\tau=-1$ (2020 Q4). Columns (1) and (2) are the preferred inference because treatment varies at the triad level; they use wild cluster bootstrap method with Webb weights and 9999 replications, and the numbers in parentheses are bootstrap p-values. The estimation sample contains 3 triads, so these p-values should be interpreted as small-cluster robustness checks rather than as asymptotic approximations.  Columns (3) and (4) report neighborhood-clustered standard errors in parentheses for comparability with the baseline specification and show that point estimates are similar even though precision differs. For all specifications, we do not report here the dynamic treatment effects for the periods before -4 to save space. *** p$<$0.01, ** p$<$0.05, * p$<$0.1.
\end{minipage}
\end{table}

A related, broader concern is that our estimates may be driven by differential neighborhood-specific housing trends over the years (which may predate Covid). We again argue that this underlying trend cannot be fully responsible for the sharp, discontinuous nature of the effect estimates we see here. Nonetheless, we address this concern by a specification that allows baseline regressivity, value slopes, and treatment effects to vary with census-tract-level gentrification (Table \ref{tab:regressivity_effect_robustness_gentrification}). We measure neighborhood gentrification using Census-tract-level classifications from the Urban Displacement Project.\footnote{See \url{https://www.urbandisplacement.org/}, last accessed 04/27/2026. The UDP gentrification measure we use is the National/Regional Displacement \& Gentrification Typology, last updated in 2018, which categorizes neighborhoods into stages such as low-income/stable, at risk of gentrification, early gentrification, ongoing gentrification, and advanced gentrification based on Census and ACS data on income, housing values, and demographic change. This measure is widely used in urban economics and policy work. In our analysis, we use a gentrification dummy for Census tracts that are categorized as one of the following: ``Ongoing displacement (non-gentr)", ``At risk of gentrification", and ``Gentrifying (early/ongoing)".} The results are robust to this flexible heterogeneity, suggesting that differential gentrification is not driving our main findings.

%%%%%%%%%%%%%%%%%%%%%%%%%%%
%% gentrification
%%%%%%%%%%%%%%%%%%%%%%%%%%%

\begin{table}[!htbp]
\footnotesize
\centering
\caption{Robustness: Price and Race Slopes --- Census-Tract level Gentrification}
\label{tab:regressivity_effect_robustness_gentrification}
\begin{tabular}{lcccc}
\toprule
 & \multicolumn{2}{c}{Preferred inference:} & \multicolumn{2}{c}{Secondary inference:} \\
  & \multicolumn{2}{c}{triad-level wild bootstrap} & \multicolumn{2}{c}{township-clustered SE} \\
    & \multicolumn{2}{c}{\emph{(p values in parentheses)}} & \multicolumn{2}{c}{\emph{(s.e. in parentheses)}} \\
\cmidrule(lr){2-3} \cmidrule(lr){4-5}
 & (1) & (2) & (3) & (4) \\
 & Price & Price + Race & Price & Price + Race \\
\midrule
Treat $\times$ Pre4 $\times$ ln(price) &    -0.030 ( 0.365) &    -0.020 ( 0.624) &    -0.030 (    0.021) &    -0.020 (    0.030) \\
Treat $\times$ Pre3 $\times$ ln(price) &     0.030 ( 0.315) &     0.026 ( 0.382) &     0.030 (    0.023) &     0.026 (    0.032) \\
Treat $\times$ Pre2 $\times$ ln(price) &     0.022 ( 0.129) &     0.018 ( 0.121) &     0.022* (    0.012) &     0.018 (    0.016) \\
Treat $\times$ Post0 $\times$ ln(price) &     0.159* ( 0.065) &     0.192** ( 0.045) &     0.159*** (    0.038) &     0.192*** (    0.036) \\
Treat $\times$ Post1 $\times$ ln(price) &     0.183** ( 0.038) &     0.199** ( 0.027) &     0.183*** (    0.028) &     0.199*** (    0.028) \\
Treat $\times$ Post2 $\times$ ln(price) &     0.195 ( 0.123) &     0.216 ( 0.109) &     0.195*** (    0.028) &     0.216*** (    0.031) \\
Treat $\times$ Post3 $\times$ ln(price) &     0.241 ( 0.145) &     0.275 ( 0.140) &     0.241*** (    0.038) &     0.275*** (    0.038) \\
Treat $\times$ Post4 $\times$ ln(price) &     0.188 ( 0.121) &     0.219 ( 0.120) &     0.188*** (    0.047) &     0.219*** (    0.047) \\
Treat $\times$ Post5 $\times$ ln(price) &     0.186 ( 0.161) &     0.218 ( 0.128) &     0.186*** (    0.033) &     0.218*** (    0.033) \\
Treat $\times$ Post6 $\times$ ln(price) &     0.157 ( 0.120) &     0.181 ( 0.147) &     0.157*** (    0.042) &     0.181*** (    0.044) \\
Treat $\times$ Post7 $\times$ ln(price) &     0.176 ( 0.132) &     0.221 ( 0.154) &     0.176*** (    0.049) &     0.221*** (    0.052) \\
Treat $\times$ Pre4 $\times$ minority &  &    -0.046 ( 0.543) &  &    -0.046 (    0.073) \\
Treat $\times$ Pre3 $\times$ minority &  &    -0.044 ( 0.108) &  &    -0.044 (    0.049) \\
Treat $\times$ Pre2 $\times$ minority &  &    -0.003 ( 0.641) &  &    -0.003 (    0.037) \\
Treat $\times$ Post0 $\times$ minority &  &     0.105 ( 0.216) &  &     0.105*** (    0.030) \\
Treat $\times$ Post1 $\times$ minority &  &     0.104 ( 0.121) &  &     0.104** (    0.040) \\
Treat $\times$ Post2 $\times$ minority &  &     0.128 ( 0.131) &  &     0.128** (    0.052) \\
Treat $\times$ Post3 $\times$ minority &  &     0.155 ( 0.107) &  &     0.155*** (    0.040) \\
Treat $\times$ Post4 $\times$ minority &  &     0.160 ( 0.445) &  &     0.160** (    0.061) \\
Treat $\times$ Post5 $\times$ minority &  &     0.137 ( 0.438) &  &     0.137*** (    0.036) \\
Treat $\times$ Post6 $\times$ minority &  &     0.106 ( 0.532) &  &     0.106** (    0.042) \\
Treat $\times$ Post7 $\times$ minority &  &     0.209 ( 0.377) &  &     0.209*** (    0.050) \\
\midrule
2020 price slope &    -0.379* ( 0.063) &    -0.441 ( 0.100) &    -0.379*** (    0.042) &    -0.441*** (    0.048) \\
2020 race slope &  &    -0.412 ( 0.126) &  &    -0.412*** (    0.065) \\
\midrule
Pooled post effect on price slope &     0.187* ( 0.072) &     0.215** ( 0.045) &     0.187*** (    0.034) &     0.215*** (    0.035) \\
Pooled post effect on minority slope &  &     0.135 ( 0.132) &  &     0.135*** (    0.036) \\
Joint post p-value: price slope &  0.091 &  0.055 &  0.000 &  0.000 \\
Joint post p-value: minority slope &  &  0.112 &  &  0.000 \\
\midrule
Observations &      169,499 &      169,499 &      169,499 &      169,499 \\
R-squared &     0.479 &     0.534 &     0.479 &     0.534 \\
Town-by-quarter FE & YES & YES & YES & YES \\
Town-specific slope baselines & YES & YES & YES & YES \\
Quarter-specific slope baselines & YES & YES & YES & YES \\
\bottomrule
\end{tabular}

\begin{minipage}{0.98\textwidth}
\footnotesize \textit{Notes:} This table presents robustness estimates to the main DiD by allowing for baseline, slopes, and effects to vary across Census-tract level gentrification (gentrification-related estimates not reported here since they are not of the primary interest). Across specifications, the outcome variable is ln(Assessed/Sale Price), where Assessed is the initial assessed value. The reference period is $\tau=-1$ (2020 Q4). Columns (1) and (2) are the preferred inference because treatment varies at the triad level; they use wild cluster bootstrap method with Webb weights and 9999 replications, and the numbers in parentheses are bootstrap p-values. The estimation sample contains 3 triads, so these p-values should be interpreted as small-cluster robustness checks rather than as asymptotic approximations.  Columns (3) and (4) report township-clustered standard errors in parentheses for comparability with the baseline specification and show that point estimates are similar even though precision differs. For all specifications, we do not report here the dynamic treatment effects for the periods before -4 to save space. *** p$<$0.01, ** p$<$0.05, * p$<$0.1.
\end{minipage}
\end{table}

\subsection{Alternative minority measures}
\label{app:rob_minority}

Our main measure of racial composition is CBG-level minority share, which matches our focus on neighborhood-level differences in assessment. To show that the findings do not depend on this specific definition, we replace it with Census Block minority share, a more spatially granular measure. The results remain similar (Table \ref{tab:regressivity_effect_robustness_block_minority}).

We also use inferred homeowner race based on the seller's name (Table \ref{tab:regressivity_effect_robustness_seller_minority}).\footnote{Specifically, we use Python's ``surgeo'' package, a standard and widely used method, to infer seller race from surnames using Census surname frequencies, generating probabilistic predictions for White, Black, Hispanic, and Other sellers. This measure is available for about 60\% of valid surnames. Consistent with the idea that seller race captures related but distinct information from neighborhood composition, the inferred seller-minority measure is positively correlated with census-block-group minority share, but only moderately so (corr = 0.3424).} This robustness check is useful because it replaces neighborhood racial composition with an owner-level proxy based on inferred seller race. We find that the price-slope and race-slope results remain qualitatively similar, though less precisely estimated. We do not view this as a contradiction of the main analysis, because inferred seller race identifies a different estimand from our preferred neighborhood-based measure. As previously explained, our main object of interest is not owner-level disparate treatment, but neighborhood-level disparate impact arising from historically segregated housing markets. That is the more relevant margin in this setting: the post-reform assessment process is largely automated and leaves relatively limited scope for case-by-case human intervention, but automated valuations can still inherit spatial patterns shaped by long-run racial segregation. For that reason, neighborhood minority share is our preferred measure, while the seller-race exercise serves as a useful but secondary robustness check.

%%%%%%%%%%%%%%%%%%%%%%%%%%%
%% Block minority
%%%%%%%%%%%%%%%%%%%%%%%%%%%
\begin{table}[!htbp]
\footnotesize
\centering
\caption{Robustness: Price and Race Slopes --- Census-Block-level Minority}
\label{tab:regressivity_effect_robustness_block_minority}
\begin{tabular}{lcccc}
\toprule
 & \multicolumn{2}{c}{Preferred inference:} & \multicolumn{2}{c}{Secondary inference:} \\
  & \multicolumn{2}{c}{triad-level wild bootstrap} & \multicolumn{2}{c}{township-clustered SE} \\
  & \multicolumn{2}{c}{\emph{(p values in parentheses)}} & \multicolumn{2}{c}{\emph{(s.e. in parentheses)}} \\
\cmidrule(lr){2-3} \cmidrule(lr){4-5}
 & (1) & (2) & (3) & (4) \\
 & Price & Price + Race & Price & Price + Race \\
\midrule
Treat $\times$ Pre4 $\times$ ln(price) &    -0.034 ( 0.134) &    -0.016 ( 0.614) &    -0.034 (    0.021) &    -0.016 (    0.028) \\
Treat $\times$ Pre3 $\times$ ln(price) &     0.023 ( 0.344) &     0.029 ( 0.153) &     0.023 (    0.021) &     0.029 (    0.029) \\
Treat $\times$ Pre2 $\times$ ln(price) &     0.018 ( 0.148) &     0.007 ( 0.412) &     0.018 (    0.014) &     0.007 (    0.020) \\
Treat $\times$ Post0 $\times$ ln(price) &     0.154 ( 0.113) &     0.186* ( 0.079) &     0.154*** (    0.040) &     0.186*** (    0.043) \\
Treat $\times$ Post1 $\times$ ln(price) &     0.182** ( 0.047) &     0.211** ( 0.036) &     0.182*** (    0.026) &     0.211*** (    0.028) \\
Treat $\times$ Post2 $\times$ ln(price) &     0.200 ( 0.121) &     0.223 ( 0.108) &     0.200*** (    0.027) &     0.223*** (    0.037) \\
Treat $\times$ Post3 $\times$ ln(price) &     0.234 ( 0.127) &     0.278 ( 0.160) &     0.234*** (    0.035) &     0.278*** (    0.041) \\
Treat $\times$ Post4 $\times$ ln(price) &     0.181 ( 0.122) &     0.214 ( 0.123) &     0.181*** (    0.047) &     0.214*** (    0.049) \\
Treat $\times$ Post5 $\times$ ln(price) &     0.187 ( 0.142) &     0.220 ( 0.126) &     0.187*** (    0.032) &     0.220*** (    0.037) \\
Treat $\times$ Post6 $\times$ ln(price) &     0.160 ( 0.156) &     0.189 ( 0.112) &     0.160*** (    0.044) &     0.189*** (    0.049) \\
Treat $\times$ Post7 $\times$ ln(price) &     0.184 ( 0.133) &     0.237 ( 0.117) &     0.184*** (    0.050) &     0.237*** (    0.056) \\
Treat $\times$ Pre4 $\times$ minority &  &    -0.031 ( 0.524) &  &    -0.031 (    0.067) \\
Treat $\times$ Pre3 $\times$ minority &  &     0.004 ( 0.686) &  &     0.004 (    0.040) \\
Treat $\times$ Pre2 $\times$ minority &  &    -0.031 ( 0.118) &  &    -0.031 (    0.025) \\
Treat $\times$ Post0 $\times$ minority &  &     0.087 ( 0.277) &  &     0.087** (    0.033) \\
Treat $\times$ Post1 $\times$ minority &  &     0.133 ( 0.122) &  &     0.133*** (    0.040) \\
Treat $\times$ Post2 $\times$ minority &  &     0.122 ( 0.141) &  &     0.122* (    0.072) \\
Treat $\times$ Post3 $\times$ minority &  &     0.158 ( 0.111) &  &     0.158*** (    0.047) \\
Treat $\times$ Post4 $\times$ minority &  &     0.140 ( 0.534) &  &     0.140** (    0.058) \\
Treat $\times$ Post5 $\times$ minority &  &     0.103 ( 0.547) &  &     0.103** (    0.050) \\
Treat $\times$ Post6 $\times$ minority &  &     0.091 ( 0.530) &  &     0.091* (    0.053) \\
Treat $\times$ Post7 $\times$ minority &  &     0.211 ( 0.457) &  &     0.211*** (    0.059) \\
\midrule
2020 price slope &    -0.362* ( 0.063) &    -0.429 ( 0.112) &    -0.362*** (    0.042) &    -0.429*** (    0.049) \\
2020 race slope &  &    -0.394 ( 0.117) &  &    -0.394*** (    0.068) \\
\midrule
Pooled post effect on ln(price) slope &     0.186* ( 0.075) &     0.220* ( 0.064) &     0.186*** (    0.033) &     0.220*** (    0.038) \\
Pooled post effect on minority slope &  &     0.128 ( 0.131) &  &     0.128*** (    0.046) \\
Joint post p-value: ln(price) slope &  0.091 &  0.063 &  0.000 &  0.000 \\
Joint post p-value: minority slope &  &  0.130 &  &  0.000 \\
\midrule
Observations &      176,567 &      176,567 &      176,567 &      176,567 \\
R-squared &     0.466 &     0.527 &     0.466 &     0.527 \\
Town-by-quarter FE & YES & YES & YES & YES \\
Town-specific slope baselines & YES & YES & YES & YES \\
Quarter-specific slope baselines & YES & YES & YES & YES \\
\bottomrule
\end{tabular}

\begin{minipage}{0.98\textwidth}
\footnotesize \textit{Notes:} This table presents robustness estimates to the main DiD by using Census-block-level minority measure. Across specifications, the outcome variable is ln(Assessed/Sale Price), where Assessed is the initial assessed value. The reference period is $\tau=-1$ (2020 Q4). Columns (1) and (2) are the preferred inference because treatment varies at the triad level; they use wild cluster bootstrap method with Webb weights and 9999 replications, and the numbers in parentheses are bootstrap p-values. The estimation sample contains 3 triads, so these p-values should be interpreted as small-cluster robustness checks rather than as asymptotic approximations.  Columns (3) and (4) report township-clustered standard errors in parentheses for comparability with the baseline specification and show that point estimates are similar even though precision differs. For all specifications, we do not report here the dynamic treatment effects for the periods before -4 to save space. *** p$<$0.01, ** p$<$0.05, * p$<$0.1.
\end{minipage}
\end{table}

%%%%%%%%%%%%%%%%%%%%%%%%%%%
%% inferred owner minority
%%%%%%%%%%%%%%%%%%%%%%%%%%%

\begin{table}[!htbp]
\footnotesize
\centering
\caption{Robustness: Price and Race Slopes --- Inferred Owner Minority from Seller Name} 
    \label{tab:regressivity_effect_robustness_seller_minority}
\begin{tabular}{lcccc}
\toprule
 & \multicolumn{2}{c}{Preferred inference:} & \multicolumn{2}{c}{Secondary inference:} \\
  & \multicolumn{2}{c}{triad-level wild bootstrap} & \multicolumn{2}{c}{township-clustered SE} \\
  & \multicolumn{2}{c}{\emph{(p values in parentheses)}} & \multicolumn{2}{c}{\emph{(s.e. in parentheses)}} \\
\cmidrule(lr){2-3} \cmidrule(lr){4-5}
 & (1) & (2) & (3) & (4) \\
 & Price & Price + Race & Price & Price + Race \\
\midrule
Treat $\times$ Pre4 $\times$ ln(price) &    -0.021 ( 0.613) &    -0.005 ( 0.724) &    -0.021 (    0.022) &    -0.005 (    0.022) \\
Treat $\times$ Pre3 $\times$ ln(price) &    -0.002 ( 0.830) &     0.008 ( 0.654) &    -0.002 (    0.035) &     0.008 (    0.029) \\
Treat $\times$ Pre2 $\times$ ln(price) &    -0.010 ( 0.389) &    -0.005 ( 0.630) &    -0.010 (    0.015) &    -0.005 (    0.019) \\
Treat $\times$ Post0 $\times$ ln(price) &     0.128 ( 0.159) &     0.140 ( 0.111) &     0.128*** (    0.040) &     0.140*** (    0.033) \\
Treat $\times$ Post1 $\times$ ln(price) &     0.157 ( 0.161) &     0.167 ( 0.129) &     0.157*** (    0.035) &     0.167*** (    0.031) \\
Treat $\times$ Post2 $\times$ ln(price) &     0.159 ( 0.124) &     0.175 ( 0.117) &     0.159*** (    0.045) &     0.175*** (    0.037) \\
Treat $\times$ Post3 $\times$ ln(price) &     0.213 ( 0.135) &     0.225 ( 0.140) &     0.213*** (    0.037) &     0.225*** (    0.033) \\
Treat $\times$ Post4 $\times$ ln(price) &     0.139 ( 0.132) &     0.150 ( 0.149) &     0.139** (    0.061) &     0.150** (    0.056) \\
Treat $\times$ Post5 $\times$ ln(price) &     0.155 ( 0.134) &     0.168 ( 0.140) &     0.155*** (    0.045) &     0.168*** (    0.039) \\
Treat $\times$ Post6 $\times$ ln(price) &     0.159 ( 0.135) &     0.172 ( 0.135) &     0.159*** (    0.056) &     0.172*** (    0.050) \\
Treat $\times$ Post7 $\times$ ln(price) &     0.171 ( 0.155) &     0.184 ( 0.138) &     0.171*** (    0.057) &     0.184*** (    0.051) \\
Treat $\times$ Pre4 $\times$ minority &  &    -0.001 ( 0.867) &  &    -0.001 (    0.045) \\
Treat $\times$ Pre3 $\times$ minority &  &     0.046 ( 0.379) &  &     0.046 (    0.031) \\
Treat $\times$ Pre2 $\times$ minority &  &     0.045** ( 0.044) &  &     0.045 (    0.038) \\
Treat $\times$ Post0 $\times$ minority &  &     0.093 ( 0.143) &  &     0.093* (    0.052) \\
Treat $\times$ Post1 $\times$ minority &  &     0.081 ( 0.102) &  &     0.081 (    0.054) \\
Treat $\times$ Post2 $\times$ minority &  &     0.111 ( 0.143) &  &     0.111* (    0.059) \\
Treat $\times$ Post3 $\times$ minority &  &     0.051* ( 0.060) &  &     0.051 (    0.051) \\
Treat $\times$ Post4 $\times$ minority &  &     0.065 ( 0.367) &  &     0.065 (    0.064) \\
Treat $\times$ Post5 $\times$ minority &  &     0.105 ( 0.127) &  &     0.105* (    0.054) \\
Treat $\times$ Post6 $\times$ minority &  &     0.091 ( 0.194) &  &     0.091* (    0.054) \\
Treat $\times$ Post7 $\times$ minority &  &     0.091 ( 0.359) &  &     0.091* (    0.051) \\
\midrule
2020 price slope &    -0.317* ( 0.081) &    -0.322* ( 0.066) &    -0.317*** (    0.042) &    -0.322*** (    0.041) \\
2020 race slope &  &    -0.105 ( 0.117) &  &    -0.105*** (    0.020) \\
\midrule
Pooled post effect on price slope &     0.160 ( 0.139) &     0.173 ( 0.141) &     0.160*** (    0.042) &     0.173*** (    0.035) \\
Pooled post effect on minority slope &  &     0.087 ( 0.158) &  &     0.087 (    0.052) \\
Joint post p-value: price slope &  0.110 &  0.117 &  0.000 &  0.000 \\
Joint post p-value: minority slope &  &  0.111 &  &  0.083 \\
\midrule
Observations &      101,256 &      101,256 &      101,256 &      101,256 \\
R-squared &     0.427 &     0.436 &     0.427 &     0.436 \\
Town-by-quarter FE & YES & YES & YES & YES \\
Town-specific slope baselines & YES & YES & YES & YES \\
Quarter-specific slope baselines & YES & YES & YES & YES \\
\bottomrule
\end{tabular}

\begin{minipage}{0.98\textwidth}
\footnotesize \textit{Notes:} This table presents robustness estimates to the main DiD by using the inferred owner minority probability using their names. Across specifications, the outcome variable is ln(Assessed/Sale Price), where Assessed is the initial assessed value. The reference period is $\tau=-1$ (2020 Q4). Columns (1) and (2) are the preferred inference because treatment varies at the triad level; they use wild cluster bootstrap method with Webb weights and 9999 replications, and the numbers in parentheses are bootstrap p-values. The estimation sample contains 3 triads, so these p-values should be interpreted as small-cluster robustness checks rather than as asymptotic approximations.  Columns (3) and (4) report township-clustered standard errors in parentheses for comparability with the baseline specification and show that point estimates are similar even though precision differs. For all specifications, we do not report here the dynamic treatment effects for the periods before -4 to save space. *** p$<$0.01, ** p$<$0.05, * p$<$0.1.
\end{minipage}
\end{table}

%%%%%%%%%%%%%%%%%%%%
%% Randomization Inference
%%%%%%%%%%%%%%%%%%%%

\subsection{Randomization inference}
\label{app:rob_ri}

As an additional design-based check, we implement exact randomization inference exploiting the fact that treatment is assigned at the triad level. Restricting the sample to 2018--2021, we re-estimate the main event-study specification after reassigning treatment in turn to each of the three triads while holding the 2021 reform timing fixed. This avoids reliance on large-cluster asymptotics and asks whether Chicago's estimated effect is unusually large relative to the other admissible assignments. As reported in Table \ref{tab:randomization_inference_main}, the results are directionally consistent with the main estimates. The exact p-values are 0.333 throughout, which is necessarily coarse given that there are only three possible assignments, so we interpret this exercise as a low-powered but reassuring robustness check.

\begin{table}[h!]
\small
\centering
\caption{Exact Randomization Inference for Main DiD Specifications}
\label{tab:randomization_inference_main}
\begin{tabular}{lcccc}
\toprule
 & \multicolumn{2}{c}{Price-only} & \multicolumn{2}{c}{Price + Race} \\
\cmidrule(lr){2-3} \cmidrule(lr){4-5}
 & Coef. & Exact RI p-value & Coef. & Exact RI p-value \\
\midrule
\multicolumn{5}{l}{\textit{Panel A. Price slope}} \\
Treat $\times$ Post0 $\times$ ln(price) &     0.155 &  0.333 &     0.187 &  0.333 \\
Treat $\times$ Post1 $\times$ ln(price) &     0.180 &  0.333 &     0.197 &  0.333 \\
Treat $\times$ Post2 $\times$ ln(price) &     0.199 &  0.333 &     0.222 &  0.333 \\
Treat $\times$ Post3 $\times$ ln(price) &     0.235 &  0.333 &     0.273 &  0.333 \\
Average post effect on ln(price) slope &     0.192 &  0.333 &     0.220 &  0.333 \\
\midrule
\multicolumn{5}{l}{\textit{Panel B. Within-price race slope}} \\
Treat $\times$ Post0 $\times$ minority &  &  &     0.098 &  0.333 \\
Treat $\times$ Post1 $\times$ minority &  &  &     0.109 &  0.333 \\
Treat $\times$ Post2 $\times$ minority &  &  &     0.138 &  0.333 \\
Treat $\times$ Post3 $\times$ minority &  &  &     0.153 &  0.333 \\
Average post effect on minority slope &  &  &     0.125 &  0.333 \\
\bottomrule
\end{tabular}

\begin{minipage}{0.98\textwidth}
\footnotesize \textit{Notes:} This table reports exact reassignment-based randomization-inference p-values for the main Cook County DiD specifications. We keep the event-time timing fixed at the actual 2021 rollout date and reassign the treated triad across the three admissible triads in the estimation sample. Because the North triad begins its own rollout in 2022, the randomization-inference sample is truncated at 2021Q4 so that all three triads remain admissible assignments over a common untreated-control window. The observed statistic is the estimate under the actual Chicago assignment. For the price-slope rows, the reported exact RI p-values are right-tail p-values, computed as the share of the three admissible assignments whose estimated coefficient is at least as large as Chicago's estimated coefficient. For the within-price race-slope rows, the reported exact RI p-values are two-sided absolute-value p-values. Because there are only three admissible treated-triad assignments, these p-values are necessarily coarse. 
\end{minipage}
\end{table}

\newpage
%%%%%%%%%%%%%%%%%%%%%%%%%%%%%%%%%%%%%%%%
\section{Do Appeals Undo the Algorithm's Equity Improvements?}
\label{app:appeals}
%%%%%%%%%%%%%%%%%%%%%%%%%%%%%%%%%%%%%%%%

Our main analysis focuses on the assessor’s initial output---the mailed assessed value. Cook County owners can appeal it twice: first to the Assessor's Office, which issues a \emph{certified} value, and then to the Board of Review, whose value is the basis for the tax bill.  This raises a natural concern: if appeals are disproportionately used (and successfully used) by higher-value or lower-minority areas, the appeals stage could partially (or fully) undo the equity gains generated by the algorithmic reform at the mailing stage.

We address this concern in two steps. First, we document who appeals and what they obtain. Second, we re-estimate our main DiD design using the post-appeal certified value and Board value as the outcomes. The short answer is that appeals are indeed socioeconomically patterned, but they do not materially attenuate the reform’s improvement in the price gradient, and the post-appeal results are extremely similar to the mailed-value results.

\paragraph{Who appeals and what they obtain.} Table~\ref{tab:appeals_regs} uses the CCAO appeals records, which are populated only from 2021 onward, and measures outcomes with the Board of Review value, so that reductions won at either stage are counted. Two patterns stand out. First, appeals and their rewards rise steeply with value. A one-log-point higher assessed value is associated with a 6.5 percentage point higher probability of appealing, against a mean of 13 percent, and, among appellants, with a 23 point higher chance of ending up with a lower value and a 3 percentage point larger reduction, against means of 50 percent and 4.6 percent. Second, homes in higher-minority neighborhoods are less likely to appeal and less likely to hire professional representation, although those who do appeal obtain somewhat larger reductions, a pattern consistent with owners appealing when over-assessment is salient.\footnote{We do not observe over-assessment among non-appellants, so we read the conditional patterns descriptively. The appeals file records appeals to the Assessor only; owners may also appeal directly to the Board of Review, and about 4 percent of parcels with no Assessor appeal receive a Board reduction each year. Column (2) counts those reductions; columns (3)--(5) condition on an Assessor appeal because Board-only appellants who obtained nothing cannot be identified.} Both patterns could, in principle, reintroduce regressivity after the mailing stage.

\begin{table}[h!]
\centering
\caption{Appeals Analysis: Propensity, Outcomes, and Representation}
\label{tab:appeals_regs}
\small
\begin{tabular}{lccccc}
\toprule
 & (1) & (2) & (3) & (4) & (5) \\
 & Appealed & Pct Reduction & Success & Pct Reduction & Has Agent \\
  & (0/1) & ($\times$100) & (0/1) & ($\times$100) &  (0/1) \\
 & (All) & (All) & (Appeal=1) & (Appeal=1) & (Appeal=1) \\
\midrule
ln(Assessment) & 0.065*** & 0.856*** & 0.229*** & 3.028*** & 0.016 \\
 & (0.017) & (0.203) & (0.015) & (0.296) & (0.025) \\
minority & -0.054*** & -0.116 & 0.077*** & 2.078*** & -0.124*** \\
 & (0.011) & (0.154) & (0.019) & (0.273) & (0.042) \\
\midrule
DV Sample Mean & 0.134 & 0.929 & 0.504 & 4.564 & 0.619 \\
Observations & 3,703,054 & 3,703,054 & 497,908 & 497,908 & 497,908 \\
R-squared & 0.108 & 0.073 & 0.069 & 0.091 & 0.107 \\
Township $\times$ Year FE & Yes & Yes & Yes & Yes & Yes \\
Class $\times$ Year FE & Yes & Yes & Yes & Yes & Yes \\
\bottomrule
\end{tabular}
\begin{minipage}{0.95\textwidth}
\vspace{0.3em}
\footnotesize
\textit{Notes:} Regressions of appeal behavior and outcomes on log mailed assessment and the block-group minority share (Black plus Hispanic, as a fraction between 0 and 1), for all residential parcels in tax years 2021--2024, the years for which the CCAO appeals records are populated and the Board of Review has ruled. Column (1): indicator for filing an appeal with the Assessor's Office. Column (2): percentage reduction between the mailed value and the Board of Review value, $\frac{\text{Mailed} - \text{Board}}{\text{Mailed}} \times 100$, which is zero for parcels whose value was not reduced at either stage and includes reductions won at the Board by owners who did not appeal to the Assessor. Columns (3)--(5) restrict to parcels that appealed to the Assessor. Column (3): the Board of Review value is below the mailed value. Column (4): the same percentage reduction as column (2). Column (5): professional representation (agent hired). The Board of Review value equals the certified value unless the Board reduced it, which it did for 36 percent of Assessor-stage appellants and 4 percent of other parcels; it exceeds the certified value for 0.01 percent of parcels. Reductions are winsorized at the 0.5 and 99.5 percentiles. All specifications include township-by-year and property-class-by-year fixed effects. Standard errors clustered at township level in parentheses. *** $p<0.01$, ** $p<0.05$, * $p<0.1$.
\end{minipage}
\end{table}

\paragraph{The reform's effect at each stage.} Table~\ref{tab:regressivity_effect_by_stage} re-estimates the joint specification of Equation~\eqref{eq:did_slopes} on the 2018--2022 sample of Table~\ref{tab:regressivity_effect} with each of the three values as the outcome. Appeals touch a meaningful share of the sample: the certified value differs from the mailed value for about 10 percent of sales and the Board of Review value for about 15 percent, with mean reductions of 0.5 and 1.2 percent across all sales.\footnote{The 2020 shares are higher because the COVID adjustment of Section~\ref{sec:setting} was applied between the mailed and certified stages.} The estimates do not move. The pooled post effect on the price slope is 0.216 with mailed values, 0.219 with certified values, and 0.216 with Board of Review values; the effect on the within-price minority slope is 0.130, 0.137, and 0.139. The quarterly paths agree to within 0.01 at every horizon, the pre-period coefficients are near zero at every stage, and the 2020 baseline slopes are the same. Appeals redistribute some gains at the margin, but they do not offset the reform: the values on which tax bills are based are as much less regressive as the mailed values.

\begin{table}[t!]\centering\footnotesize
\caption{The Reform's Effect at Each Stage of the Assessment Process}\label{tab:regressivity_effect_by_stage}
\begin{tabular}{lccc}\toprule
 & Mailed & Certified & Board of Review \\
 & (initial value) & (after Assessor appeals) & (after both appeal stages) \\
\midrule
Treat $\times$ Pre4 $\times$ ln(price) & $-0.022$ (0.580) [0.030] & $-0.013$ (0.615) [0.033] & $-0.012$ (0.621) [0.032] \\
Treat $\times$ Pre3 $\times$ ln(price) & 0.018 (0.374) [0.031] & 0.021 (0.366) [0.033] & 0.021 (0.380) [0.032] \\
Treat $\times$ Pre2 $\times$ ln(price) & 0.013 (0.146) [0.017] & 0.015 (0.120) [0.020] & 0.018 (0.128) [0.020] \\
Treat $\times$ Post0 $\times$ ln(price) & 0.187 (0.122) [0.039] & 0.189 (0.035) [0.036] & 0.188 (0.059) [0.038] \\
Treat $\times$ Post1 $\times$ ln(price) & 0.200 (0.036) [0.028] & 0.202 (0.046) [0.026] & 0.199 (0.038) [0.027] \\
Treat $\times$ Post2 $\times$ ln(price) & 0.223 (0.100) [0.033] & 0.222 (0.056) [0.032] & 0.218 (0.067) [0.033] \\
Treat $\times$ Post3 $\times$ ln(price) & 0.272 (0.160) [0.038] & 0.277 (0.140) [0.038] & 0.269 (0.136) [0.038] \\
Treat $\times$ Post4 $\times$ ln(price) & 0.216 (0.159) [0.050] & 0.222 (0.131) [0.046] & 0.214 (0.126) [0.047] \\
Treat $\times$ Post5 $\times$ ln(price) & 0.218 (0.128) [0.035] & 0.223 (0.136) [0.032] & 0.222 (0.135) [0.033] \\
Treat $\times$ Post6 $\times$ ln(price) & 0.185 (0.138) [0.047] & 0.188 (0.121) [0.045] & 0.188 (0.141) [0.045] \\
Treat $\times$ Post7 $\times$ ln(price) & 0.226 (0.145) [0.055] & 0.234 (0.135) [0.055] & 0.232 (0.134) [0.055] \\
Treat $\times$ Pre4 $\times$ minority & $-0.035$ (0.570) [0.068] & $-0.032$ (0.524) [0.069] & $-0.032$ (0.594) [0.068] \\
Treat $\times$ Pre3 $\times$ minority & $-0.034$ (0.126) [0.042] & $-0.028$ (0.139) [0.045] & $-0.029$ (0.121) [0.044] \\
Treat $\times$ Pre2 $\times$ minority & 0.001 (0.696) [0.032] & 0.007 (0.587) [0.036] & 0.006 (0.535) [0.036] \\
Treat $\times$ Post0 $\times$ minority & 0.098 (0.125) [0.031] & 0.104 (0.115) [0.029] & 0.113 (0.149) [0.031] \\
Treat $\times$ Post1 $\times$ minority & 0.110 (0.140) [0.044] & 0.117 (0.120) [0.042] & 0.118 (0.148) [0.041] \\
Treat $\times$ Post2 $\times$ minority & 0.137 (0.140) [0.056] & 0.145 (0.130) [0.053] & 0.150 (0.170) [0.053] \\
Treat $\times$ Post3 $\times$ minority & 0.151 (0.145) [0.042] & 0.159 (0.128) [0.042] & 0.160 (0.139) [0.040] \\
Treat $\times$ Post4 $\times$ minority & 0.150 (0.397) [0.064] & 0.160 (0.388) [0.061] & 0.156 (0.392) [0.056] \\
Treat $\times$ Post5 $\times$ minority & 0.125 (0.419) [0.036] & 0.129 (0.456) [0.035] & 0.126 (0.439) [0.035] \\
Treat $\times$ Post6 $\times$ minority & 0.094 (0.506) [0.042] & 0.106 (0.533) [0.041] & 0.108 (0.491) [0.040] \\
Treat $\times$ Post7 $\times$ minority & 0.188 (0.395) [0.050] & 0.194 (0.377) [0.050] & 0.198 (0.392) [0.051] \\
\midrule
2020 price slope & $-0.430$ (0.092) [0.048] & $-0.432$ (0.092) [0.048] & $-0.433$ (0.092) [0.047] \\
2020 race slope & $-0.415$ (0.117) [0.067] & $-0.416$ (0.117) [0.067] & $-0.413$ (0.136) [0.067] \\
\midrule
Pooled post effect on ln(price) slope & 0.216 (0.029) [0.037] & 0.219 (0.047) [0.035] & 0.216 (0.036) [0.036] \\
Pooled post effect on minority slope & 0.130 (0.122) [0.038] & 0.137 (0.112) [0.037] & 0.139 (0.103) [0.037] \\
Joint post $p$: ln(price) slope & 0.064 / 0.000 & 0.063 / 0.000 & 0.075 / 0.000 \\
Joint post $p$: minority slope & 0.130 / 0.001 & 0.111 / 0.000 & 0.107 / 0.000 \\
\midrule
Share of sales whose value  value & -- & 9.8\% & 15.1\% \\
differs from the mailed & & & \\
Mean reduction from the mailed value (\%) & -- & 0.47 & 1.19 \\
Observations & 176,567 & 176,567 & 176,567 \\
R-squared & 0.525 & 0.516 & 0.516 \\
\bottomrule\end{tabular}

\begin{minipage}{0.96\textwidth}\footnotesize \textit{Notes:} Each column estimates the joint (price + race) specification of Equation \eqref{eq:did_slopes} on the 2018--2022 sample of Table \ref{tab:regressivity_effect}, with the outcome $\ln(\text{Assessed}/\text{Sale Price})$ built from the value indicated in the column header. The mailed column reproduces columns (2) and (4) of Table \ref{tab:regressivity_effect}. Each cell reports the coefficient, the triad-level wild-cluster bootstrap p-value (Webb weights, 9999 replications) in parentheses, and the township-clustered standard error in brackets. The reference quarter is $\tau=-1$ (2020 Q4); coefficients for quarters before Pre4 are omitted. The pooled post effect replaces the eight post-quarter interactions with a single Treat $\times$ Post interaction; the joint post $p$ tests that all eight post-quarter interactions are zero, with the bootstrap p-value before the slash and the township-clustered p-value after it. The last rows describe the sample: the share of sales whose certified (Board of Review) value differs from the mailed value, and the mean percent reduction from the mailed value, counting untouched parcels as zero. All specifications include township-by-quarter fixed effects, town-specific slopes, and quarter-specific slopes.\end{minipage}
\end{table}

\newpage

\newpage
\section{Heterogeneous Error Reduction and the North's 2022 Reassessment}
\label{app:hetero}

\subsection{Heterogeneous error reduction: specification and estimates}
\label{app:hetero_spec}

Section~\ref{subsec:hetero_main} tests whether the reform's accuracy gains concentrated in the tails of the price and minority-share distributions. The sample is a balanced census-tract-by-year panel for 2018 and 2021, restricted to tracts with at least eight sales in each year, which removes about 25 percent of tracts but only 5 percent of sales. We estimate
\begin{equation}
\label{eq:error_heterogeneity}
\begin{aligned}
y_{ct}
&= \alpha_c + \lambda_t
+ \beta \left(\text{Treat}_c \times \text{Post}_t\right) \\
&\quad + \theta_H \left(\text{Treat}_c \times \text{Post}_t \times \text{HighPrice}_c\right) 
+ \theta_L \left(\text{Treat}_c \times \text{Post}_t\times \text{LowPrice}_c\right) \\
&\quad + \phi_H \left(\text{Treat}_c \times \text{Post}_t \times \text{HighMinority}_c\right)
+ \phi_L \left(\text{Treat}_c \times \text{Post}_t \times \text{LowMinority}_c\right)
+ \varepsilon_{ct}.
\end{aligned}
\end{equation}

Here, $c$ indexes census tracts and $t$ indexes year (2018 or 2021). The outcome $y_{ct}$ is the logged tract-year mean absolute assessment error plus 1, constructed from property-level absolute assessment errors (in percentage terms) and aggregated after trimming at the 99th percentile. We use $\alpha_c$ and $\lambda_t$ to denote tract and year fixed effects. $\text{Treat}_c$ indicates tracts in Chicago, the treated triad, and $\text{Post}_t$ indicates the post-reform year (i.e., 2021). The indicators $\text{HighPrice}_c$ and $\text{LowPrice}_c$ equal one for tracts in the top and bottom terciles of the tract median sale-price distribution, while $\text{HighMinority}_c$ and $\text{LowMinority}_c$ are defined analogously using terciles of tract minority share. The omitted categories are the middle price tercile and the middle minority-share tercile. Thus, $\beta$ captures the post-reform error reduction for middle-tercile tracts, while $\theta_H$, $\theta_L$, $\phi_H$, and $\phi_L$ measure whether that reduction is larger or smaller in the tails. Standard errors are clustered at the tract level.

Table~\ref{tab:error_reduction_where} reports the estimates. In Panel A the coefficient on Treat $\times$ Post is negative, so absolute error fell in the omitted middle group after the reform, and the four tail interactions are also negative, the low-price one imprecisely. Panel B estimates separate regressions within each tercile group with the mean absolute error in percentage points as the outcome and reports the implied percentage reduction; the reductions are largest in the low- and high-price terciles and in the low- and high-minority terciles, and smallest in the middle terciles.

\begin{table}[h!]
\centering
\footnotesize
\caption{Where the Gains Appear: Heterogeneous Error Reduction Across Tracts}
\label{tab:error_reduction_where}
\begin{tabular}{lcccccc}
\toprule
& \multicolumn{6}{c}{\textit{Panel A: Three-way Interaction (Outcome: log(absolute error $\times$ 100 + 1))}} \\
\cmidrule(lr){2-7}
& \multicolumn{6}{c}{(1)} \\
\midrule
Treat$\times$Post & \multicolumn{6}{c}{-0.151***} \\
& \multicolumn{6}{c}{( 0.037)} \\
Treat$\times$Post$\times$High price & \multicolumn{6}{c}{-0.098*} \\
& \multicolumn{6}{c}{( 0.053)} \\
Treat$\times$Post$\times$Low price & \multicolumn{6}{c}{-0.028} \\
& \multicolumn{6}{c}{( 0.074)} \\
Treat$\times$Post$\times$High minority & \multicolumn{6}{c}{-0.131*} \\
& \multicolumn{6}{c}{( 0.075)} \\
Treat$\times$Post$\times$Low minority & \multicolumn{6}{c}{-0.118**} \\
& \multicolumn{6}{c}{( 0.052)} \\
\midrule
Observations & \multicolumn{6}{c}{       1,774} \\
Tract FE & \multicolumn{6}{c}{YES} \\
Year FE & \multicolumn{6}{c}{YES} \\
\% tracts kept & \multicolumn{6}{c}{74.2\%} \\
\% sales kept & \multicolumn{6}{c}{95.6\%} \\
\midrule
&\multicolumn{6}{c}{\textit{Panel B: Separate Regressions (Outcome: Absolute error $\times$ 100)}} \\
\cmidrule(lr){2-7}
& (1) & (2) & (3) & (4) & (5) & (6) \\
& Low Price & Mid Price & High Price & Low Minority & Mid Minority & High Minority \\
\midrule
Treat$\times$Post & -10.348*** & -4.534*** & -7.321*** & -7.169*** & -3.692*** & -10.530*** \\
& ( 1.612) & ( 0.758) & ( 0.714) & ( 0.620) & ( 0.981) & ( 1.526) \\
\midrule
\% error reduction & -32.17\% & -20.57\% & -39.17\% & -39.53\% & -16.06\% & -32.42\% \\
Observations &          574 &          646 &          554 &          632 &          590 &          552 \\
Tract FE & YES & YES & YES & YES & YES & YES \\
Year FE & YES & YES & YES & YES & YES & YES \\
\bottomrule
\end{tabular}
\begin{minipage}{0.96\textwidth}
\footnotesize \textit{Notes:} Panel A reports a single tract-level difference-in-differences regression where the dependent variable is $\log(\text{absolute error}\times 100+1)$, and heterogeneity is captured by interactions of Treat $\times$ Post with terciles of tract median price and tract minority share; the omitted groups are the middle terciles. Panel B reports separate tract-level regressions by price tercile and minority-share tercile, with the dependent variable equal to mean absolute assessment error in percentage points. The sample is a balanced tract-year panel for 2018 and 2021. Property-level absolute assessment errors are trimmed at the 99th percentile before aggregation, and tracts are required to have at least 8 sales in each year. All specifications include tract and year fixed effects. Standard errors clustered at the tract level are shown in parentheses. *** p$<$0.01, ** p$<$0.05, * p$<$0.10.
\end{minipage}
\end{table}

\subsection{The North's 2022 reassessment}
\label{app:north2022}

The North triad was reassessed under the reformed algorithm in 2022, one year after Chicago. This gives a second event on which to run the identical design, with the South triad, not reassessed until 2023, serving as a not-yet-treated control. Table~\ref{tab:north_did_2022} reports the results. The average post-period effect on the price slope is 0.033 (s.e.\ 0.030) in the price-only specification and 0.042 (0.029) in the joint specification, and it is unchanged when we extend the pre-period to 2020q1 (column 3) or add Chicago to the comparison group (column 4, 0.045). None of these is distinguishable from zero at the 5 percent level, and all are less than a quarter of the Chicago benchmark in column (5). The effect on the minority slope, 0.084 (0.044), is positive and marginally significant, about 60 percent of the Chicago estimate. We note that the pre-period coefficients are jointly different from zero in columns (3) and (4), so the longer-window and three-triad versions should be read with more caution than columns (1) and (2), where the pre-trend test does not reject ($p=0.45$ and $0.49$).

We draw two conclusions. First, the design does not return a positive price-slope estimate simply because a triad is reassessed under the reformed algorithm, which supports a causal reading of the Chicago estimate. Second, the null is what one would expect given where the North triad started. In the year before its algorithmic reassessment the North triad's price slope was about $-0.22$, half of Chicago's pre-reform $-0.43$, and its mean absolute assessment error at its previous reassessment was 17.1 percent against Chicago's 40.0 percent. The algorithmic reassessment lowered the North's error only slightly, to 15.9 percent, while it lowered Chicago's to 28.0 percent. Where assessments were already accurate, a more accurate model has little regressivity to remove. Consistent with this, Section~\ref{sec:mechanism} finds no shift from coarse neighborhood averages to fine geography in the North's assessed values (Table~\ref{tab:mechanism_partial_r2}).

% ---------- NEW TABLE: North 2022 (tables/stacked_north_did_2022.tex) ----------
\begin{table}[!htbp]\centering\small
\caption{The Same Design Applied to the North Suburbs' 2022 Algorithmic Reassessment}\label{tab:north_did_2022}
\resizebox{\textwidth}{!}{%
\begin{tabular}{lccccc}\toprule
 & (1) & (2) & (3) & (4) & (5) \\
Event & North 2022 & North 2022 & North 2022 & North 2022 & Chicago 2021 \\
Comparison triad(s) & South & South & South & South + Chicago & North + South \\
Sample window & 2021q1--22q4 & 2021q1--22q4 & 2020q1--22q4 & 2021q1--22q4 & 2020q1--21q4 \\
Specification & Price & Price + race & Price & Price & Price + race \\
\midrule
Average post effect: ln(price) slope &  0.033 &  0.042 &  0.033 &  0.045* &  0.220*** \\
 & (0.030) & (0.029) & (0.030) & (0.026) & (0.035) \\
%\quad wild bootstrap p (triad) & -- & -- & -- & [0.596] & -- \\
\addlinespace
Average post effect: minority slope & &  0.084* & & &  0.138*** \\
 & & (0.044) & & & (0.041) \\

%\quad wild bootstrap p (triad) & & -- & & & -- \\
\addlinespace
Joint pre-trend p: ln(price) slope & 0.452 & 0.489 & 0.037 & 0.069 & 0.165 \\
Observations &    50,548 &    50,548 &    78,105 &    75,476 &    78,449 \\
Town-by-quarter FE; town- and quarter-specific slopes & YES & YES & YES & YES & YES \\
\bottomrule\end{tabular}}

\begin{minipage}{0.98\textwidth}\footnotesize \textit{Notes:} Each column estimates Equation \eqref{eq:did_slopes} around the indicated event and reports the average of the four post-quarter Treat $\times$ Post $\times$ slope interactions; the reference period is the quarter before the event. Every window ends before the next reassessment of any triad in the sample. In columns (1)--(3) the comparison group is the South triad, which is not reassessed under the reformed algorithm until 2023 and is therefore a not-yet-treated control. Column (4) adds Chicago, which was already treated in 2021. Column (5) is a benchmark: the Chicago 2021 event estimated on a window of the same length. The outcome is ln(Assessed/Sale Price) using initial (mailed) values. Township-clustered standard errors in parentheses. *** p$<$0.01, ** p$<$0.05, * p$<$0.1 based on township-clustered inference.\end{minipage}
\end{table}

\newpage
\section{The South 2023 Reassessment and the Chicago--South Comparison}
\label{app:south}

Table~\ref{tab:south_did_2023} reports the estimates discussed in Section~\ref{sec:south_null}: the paper's event-study specification around the South triad's 2023 reassessment, on a 2022q1--2024q4 window, against Chicago, the North, and both pooled, with eight and four post-quarters.

\begin{table}[!htbp]\centering
\small
\caption{The South 2023 Algorithmic Reassessment: No Detectable Effect on Price Regressivity}\label{tab:south_did_2023}
\begin{tabular}{lccccc}\toprule
 & (1) & (2) & (3) & (4) & (5) \\
Comparison group & Chicago + North & North only & Chicago only & Chicago + North & Chicago + North \\
Specification & Price & Price & Price & Price + race & Price \\
Post quarters & 8 & 8 & 8 & 8 & 4 \\
\midrule
Average post effect on & & & & & \\
\quad ln(price) slope & -0.012 & -0.013 & -0.012 &  0.007 & -0.029 \\
 & (0.027) & (0.027) & (0.031) & (0.033) & (0.024) \\
\addlinespace
Joint pre-trend p-value & 0.335 & 0.013 & 0.716 & 0.161 & 0.341 \\
Observations &    93,757 &    63,717 &    65,142 &    93,757 &    65,657 \\
Town-by-quarter FE & YES & YES & YES & YES & YES \\
\bottomrule\end{tabular}

\begin{minipage}{0.98\textwidth}\footnotesize \textit{Notes:} Sample is 2022q1--2024q4 (2022q1--2023q4 in column 5), so the North suburbs' 2025 reassessment never enters the comparison window. The South triad is reassessed in 2023q1; the comparison triads were reassessed under the same reformed algorithm in 2021 (Chicago) and 2022 (North) and are therefore already treated, not untreated controls. Reference period is 2022q4. Each column reports the average of the post-quarter Treat $\times$ Post $\times$ ln(price) interactions from the paper's event-study specification, with township-clustered standard errors in parentheses. The joint pre-trend p-value tests that all pre-period interactions are zero. Column (4) additionally interacts neighborhood minority share with township, quarter, and event time. *** p$<$0.01, ** p$<$0.05, * p$<$0.1.\end{minipage}
\end{table}

Table~\ref{tab:chi_vs_south} tests the Chicago 2021 and South 2023 effects against each other in one stacked regression on windows of equal length, four pre- and four post-quarters each, with every fixed effect and slope baseline of Equation~\eqref{eq:did_slopes} interacted with the event. The difference is 0.22 in the price-only and 0.24 in the joint specification, with township-clustered standard errors of 0.03 ($p<0.01$) and $p = 0.11$ under the triad-level bootstrap. On this short window the Chicago 2021 pre-trend test rejects in the price-only column ($p=0.013$) but not in the joint column ($p=0.165$), which is the specification reported in Table~\ref{tab:five_events}.

\begin{table}[!htbp]\centering\small
\caption{Direct Test: Chicago 2021 vs.\ South 2023 on Matched Windows (Stacked Event Study)}\label{tab:chi_vs_south}
\begin{tabular}{lcc}\toprule
 & (1) Price & (2) Price + race \\
\midrule
Chicago 2021: average post effect on ln(price) slope &  0.189*** (0.033) &  0.220*** (0.035) \\
\quad triad wild bootstrap $p$ & [0.144] & [0.143] \\
South 2023: average post effect on ln(price) slope & -0.029 (0.024) & -0.018 (0.028) \\
\quad triad wild bootstrap $p$ & [0.116] & [0.128] \\
Difference (Chicago $-$ South) &  0.219*** (0.031) &  0.238*** (0.033) \\
\quad triad wild bootstrap $p$ & [0.110] & [0.105] \\
Joint pre-trend $p$: Chicago 2021 / South 2023 & 0.013 / 0.341 & 0.165 / 0.163 \\
Observations &   144,106 &   144,106 \\
Event-specific town-by-quarter FE; event-specific town and quarter slopes & YES & YES \\
\bottomrule\end{tabular}

\begin{minipage}{0.92\textwidth}\footnotesize \textit{Notes:} Two events are stacked: Chicago 2021 (treated Chicago, comparison North and South, window 2020q1--2021q4, reference 2020q4) and South 2023 (treated South, comparison Chicago and North, both already reassessed under the reformed algorithm, window 2022q1--2023q4, reference 2022q4). Each event has four pre- and four post-quarters, so the two effects are measured on windows of equal length. All fixed effects and slope baselines of Equation~\eqref{eq:did_slopes} are interacted with the event indicator, so each event's estimate equals its separate estimate (Tables~\ref{tab:early_reassess_did} and \ref{tab:south_did_2023}) and the difference can be tested directly. Township-clustered standard errors in parentheses (38 townships); triad-level wild-cluster bootstrap p-values (Webb weights, 9,999 replications, three clusters) in brackets. *** p$<$0.01, ** p$<$0.05, * p$<$0.1 based on township-clustered inference.\end{minipage}
\end{table}

\newpage
\section{Fairness Drift: Additional Evidence}
\label{app:drift}

This appendix supports Section~\ref{sec:market} with four exhibits: the price slope and its two components at every reassessment; the segment-divergent appreciation of Figure~\ref{fig:competing} measured on the same homes; the robustness of the tilt of fresh values against fast-appreciating neighborhoods in Figure~\ref{fig:tilt}; and the accuracy of each reassessment's values by neighborhood price tier.

\paragraph{The price slope and its components at every reassessment.} Table~\ref{tab:slope_decomposition} reports, for every reassessment in all three triads, the within-township price slope used throughout the paper and its between- and within-neighborhood components, each estimated three times on the same sales: with the previous reassessment's values on that year's sales, with the previous values applied to the reassessment year's sales, and with the reassessment's own values. The first difference is the drift the market produced while values were frozen; the second is the jump the reassessment made holding the sales fixed. Section~\ref{sec:competing} uses the Chicago and South rows; the North rows show a triad with no between-neighborhood regressivity for the market or the model to remove.

\begin{table}[!htbp]\centering\footnotesize
\caption{Where Regressivity Sits and What Moves It: Between- and Within-Neighborhood Price Slopes at Every Reassessment, 2018--2025}\label{tab:slope_decomposition}
\resizebox{\textwidth}{!}{%
\begin{tabular}{llcccccc}\toprule
 & & \multicolumn{3}{c}{Price slope on the same sales} & \multicolumn{2}{c}{Change in slope} & Mean abs. \\
\cmidrule(lr){3-5}\cmidrule(lr){6-7}
Reassessment & Valuation & Year $R-3$, & Year $R$, & Year $R$, & Drift & Reassessment & error (\%), \\
year $R$ & workflow & fresh values & previous values & fresh values & over cycle & jump & year $R-3$ / $R$ \\
\midrule
\multicolumn{8}{l}{\textit{Panel A. Between neighborhoods: slope of neighborhood mean ratios on neighborhood mean prices}} \\
Chicago 2018 & legacy & -0.348 (0.049) & -0.322 (0.079) & -0.189 (0.034) &  0.025 &  0.133 &  \\
Chicago 2021 & reformed algorithm & -0.189 (0.034) & -0.121 (0.050) &  0.101 (0.017) &  0.068 &  0.221 &  \\
Chicago 2024 & reformed algorithm &  0.101 (0.017) &  0.079 (0.020) &  0.004 (0.019) & -0.021 & -0.075 &  \\
North 2019 & earlier workflow & -0.059 (0.029) &  0.062 (0.023) &  0.056 (0.009) &  0.120 & -0.005 &  \\
North 2022 & reformed algorithm &  0.056 (0.009) &  0.037 (0.020) & -0.007 (0.016) & -0.020 & -0.044 &  \\
North 2025 & reformed algorithm & -0.007 (0.016) & -0.079 (0.010) &  0.006 (0.015) & -0.071 &  0.085 &  \\
South 2020 & earlier workflow & -0.226 (0.055) & -0.049 (0.057) & -0.022 (0.033) &  0.177 &  0.027 &  \\
South 2023 & reformed algorithm & -0.022 (0.033) &  0.047 (0.023) & -0.078 (0.042) &  0.069 & -0.125 &  \\
\midrule
\multicolumn{8}{l}{\textit{Panel B. Within neighborhoods: sale-level slope with neighborhood-by-quarter fixed effects}} \\
Chicago 2018 & legacy & -0.741 (0.063) & -0.764 (0.058) & -0.743 (0.070) & -0.023 &  0.022 &  \\
Chicago 2021 & reformed algorithm & -0.743 (0.070) & -0.703 (0.067) & -0.549 (0.047) &  0.039 &  0.155 &  \\
Chicago 2024 & reformed algorithm & -0.549 (0.047) & -0.614 (0.030) & -0.590 (0.059) & -0.065 &  0.024 &  \\
North 2019 & earlier workflow & -0.400 (0.041) & -0.430 (0.031) & -0.417 (0.037) & -0.030 &  0.013 &  \\
North 2022 & reformed algorithm & -0.417 (0.037) & -0.404 (0.036) & -0.342 (0.031) &  0.013 &  0.062 &  \\
North 2025 & reformed algorithm & -0.342 (0.031) & -0.427 (0.037) & -0.318 (0.037) & -0.085 &  0.108 &  \\
South 2020 & earlier workflow & -0.706 (0.043) & -0.690 (0.049) & -0.673 (0.051) &  0.015 &  0.017 &  \\
South 2023 & reformed algorithm & -0.673 (0.051) & -0.740 (0.041) & -0.726 (0.046) & -0.067 &  0.014 &  \\
\midrule
\multicolumn{8}{l}{\textit{Panel C. Within townships (the paper's slope): sale-level slope with township-by-quarter fixed effects}} \\
Chicago 2018 & legacy & -0.524 (0.064) & -0.525 (0.075) & -0.443 (0.063) & -0.000 &  0.082 & 46.4 / 40.0 \\
Chicago 2021 & reformed algorithm & -0.443 (0.063) & -0.394 (0.069) & -0.204 (0.046) &  0.049 &  0.190 & 40.0 / 28.0 \\
Chicago 2024 & reformed algorithm & -0.204 (0.046) & -0.263 (0.053) & -0.288 (0.037) & -0.059 & -0.025 & 28.0 / 28.7 \\
North 2019 & earlier workflow & -0.249 (0.032) & -0.233 (0.029) & -0.228 (0.026) &  0.016 &  0.006 & 17.5 / 17.1 \\
North 2022 & reformed algorithm & -0.228 (0.026) & -0.230 (0.024) & -0.210 (0.022) & -0.002 &  0.020 & 17.1 / 15.9 \\
North 2025 & reformed algorithm & -0.210 (0.022) & -0.295 (0.023) & -0.194 (0.024) & -0.086 &  0.101 & 15.9 / 16.5 \\
South 2020 & earlier workflow & -0.463 (0.056) & -0.379 (0.063) & -0.356 (0.059) &  0.083 &  0.023 & 35.7 / 26.7 \\
South 2023 & reformed algorithm & -0.356 (0.059) & -0.385 (0.051) & -0.432 (0.043) & -0.029 & -0.047 & 26.7 / 26.7 \\
\bottomrule\end{tabular}}

\begin{minipage}{0.98\textwidth}\footnotesize \textit{Notes:} Each row is one reassessment $R$ of one triad (legacy: previous administration; earlier workflow: the new administration's township-level reassessments before the reformed algorithm; reformed: the reformed algorithm, with the post-adjustment step in Chicago 2021 only). Every cell is a slope of ln(Assessed/Sale Price) on ln(sale price) estimated on the triad's sales in the stated year, with the assessed values indicated by the column: \emph{Year $R-3$, fresh values} uses the previous reassessment's values on that year's sales; \emph{Year $R$, previous values} applies each parcel's year-$(R-1)$ value to the year-$R$ sales, the slope the frozen values would have had; \emph{Year $R$, fresh values} uses the reassessment's own values on the same sales. Drift over cycle is the second column minus the first, the movement in the slope produced by prices and the mix of homes sold while values were frozen; reassessment jump is the third minus the second, the change the reassessment made holding the sales fixed. Panel A regresses neighborhood-year mean log ratios on neighborhood-year mean log prices with township fixed effects, weighted by the number of sales; it measures regressivity across neighborhoods of different price levels, the dimension along which neighborhoods appreciate at different rates. Panel B regresses sale-level log ratios on log prices with neighborhood-by-quarter fixed effects; it measures regressivity among homes in the same neighborhood, the dimension that depends on how well individual homes are valued, and its level includes the mechanical regressivity that idiosyncratic sale prices produce, so its changes are more informative than its level. Panel C is the within-township slope used throughout the paper, a blend of the two; its last column gives the mean absolute error of the fresh values in years $R-3$ and $R$ (100 $\times$ the mean of $|$Assessed/sale price $-1|$, winsorized at the 99.5th percentile). Township-clustered standard errors in parentheses. Higher (less negative) values are less regressive. The jumps in Panel C differ from the event-study estimates in Table~\ref{tab:five_events} in being price-only, using the reassessment year's sales, and needing no comparison triad; they agree closely where the comparison triads' slopes were stable (Chicago 2021: 0.19 against 0.19) and differ where those slopes were drifting (South 2023: $-0.05$ against $-0.03$).\end{minipage}
\end{table}

\begin{figure}[!htbp]
\centering
\includegraphics[width=0.92\textwidth]{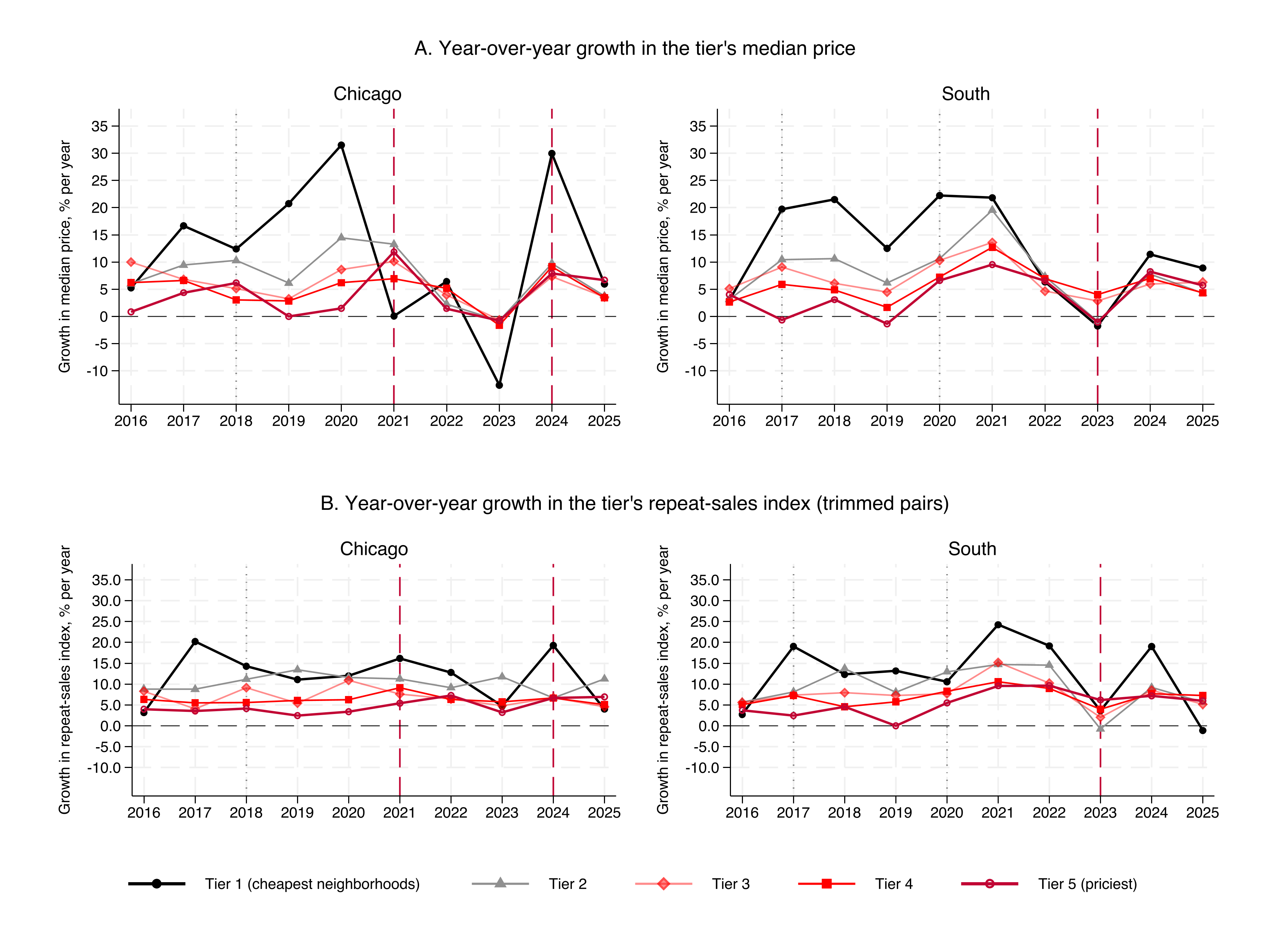}
\caption{Year-over-year growth of the tier price series in Figure~\ref{fig:competing}: median prices (Panel A) and the repeat-sales index (Panel B).}
\label{fig:tier_growth_repeat}
\begin{minipage}{0.96\textwidth}
\footnotesize \textit{Notes:} Same neighborhoods, tiers, and series as Figure~\ref{fig:competing}; each point is the percent change of the series from the previous year. Panel A is the growth of the tier's median sale price and is affected by the mix of homes sold; Panel B is the growth of the trimmed repeat-sales index for the same tiers and is not. Gray dotted vertical lines mark reassessments before the reformed algorithm, red dashed lines reassessments under it.
\end{minipage}
\end{figure}

\paragraph{Segment-divergent appreciation on the same homes.} Figure~\ref{fig:competing} follows five neighborhood price tiers with two series, the tier's median sale price and a repeat-sales index of the same homes. Figure~\ref{fig:tier_growth_repeat} plots the year-over-year growth of each. Two facts carry over to the text. In the South the cheapest tier outgrew the priciest by 15 points in 2021 and 10 in 2022 on the same homes, and by $-3$ in 2023, the first negative gap since the panel begins. In Chicago the same-home gap was 11 points in 2021, 6 in 2022, and 2 in 2023, so the divergence slowed but did not reverse before 2024.

\begin{figure}[h!]
\centering
\includegraphics[width=0.96\textwidth]{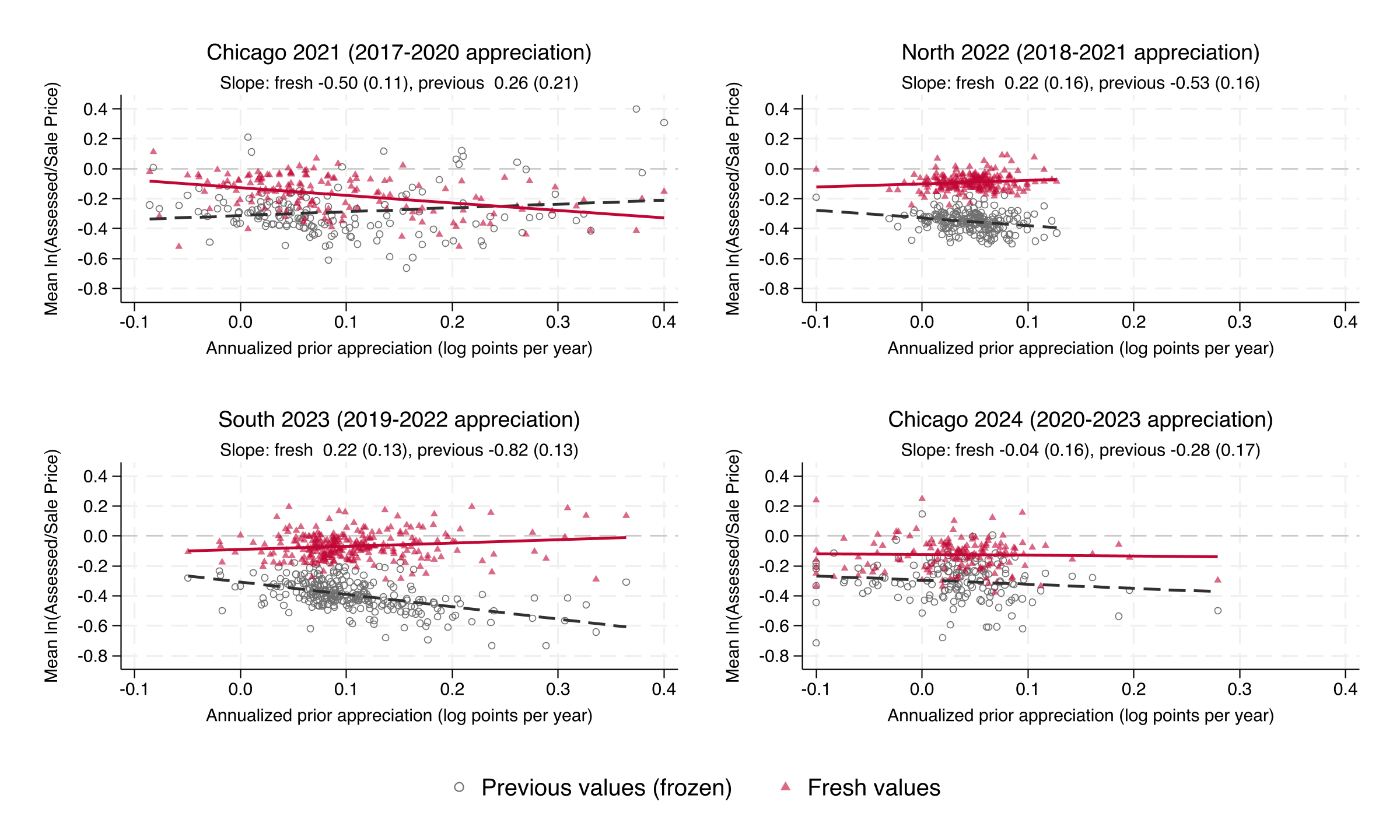}
\caption{Figure~\ref{fig:tilt} with a four-year lookback window.}
\label{fig:tilt_4yr}
\begin{minipage}{0.96\textwidth}
\footnotesize \textit{Notes:} As Figure~\ref{fig:tilt}, with prior appreciation measured over the four calendar years ending the year before the reassessment (divided by three) and neighborhoods required to have at least eight sales in the reassessment year, the year before, and the first year of that window (159, 212, 273, and 154 neighborhoods). Axes are wider than in Figure~\ref{fig:tilt}.
\end{minipage}
\end{figure}

\begin{figure}[h!]
\centering
\includegraphics[width=0.96\textwidth]{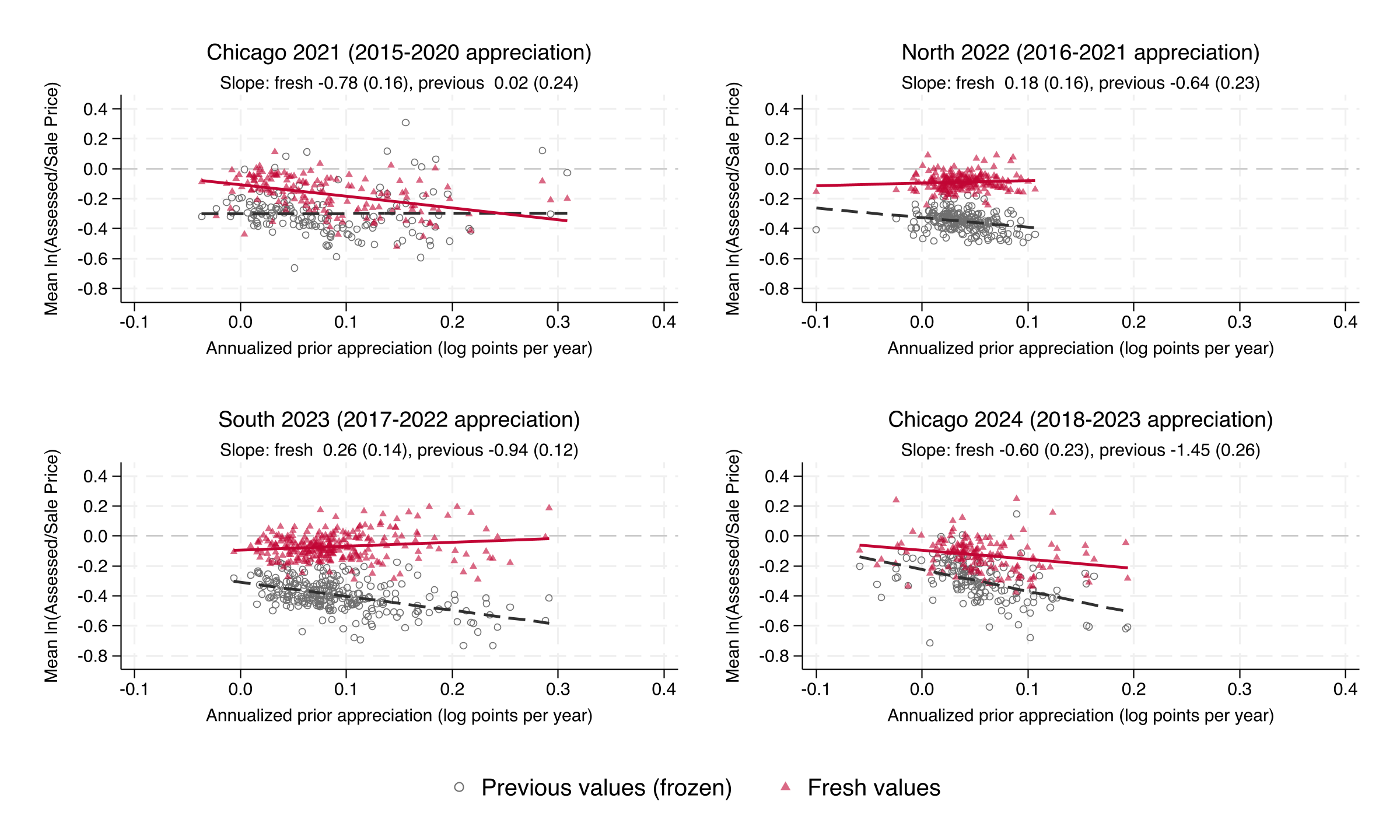}
\caption{Figure~\ref{fig:tilt} with a six-year lookback window.}
\label{fig:tilt_6yr}
\begin{minipage}{0.96\textwidth}
\footnotesize \textit{Notes:} As Figure~\ref{fig:tilt}, with prior appreciation measured over the six calendar years ending the year before the reassessment (divided by five) and neighborhoods required to have at least eight sales in the reassessment year, the year before, and the first year of that window (155, 211, 272, and 154 neighborhoods). Axes are wider than in Figure~\ref{fig:tilt}.
\end{minipage}
\end{figure}

\paragraph{The tilt by prior appreciation.} Figures~\ref{fig:tilt_4yr} and \ref{fig:tilt_6yr} draw the four- and six-year versions of Figure~\ref{fig:tilt} for the training data period. The South 2023 slope of the fresh values is positive at every training-data window.

\paragraph{Accuracy by neighborhood price tier.} Table~\ref{tab:error_by_tier} reports the mean absolute error of the values at each reassessment by neighborhood price tier, first with the previous year's values applied to that year's sales and then with the reassessment's own values on the same sales. It locates the accuracy gains that Section~\ref{sec:mechanism} converts into fairness gains. Chicago's 2021 gain of twelve points came almost entirely from the cheapest two tiers, where the legacy values had been worst (an error of 86 percent in the cheapest fifth of neighborhoods in 2018, against 16 in the priciest), and the 2024 values hold that gain. The South's 2023 values are between one and two points more accurate than the 2020 values in every tier, but the aggregate is unchanged because the cheapest tier's share of sales rose from 21 to 28 percent between the two years and its error is twice the average. The reformed algorithm's error in the cheapest tier is 42 to 45 percent wherever it has been deployed.

\begin{table}[!htbp]\centering\small
\caption{Mean Absolute Assessment Error by Neighborhood Price Tier at Each Reassessment}\label{tab:error_by_tier}
\begin{tabular}{llcccccc}\toprule
 & & \multicolumn{6}{c}{Mean absolute error (\%)} \\
\cmidrule(lr){3-8}
Reassessment & Values & Tier 1 (cheapest) & Tier 2 & Tier 3 & Tier 4 & Tier 5 (priciest) & All tiers \\
\midrule
Chicago 2018 & previous & 86.7 & 45.9 & 25.7 & 24.2 & 23.1 & 41.8 \\
Chicago 2018 & fresh    & 86.1 & 46.7 & 22.1 & 20.6 & 16.3 & 39.5 \\
Chicago 2021 & previous & 49.5 & 42.1 & 31.4 & 28.1 & 24.2 & 35.3 \\
Chicago 2021 & fresh    & 43.9 & 36.2 & 23.2 & 19.2 & 15.5 & 27.9 \\
Chicago 2024 & previous & 45.7 & 38.8 & 29.1 & 26.5 & 25.5 & 34.2 \\
Chicago 2024 & fresh    & 45.4 & 34.1 & 20.8 & 19.4 & 16.5 & 28.7 \\
\midrule
North 2019 & previous & 19.8 & 17.5 & 17.9 & 17.8 & 21.8 & 18.7 \\
North 2019 & fresh    & 17.0 & 15.1 & 16.7 & 17.5 & 22.7 & 17.0 \\
North 2022 & previous & 33.7 & 29.6 & 29.2 & 27.7 & 30.6 & 30.5 \\
North 2022 & fresh    & 16.8 & 14.7 & 15.6 & 15.5 & 18.0 & 15.9 \\
\midrule
South 2020 & previous & 44.8 & 31.1 & 25.0 & 20.2 & 19.5 & 28.4 \\
South 2020 & fresh    & 42.4 & 28.5 & 23.2 & 19.7 & 18.5 & 26.6 \\
South 2023 & previous & 44.4 & 36.5 & 34.5 & 31.7 & 29.6 & 36.4 \\
South 2023 & fresh    & 41.7 & 26.2 & 21.0 & 17.4 & 17.5 & 26.6 \\
\bottomrule\end{tabular}

\begin{minipage}{0.98\textwidth}\footnotesize \textit{Notes:} Main sales sample. Neighborhoods are ranked within triad into quintiles of their pooled 2016--2017 median sale price (at least eight sales), a ranking predetermined relative to every reassessment shown. For each reassessment year, the \emph{previous} row applies each parcel's prior-year assessed value to that year's sales and the \emph{fresh} row uses the reassessment's own values on the same sales. Mean absolute error is 100 $\times$ the mean of $|$Assessed/sale price $-1|$, winsorized at the 99.5th percentile. \emph{All tiers} is the sales-weighted mean over the five tiers, so it differs slightly from the error reported in Table~\ref{tab:slope_decomposition}, which includes neighborhoods too small to be ranked.\end{minipage}
\end{table}
\newpage
\section{Reconstructed Valuation Models}
\label{app:models}

The models compared in Table~\ref{tab:ml_vs_hedonic} are our own
reconstructions, fitted to training data the Cook County Assessor has
published. They are not the production model, and the simple benchmark is not
the legacy valuation system, which was never published. The exercise holds
data, sample, features, and evaluation sales fixed and varies only the learner,
so that the comparison isolates what a flexible model adds to a simple one when
the market stays on its trend and when it leaves it.

\paragraph{Design.}
The training data are the assessor's archived model samples: an end-2022
archive, with sales from 2014 through 2022, for the 2019--2022 evaluations, and
an archive dated April 6, 2023 for the 2023 evaluation.\footnote{Both archives
are recovered from the assessor's GitLab model repository; the production
model's code is public, but its fitted workflow and raw predictions are not,
and we have not been able to obtain them.} We keep the single-family classes
of the main analysis and sales of at least \$10,000. Parcels are assigned by a
hash of their identifier to fitting (60 percent), calibration (20 percent), and
evaluation (20 percent) roles; an evaluation parcel never enters fitting,
tuning, or calibration, including through its own earlier sales. Both learners
see the same twenty or so inputs: building and land area, age, rooms, baths,
and other structural characteristics, location (township, neighborhood,
latitude and longitude), sale month, and a linear calendar trend. No assessed
value, sale price, or race measure enters any model. The simple benchmark is an
additive hedonic model, cubic splines in each numeric characteristic plus
category indicators, fitted by ridge regression. The flexible model is a
gradient-boosted tree (LightGBM), the model family the reform adopted.
Hyperparameters were tuned on earlier years and on parcels disjoint from the
evaluation set, and fixed in a written protocol before any result was
inspected. For evaluation year $t$, both models train on sales from $t-3$
through $t-1$, with the training rows subsampled to a common count and the
identical rows supplied to each learner. Predictions are scored on the same
held-out sales, so differences between learners are paired. The accuracy
measure is the one used for actual assessments throughout the paper, the mean
absolute percentage error winsorized at the 99.5th percentile; differences are
estimated on the paired errors with township-clustered standard errors.
Table~\ref{tab:ml_vs_hedonic_rmse} repeats the comparison in log-price root
mean squared error, the criterion used for tuning.

\begin{table}[!htbp]\centering
\caption{Flexible vs.\ Simple Prediction, by Whether the Learned Price Trend Continued}\label{tab:ml_vs_hedonic}
\begin{tabular}{lcrccc}\toprule
 & Price trend after & & \multicolumn{3}{c}{Mean absolute error (\%)} \\
\cmidrule(lr){4-6}
Evaluation year & the valuation date & Sales & Hedonic & LightGBM & LightGBM $-$ Hedonic \\\midrule
2019 & continued &  2,821 &  29.1 &  28.0 & $-$1.06* (0.55) \\
2020 & continued &  2,993 &  27.3 &  27.6 & +0.30 (0.43) \\
2021 & continued &  2,764 &  25.0 &  24.8 & $-$0.15 (0.31) \\
2022 & continued &  2,740 &  25.8 &  25.5 & $-$0.33 (0.27) \\
2023 & \textbf{broke} &  1,751 &  25.9 &  26.6 & +0.69*** (0.22) \\
\bottomrule\end{tabular}

\begin{minipage}{0.92\textwidth}\footnotesize \textit{Notes:} All rows describe the South triad. The models are of our own specification, fitted to the assessor's publicly archived training data; neither is the deployed production model, and the hedonic benchmark is not the legacy system. Within each row the two learners use identical three-year training windows of equal sample size, identical features, and the same held-out evaluation sales. The 2019--2022 rows predict the price at each sale's date from the end-2022 archive; the 2023 row predicts a fixed January 1, 2023 valuation from the April 2023 archive on sales after April 7, 2023. Mean absolute error is the accuracy measure used for actual assessments elsewhere in the paper, applied to each model's predicted value: 100 $\times$ the mean of $| Assessed / sale price - 1|$, winsorized at the 99.5th percentile. The last column is the paired difference on the same sales, with township-clustered standard errors in parentheses (17 townships); a negative value favors LightGBM. *** p$<$0.01, ** p$<$0.05, * p$<$0.1.
\end{minipage}
\end{table}

\begin{table}[!htbp]\centering
\caption{Flexible vs.\ Simple Prediction in Log-Price Root Mean Squared Error, South}\label{tab:ml_vs_hedonic_rmse}
\begin{tabular}{lcrcc}\toprule
Evaluation year & Price trend & Sales & Hedonic & LightGBM \\\midrule
2019 & continued &  2,821 & 0.3764 & 0.3720 \\
2020 & continued &  2,993 & 0.3682 & 0.3796 \\
2021 & continued &  2,764 & 0.3724 & 0.3402 \\
2022 & continued &  2,740 & 0.3431 & 0.3390 \\
2023 & broke &  1,751 & 0.3355 & 0.3407 \\
\bottomrule\end{tabular}

\begin{minipage}{0.75\textwidth}\footnotesize \textit{Notes:} Same models, samples, evaluation sales, and prediction targets as Table \ref{tab:ml_vs_hedonic} (sale-date predictions for 2019--2022, a January 1 valuation for 2023); root mean squared error of ln(predicted value / sale price). Squared error weights the largest misses more heavily than the mean absolute error reported in the main text.\end{minipage}
\end{table}

\paragraph{Results.}
When the price trend continued, the two models are close: the flexible model's
error is lower in three of the four years, never by more than about one
percentage point on a base of 25 to 29, and only the 2019 difference is even
marginally distinguishable from zero. In 2023, when the trend broke, the
ranking reverses, and the hedonic model's 0.7-point advantage is the most
precisely estimated difference in the table. The magnitudes are modest. What
they show is that the accuracy premium a flexible learner earns over a simple
one is small on these data, and that it turns into a deficit in the one year
the market left the path the learner had been shown. The 2023 row also changes
the archive and the prediction target, so the reversal across years is not a
single controlled comparison; the controlled comparison is within 2023.

\paragraph{The 2023 test.}
The 2023 row is the one evaluation in which no input postdates the outcomes.
Models are trained on 2020--2022 sales from the April 2023 archive and
evaluated on the 1,751 South sales that occurred after the archive date. Two
targets are computed. For the fixed January 1, 2023 valuation the assessor
targets, the hedonic model is more accurate, as the table reports. For a
forecast to each sale's own date, the two models are statistically
indistinguishable. The reversal is therefore a statement about valuing every
home as of a date the market had just left, not about predicting prices at the
moment of sale. Withholding the last 90 days of sales before the valuation date
from fitting leaves the ranking unchanged.

\paragraph{Limits.}
The 2019--2022 evaluations use characteristics from the end-2022 archive,
which may include later corrections, so they are retrospective backtests
rather than replays of the assessor's information set. The models omit
production features and post-processing and use a shorter training history
than the production model. In a partition that withholds entire neighborhoods
from fitting, the tree predicts less accurately than the hedonic benchmark, so
its advantage within a market does not extend to unseen locations. None of
this identifies the causal effect of the production system. The reconstruction
shows how a flexible and a simple learner behave on the same data when the
market continues and when it turns.

\end{document}